\documentclass[11pt]{article}
\usepackage[left=1.5cm,right=1.5cm]{geometry}
\usepackage[numbers,square]{natbib}
\usepackage[T1]{fontenc}
\usepackage{tikz-cd}
\usepackage[british]{babel}
\usepackage{layout}
\usepackage{amsmath}
\usepackage{amssymb,amstext, amsthm, simplewick, amsfonts}
\usepackage{framed}
\usepackage{amsfonts}
\usepackage{color} 
\usepackage{braket}
\usepackage{multicol} 
\usepackage{epsfig}
\usepackage{cancel}
\usepackage{caption}
\usepackage{epsf}
\usepackage{feynmf} 
\usepackage{frontespizio}
\usepackage{rotating}
\usepackage{subfigure}
\usepackage{pstricks}
\usepackage{type1ec}
\usepackage{lettrine}
\usepackage{bbold}
\usepackage{calligra}
\usepackage{tikz}
\usepackage{float}
\usepackage{subfigure}
\usepackage{slashed}
\usepackage{mathrsfs}
\usepackage{curve2e}
\usepackage{setspace}
\usepackage{indentfirst}
\usepackage{emptypage}
\usepackage{relsize}
\usepackage{mathrsfs}
\usepackage{stackengine}
\usepackage{calc}
\usepackage{hyperref}
\hypersetup{
	colorlinks,
	citecolor=green,
	filecolor=black,
	linkcolor=blue,
	urlcolor=black
}

\let\ul=\underline
\newcommand{\3}[1]{C_{
		\ifthenelse{\equal{\ThreePt}{\empty}}{#1}{
			\ifthenelse{\equal{#1}{\empty}}{\ThreePt}{\ThreePt,#1}}}}
\newcommand{\redef}[1]{{C'}_{
		\ifthenelse{\equal{\ThreePt}{\empty}}{#1}{
			\ifthenelse{\equal{#1}{\empty}}{\ThreePt}{\ThreePt,#1}}}}
\newcommand{\ren}[1]{C_{
		\ifthenelse{\equal{\ThreePt}{\empty}}{#1}{
			\ifthenelse{\equal{#1}{\empty}}{\ThreePt}{\ThreePt,#1}}}}
\newcommand{\sd}[1]{D_{
		\ifthenelse{\equal{\ThreePt}{\empty}}{#1}{
			\ifthenelse{\equal{#1}{\empty}}{\ThreePt}{\ThreePt,#1}}}}

\numberwithin{equation}{section} 
\makeatletter
\@addtoreset{equation}{section}
\newcommand{\bea}{\begin{eqnarray}}
\newcommand{\eea}{\end{eqnarray}}
\makeatother
\newcommand{\beqa}{\begin{eqnarray}}
	\newcommand{\eeqa}{\end{eqnarray}}

\newcommand{\sm}{\mathcal{S}}
\newcommand{\nn}{\nonumber}
\let\a=\alpha   \let\b=\beta   \let\g=\gamma   \let\d=\delta
\let\e=\epsilon    \let\h=\eta     
    \let\k=\kappa  \let\l=\lambda  \let\m=\mu
\let\n=\nu           \let\p=\pi      \let\r=\rho
\let\s=\sigma        
\let\c=\chi         \let\w=\omega
\def\pd{\partial}
\def\si{\sigma}
\newcommand{\bann}{\begin{eqnarray*}}
\newcommand{\eann}{\end{eqnarray*}}
\def\eps{\epsilon}

\newcommand{\bmi}{\begin{minipage}}
\newcommand{\emi}{\end{minipage}}

\newcommand{\mD}{\mathcal{D}}
\let\G=\Gamma  \let\D=\Delta

\newcommand{\la}{\langle}
\newcommand{\ra}{\rangle}
\newcommand{\bs}[1]{\boldsymbol{#1}}
\newcommand{\sq}{\square}

\newcommand{\be}{\begin{equation}}
	\newcommand{\ee}{\end{equation}}
\newcommand{\sdfrac}[2]{\mbox{\small$\displaystyle\frac{#1}{#2}$}}

\newcommand{\beq}{\begin{equation}}
	\newcommand{\eeq}{\end{equation}}
\newcommand{\figref}[1]{Fig.~\ref{#1}}			% for figures
\newcommand{\secref}[1]{Section~\ref{#1}}		% for sections
\newcommand{\Tr}{\text{Tr}}

\newcommand{\ThreePt}{\empty}
\usepackage{accents}
\makeatletter
\newcommand{\xLine}[2][]{\ext@arrow 0359\Rightarrowfill@{#1}{#2}}
\makeatother
\xdefinecolor{darkgreen}{RGB}{102, 204, 70}
\xdefinecolor{darkblue}{RGB}{0, 0, 153}
\usepackage{amssymb}
\usepackage{pifont}
\newcommand{\bes}{\begin{subequations}}
	\newcommand{\ees}{\end{subequations}}

\usepackage{tikz-feynman}
\usetikzlibrary{arrows}
\usetikzlibrary{snakes}
\usetikzlibrary{decorations,decorations.pathmorphing,decorations.markings}

\tikzset{graviton/.style={decorate, decoration={snake}, double}}
\tikzset{gluon/.style={decorate, decoration={coil, segment length=8, aspect=1.2, amplitude=3 }}}

\begin{document}
	\begin{center}
		\vspace{1.5cm}
	\begin{center}
	\vspace{1.5cm}
	{\Large \bf  The Gravitational Form Factors of Hadrons\\
	 from CFT in Momentum Space and the  Dilaton in Perturbative QCD\\}
\vspace{0.3cm} 
	
	\vspace{0.3cm}
	
		\vspace{1cm}
		{\bf Claudio Corian\`o, Stefano Lionetti,  Dario Melle and Riccardo Tommasi\\}
		\vspace{1cm}
{\it  Dipartimento di Matematica e Fisica, Universit\`{a} del Salento \\
and INFN Sezione di Lecce,Via Arnesano 73100 Lecce, Italy\\
National Center for HPC, Big Data and Quantum Computing\\}
%{\it and  \\}

\vspace{0.5cm}

		%\email{emil@lanl.gov}
	\end{center}
	%\date{\version}
	\begin{abstract}
We analyze the hard scattering amplitude of the gravitational form factors (GFFs) of hadrons at one-loop, in relation to their conformal field theory (CFT) description, within the framework of QCD factorization for hard exclusive processes at large momentum transfers. 
These form factors play an essential role in studying the quark and gluon angular momentum of the hadrons due to their relation to the Mellin moments of the Deeply Virtual Compton Scattering (DVCS) invariant amplitudes. 
Our analysis is performed using a diffeomorphism invariant approach, applying the formalism of the 
gravitational effective action and conformal symmetry in momentum space for the discussion of the quark and gluon contributions.  
The interpolating correlator in the hard scattering of any GFF is the non-Abelian $TJJ$ (stress-energy/gluon/gluon) 3-point function at $O(\alpha_s^2)$, revealing an effective dilaton interaction in the $t$-channel due to the trace anomaly, in the form of a massless anomaly pole in the QCD hard scattering. We investigate the role of quarks, gauge-fixing and ghost contributions in the reconstruction of the hard scattering amplitude mediated by this interaction, performed in terms of its transverse traceless, longitudinal, and trace decomposition, as identified from CFT in momentum space (CFT$_p$). We present a convenient parameterization of the hard scattering amplitude relevant for future experimental investigations of the DVCS/GFF amplitudes at the Electron-Ion Collider at BNL.\end{abstract}

	\end{center}
	\newpage
\section{Introduction}
Conformal symmetry imposes strong constraints on 3-point functions of scalar and tensor correlators. This symmetry allows to establish a connection between the conventional free-field theory realization of the same correlators in Lagrangian conformal field theories (CFT), and their expected general tensorial structure. This is identified by Lorentz covariance through the use of Conformal Ward identities (CWIs) and the operator product expansion in the abstract CFT formulation. When this approach is applied in momentum space, here denoted as $CFT_p$ \cite{Coriano:2013jba,Bzowski:2013sza,Bzowski:2015pba,Coriano:2020ees}, it allows to establish a link between the general expression of a certain correlator determined by solving the CWIs and its ordinary Feynman expansion, i.e. the corresponding amplitudes \cite{Coriano:2023hts,Coriano:2018bsy} (see \cite{Coriano:2020ees} for a review of the methods and \cite{Caloro:2022zuy,Bzowski:2020kfw}\cite{Marotta:2022jrp,Jain:2021wyn,Coriano:2024ssu} for recent extensions to higher point functions or parity-odd correlators).\\
The free-field theory expressions of the correlators are generally characterized by the presence of anomalous dimensions in their renormalization group evolution as well as by explicit violations of the conformal symmetry associated with dimensionful scales, such as mass parameters. Obviously, the latter are naturally absent in the abstract CFT solution. In the QCD case, however, the renormalization procedure is responsible for the inclusion of a renormalization scale and for the generation of a nonzero trace, the trace (scale) anomaly \cite{Capper:1975ig,Duff:1977ay,Christensen:1978gi,Duff:1993wm}, breaking conformal symmetry.\\
 In a mass-independent regularization scheme such as dimensional regularization (DR), the corresponding definition of the QED/QCD $\beta$ functions are such that the anomaly contribution, which is unrelated to the mass-dependent corrections of the theories, separate. This result can be inferred from explicit computations 
 in the Abelian \cite{Giannotti:2008cv,Armillis:2009pq} and non-Abelian cases \cite{Armillis:2010qk} of correlators containing the stress energy tensor.\\
  In a CFT approach, the CWIs determine the solution of a generic 3-point function in terms of few constants that are matched  in free field theory by the perturbative computation of the same correlator. The analysis of such correlators in momentum space is particularly relevant in the presence of chiral and conformal (trace) anomalies, where the emergence of an anomaly contribution is directly and uniquely associated with the exchange of an anomaly pole.  
 For both older and more recent perturbative analyses of trace anomalies in QCD and operatorial mixing, we refer to \cite{Freedman:1974gs, Tanaka:2022wrr, Hatta:2018sqd, Ahmed:2022adh}. \\
 This procedure has been applied to various correlators, including those of both even and odd parity. In the  case of the axial-vector/vector/vector vertex, where a chiral anomaly is present, the solution to the anomalous CWIs constraining the full correlator can be obtained by introducing a chiral anomaly pole in the spectrum, highlighting the central role of this contribution \cite{Coriano:2025ceu}. A similar approach has been explored in the context of other parity-odd trace anomalies \cite{Coriano:2023hts, Coriano:2023gxa, Coriano:2023cvf}.\\
For parity even trace anomalies, the pole can be identified as a dilaton state and, as we are going to show elsewhere, is characterised by the presence of a sum rule in the anomaly form factor in which it appears.\\
 At lower momentum transfers, the effects of anomalous scale breaking in hadronic matrix elements involving a $T^{\mu\nu}$ insertion must be addressed using effective models that account for chiral symmetry breaking effects \cite{Kharzeev:2021qkd} and the QCD instanton vacuum \cite{Liu:2024jno,Liu:2024vkj}. In contrast, at larger momentum transfers, the hard scattering can be analyzed using the formal approach we present.

 \subsection{The TJJ and conformal symmetry in the sector decomposition}
The correlator we are going to examine in its off-shell expansion is the non-Abelian $TJJ$, which has been previously discussed in momentum space by CFT methods in \cite{Bzowski:2018fql} and in perturbative QCD (pQCD) for on-shell external gluons in \cite{Armillis:2010qk}. Here, $J$ represents a non-Abelian vector current in four dimensions ($d=4$), and $T$ is the gauge-fixed stress-energy tensor of QCD. To include this interaction in the generalized form factors (GFFs) of hadrons, we need to extend the on-shell analysis presented in \cite{Armillis:2010qk}. This extension will be achieved by closely following the formalism of CFT$_p$, as developed in \cite{Coriano:2013jba, Bzowski:2013sza}, subsequent to the initial analysis of this correlator in \cite{Armillis:2010qk}.\\
In the Abelian case, an initial parameterization of this correlator was discussed in \cite{Giannotti:2008cv, Armillis:2009pq}, where it was expressed in terms of 13 form factors (the F-basis), with only two of them containing kinematical poles. This parameterization, which we will review and apply to the quark contributions to illustrate how the anomaly emerges in dimensional regularization (DR), differs significantly from the minimal approach proposed in CFT\(_p\) for characterizing the transverse-traceless (tt) sector \cite{Bzowski:2013sza} of a tensor correlator that includes one or more stress-energy tensors.
The general structure of CFT$_p$ introduced in \cite{Bzowski:2013sza} was matched to free field theory realizations in \cite{Coriano:2018bbe, Coriano:2018zdo}. \\
The reduction in the number of form factors in this (longitudinal/transverse)  parameterization, referred to as the LT-basis, is achieved by symmetrically incorporating all three momenta of the vertex within the transverse traceless (tt) sector. 
The F-basis mentioned above turns useful in QED, but covers only the quark sector in QCD, since in QCD the WIs of QED are replaced by Slavnov-Taylor idendities (STIs). 
In the case of Abelian theories, the relation between the F-basis and the LT-basis was discussed in \cite{Coriano:2018bbe}. \\
In CFT\(_p\), the correlator is constructed in the LT-basis around the  (tt) sector by reorganizing the CWIs such that they take the form of second-order differential equations (primary equations) for the form factors in this sector, and first-order (secondary equations) for the remaining sectors. The conformal anomaly is then directly linked to the inclusion of a single counterterm (\(\sim F^2\)) in the tensorial expansion, leading to the generation of a trace sector. \\
In this approach, the form factors in the (tt) sector are found explicitly, with the solution expressed in terms of a set of integration constants, which are then refined using secondary conformal constraints from the remaining ordinary Ward identities. We will clarify these aspects in the QCD context, which differs from the standard CFT$_p$ approach due to the presence of virtual gluons in quantum corrections. \\
While CFT$_p$ can handle correlators involving non-Abelian currents, this method cannot be directly applied to perturbative QCD (pQCD) due to gauge-fixing, which breaks the conformal symmetry of the QCD action. However, we will show how this limitation can be addressed by decomposing both quark and gluon contributions into the LT-basis, revealing additional form factors from the exchange of virtual gluons in the longitudinal sectors. \\
To prevent any confusion, throughout this work, {\em "sector decomposition"} refers specifically to the decomposition of a tensor into its longitudinal, transverse traceless, and trace components using projectors. This definition is unrelated to the decomposition of a Feynman integral into distinct momentum integration regions. \\
Our analysis will concentrate on the tensor structure of the hard scattering, derived from the CFT\(_p\) parameterization, guided by the standard CWIs satisfied by the quark contributions, which follows closely the Abelian case. The gluon sector, instead, is incorporated through a direct perturbative analysis. \\
In principle, also the reconstruction of the gluon sector within CFT$_p$ can be achieved by utilizing the broken CWIs of pQCD in a general framework, akin to the treatment of any conformal 3-point function. This approach should lead to the same decomposition presented in this work. However, for clarity and simplicity, we choose to organize the decomposition by employing a direct perturbative expansion, where gluon contributions are decomposed in the LT-basis and then added to the quark sector. 
The approach we propose allows for the direct isolation of all the form factors and, in particular, of the anomaly form factor and the QCD conformal anomaly present in the hard scattering process. This sets the stage for a future comprehensive analysis of the gravitational form factors (GFF) of the pion and proton at the hadronic level, within the framework of QCD factorization for exclusive processes, which will be explored in a forthcoming work.\\
 From our perspective, the parameterization of the hard scattering that we present provides the optimal framework for characterizing the complete hadronic matrix elements, essential for the experimental investigation of potential anomaly effects, since anomalies, their poles and conformal symmetry are closely related.\\
 The approach also allows for a straightforward formulation of higher-order perturbative corrections. Given that the longitudinal-transverse (LT) sector decomposition is always valid and minimal, it proves to be particularly valuable. In this formulation, the quark and gluon sectors combine to produce a gauge-invariant anomaly pole as a signature of the exchange of a dilaton pole due to the anomaly. \\
 The introduction of mass corrections modifies the anomaly form factor, transforming 
 it from a pole to a cut. However, the true significance of the anomaly lies not in the presence of the pole or the cut, but in the existence of a sum rule. This sum rule, satisfied by the integral of the spectral density of the form factor, precisely reflects the anomaly.

\subsection{Conformal Anomaly poles in Abelian and non-Abelian theories}
Dilaton-like $t$-channel exchanges in the $TJJ$ case, were originally investigated perturbatively in QED \cite{Giannotti:2008cv, Armillis:2009pq} and QCD \cite{Armillis:2010qk}, as previously mentioned, to characterize the coupling of gravity to the fields of the Standard Model via anomaly poles. A discussion of these features in the neutral current sector of the Standard Model can be found in \cite{Coriano:2011ti, 2013JHEP...06..077C}, and in supersymmetric models in \cite{Coriano:2014gja}. In the last case, for instance, these analysis provided explicit proofs of the existence of supersymmetric sum rules for the anomaly form factors of the superconformal anomaly supermultiplet, involving both chiral and conformal anomalies, as well as for the Konishi anomaly. \\
A sum rule for the $TJJ$ in QED was derived in \cite{Giannotti:2008cv}. As shown in \cite{Coriano:2023gxa}, such poles are also present in the form factor of the gravitational chiral anomaly when not only fermion currents, but also spin-1 Chern-Simons currents, are involved. This anomaly was originally discussed in \cite{Dolgov:1987yp, Dolgov:1988qx} and more recently in \cite{Agullo:2018nfv}.\\
 Recent analyses of correlators involving insertions of stress-energy tensors in conformal field theory (CFT), along with the study of conformal anomaly actions, have revealed that the corresponding interaction vertices can be expanded in terms of the dimensionless parameter $R\Box^{-1}$ in a general gravitational background. In QCD, coupling the stress-energy tensor to an external metric allows us to leverage results from the formalism of the conformal anomaly effective action, which is particularly useful for addressing conformal constraints. Here, $R$ represents the scalar curvature, while the nonlocal $1/\Box$ interaction that we identify characterizes the exchange of an effective dilaton-like state in the $t$-channel of the $TJJ$ vertex.\\
Unlike the phenomenological dilaton effective action, which includes a conformal breaking scale $(\Lambda)$ and features a coupling of the form $(\chi/\Lambda) FF$, where $\chi$ is a dilaton field locally coupled to the anomaly, the nonlocal action is derived directly from the ultraviolet behavior of the diagrams we compute. These diagrams define a one-particle irreducible (1PI) effective action, from which the anomaly form factor and its associated anomaly pole can be extracted.
 
\subsection{The non-Abelian $TJJ$ in exclusive processes, CFT$_p$ and the anomaly effective action}
The $TJJ$  interaction enters at next-to-leading order (NLO) in the strong coupling constant \(\alpha_s\) in the gravitational form factors (GFFs) of hadrons through factorization in pQCD. This is a key component of hard scattering processes and induces a trace anomaly. 
%In this work \(T\), specifically, will denote the QCD stress-energy tensor and \(J\) the non-Abelian gluon %current.\\
We will examine this interaction in relation to the general properties of such exchanges, as investigated in pQCD, within the framework of CFT$_p$.
The phenomenological applications of the results of this analysis will be discussed in a separate work. \\
We aim to disentangle the dilaton-like effective degree which is part of the 1PI effective vertex, and show how effective nonlocal actions can be formulated for such anomalous interactions by a suitable analysis of the hard scattering.\\
Since the stress-energy tensor can be derived by varying the QCD partition function with respect to an external gravitational field, we will formulate the CWIs for this correlator using the general framework of the gravitational effective action for a non-Abelian theory, expanded around flat spacetime. The conformal anomaly interaction that we are focusing on, is a specific case within this broader formulation. To our knowledge, this approach has not been previously applied in the context of QCD, yet it proves highly effective in analyzing the constraints governing amplitudes that include a stress-energy tensor.\\
Concerning the symmetry constraints to be imposed, the correlator is governed by a less restrictive Slavnov-Taylor identity (STI) compared to an ordinary WI. Notably, the general nonperturbative CFT solutions for the \(TJJ\) correlator for both non-Abelian and Abelian currents are quite similar for non-Lagrangian theories \cite{Bzowski:2018fql}. However, in the context of QCD, these solutions are applicable only in the quark sector, as the gauge-fixing of the QCD action alters the gluon sector and breaks its conformal symmetry at $d=4$. Indeed, the hierarchical equations satisfied by the quark sector take the form of of ordinary CWIs, derivable by a partition function where the gluons are treated as external fields. A similar approach in the gluon sector is more involved and requires the implementation of the broken CWIs discussed in previous works \cite{Sarkar:1974xh, Nielsen:1975ph} \cite{Braun:2018mxm}, which we will not pursue. This is not strictly necessary in our current analysis, being our work focused on the conformal limit of the interaction.

\subsection{Content of this work}
Our work is organized as follows: In \secref{gen1}, after a brief overview of the role of the GFFs of hadrons in the context of QCD factorization and their relation to the deeply virtual Compton scattering (DVCS) amplitude, we discuss the partonic \($TJJ$\) interaction as it emerges in the hard scattering of the GFFs. In \secref{three1} the correlator is discussed in full generality within the formalism of the anomaly effective action, computed in the presence of both gravitational and gauge backgrounds, of which this correlator is part. \\
In \secref{four1} we discuss the symmetries of the generating functional in the quark sector, which are essential for the formal derivation of the WIs and CWIs constraints at one loop in this sector. In this section the gluons are treated as external classical fields. In \secref{BRSTX} we turn to the analysis of the symmetries of the complete QCD partition function, deriving the relevant STIs satisfied by the correlator. 
In \secref{seven}, we describe the sector decomposition, which naturally leads to the emergence of the dilaton pole. This is followed by an analysis of conformal constraints in this sector and their connection to perturbative results in \secref{eight}. The gluon contribution to the correlator is discussed in \secref{nine}, while \secref{ten} presents the final parameterization of the perturbative correlator for both massless and massive fermions. \\
In \secref{eleven}, we briefly review the on-shell case and explain how the dilaton anomaly pole arises in perturbation theory due to the renormalization of trace constraints in $d$ dimensions, taking the limit $d \to 4$. This is illustrated in the quark sector, along with a verification of a sum rule for the spectral density of the dilaton form factor, derived from the explicit expression of the on-shell correlator. \\
In \secref{dec22}, we conclude with a discussion of future extensions to hadronic systems, particularly the proton and pion. Several technical details are provided in the appendices, including explicit expressions for form factors in the massless case. Appendices A and B cover the derivation of Slavnov-Taylor identities and list the relevant Feynman rules. Appendix C summarizes the perturbative computation of the quark sector, using projectors to extract the form factors introduced in \secref{seven}. Appendices D, E, and F detail the implementation of the conformal procedure in $CFT_p$ and its modifications due to QCD gauge fixing. Appendices G and H present the transverse traceless form factors for both massless and massive cases, along with the fundamental integrals appearing in their definitions.
\section{The $TJJ$ and the gravitational form factors of hadrons }
\label{gen1}
The experimental investigation of the GFFs of the proton and the pion provide nonperturbative insights into the coupling of these hadrons to the energy-momentum tensor of QCD, revealing essential information about the distribution of their energy, spin, pressure, and shear forces. \\
GFFs are expanded in terms of the matrix elements of the EMT between hadron states. Quantum 
corrections break the conformal symmetry due to the trace anomaly of the stress energy tensor
\cite{Adler:1976zt,Nielsen:1977sy,Collins:1976yq}
\be\label{anom1}
    	\hat{T}_{\!\mu}^{\,\mu} \equiv 
	{\beta(g)}\;F^{a,\mu\nu}{F^{a,}}_{\!\!\mu\nu}
    	+(1+\gamma_m)\sum_qm_q\bar\psi_q\psi_q \;,\ee
where $\beta(g)$ is the $\beta$-function of QCD and $\gamma_m$ is 
the anomalous dimension of the mass operator. The trace anomaly
has the same form in QED.\\
These expansions reveal how the internal structure of the proton is related to its energy, momentum, and stress distributions. The matrix elements of the EMT for a spin 1/2 hadron with momentum \(P\) can be expressed in terms of the GFFs as

\begin{align}
    \langle p^\prime,s^\prime| T_{\mu\nu}(0) |p,s\rangle
    = \bar u{ }^\prime\biggl[
      A(t)\,\frac{\gamma_{\{\mu} P_{\nu\}}}{2}
    + B(t)\,\frac{i\,P_{\{\mu}\sigma_{\nu\}\rho}\Delta^\rho}{4 M}
    + D(t)\,\frac{\Delta_\mu\Delta_\nu-g_{\mu\nu}\Delta^2}{4M}
    + {M}\,\sum_{\hat{a}} {\bar c}^{\hat{a}}(t)\,g_{\mu\nu} \biggr]u\,
    \label{fund1}
\end{align}
where $u(p)$ and $\overline{u}(p')$ are the proton spinors, $ P = (p + p')/2$ is the average momentum, 
$\Delta = p' - p$ is the momentum transfer,  $t = \Delta^2$, and  $M$ is the mass of the proton. $\gamma^{(\mu} P^{\nu)}$denotes the symmetric combination $\gamma^\mu P^\nu + \gamma^\nu P^\mu$.\\
Using the Gordon identity, the separate components related to quarks ($\hat{a}\equiv q$) and gluons 
($\hat{a}\equiv g$)
 can be expressed in the form
\bea
    \la p^\prime,s^\prime| T_{\mu\nu}^{\hat{a}}(0) |p,s\rangle
    = \bar u^\prime\biggl[
      A^{\hat{a}}(t)\,\frac{P_\mu P_\nu}{M}
    + J^{\hat{a}}(t)\ \frac{i\,P_{\{\mu}\sigma_{\nu\}\rho}\Delta^\rho}{2 M}
    + D^{\hat{a}}(t)\,\frac{\Delta_\mu\Delta_\nu-g_{\mu\nu}\Delta^2}{4 M}
    +{M}\,{\bar c}^{\hat{a}}(t)g_{\mu\nu} \biggr]u \nonumber \\
    \label{Eq:EMT-FFs-spin-12-alternative} \eea
The two representations are equivalent, and the form factors are
related as $A^a(t)+B^a(t) = 2\,J^a(t)$. 
For a spin 0 hadron it takes the form 
\bea
\label{fund2}
	\la p^{\,\prime\,}|\hat{T}_{\mu\nu}(0)|p\ra = 
	\biggl[2 P_\mu P_\nu\, A(t) + 
	\frac12(\Delta_\mu\Delta_\nu - g_{\mu\nu} \Delta^2)\,D(t)
	+ {2\ m^2}\,{\bar c}(t)\,g_{\mu\nu} \biggr] \,.
\eea

The stress eenrgy tensor (see \cite{Polyakov:2018zvc} for an overview) can be investigated by studying auxiliary processes involving generalized parton distribution functions \cite{Ji:1996ek,Radyushkin:1996nd,Radyushkin:1996ru,
Ji:1996nm} \cite{Collins:1996fb,Radyushkin:1997ki,Vanderhaeghen:1998uc} in hard exclusive reactions. 
The process provides information about  the mass and the spin of hadron 
\cite{Kobzarev:1962wt,Pagels:1966zza,Ji:1998pc,Radyushkin:2000uy,Goeke:2001tz,Diehl:2003ny,Belitsky:2005qn}. 
A perturbative analysis in the context of QCD factorization has been presented in \cite{Tong:2022zax}. \\
Together with the $D$-term \cite{Polyakov:1999gs} these form factors allow to gain information on the tomography of the proton. 
The GFFs $A(t)$, $B(t)$, and $D(t)$ in \eqref{fund1} have specific physical interpretations. For example, $A(t)$ represents the distribution of the proton's momentum among its constituents (quarks and gluons). In the forward limit (\(t = 0\)), \(A(0)\) sums to 1, indicating the total momentum of the proton.\\
$B(t)$ is associated with the distribution of the proton's angular momentum. 
The form factor $B(t)$ is present only for hadrons with $J>0$ and satisfies at zero-momentum transfer the constraint $B(0) = 0$, indicating the vanishing of the anomalous gravitomagnetic moment in the same kinematical limit. The constraints at $t=0$ derive from the fact that these form factors are related to the generators of the Poincar\`e group and henceforth to the mass and spin of the hadron. \\
On the other end, the form factor $D(t)$ at zero-momentum transfer is unconstrained, and identifies the $D$-term, typical of any hadron. In contrast to $A(t)$ and $B(t)$ which are determined at $t=0$ by the mass and spin of the particles, the $D$-term is related to the stress tensor and internal forces.
The combination \(A(t) + B(t)\) at \(t = 0\) gives the total angular momentum carried by the quarks and gluons, as described by the sum rule  \cite{Ji:1996ek}
    \beq
    J_q + J_g = \frac{1}{2} [A_q(0) + B_q(0)] + \frac{1}{2} [A_g(0) + B_g(0)] = \frac{1}{2}.
    \eeq

\begin{figure}
    \centering
    \begin{tikzpicture}
\begin{feynman}
    \vertex (i1);
    \vertex[right=1cm of i1] (a1);
    \vertex[right=1cm of a1] (b1);
    \vertex[right=3cm of a1] (o1) ;
    \vertex[below=1cm of i1] (i2) ;
    \vertex[right=1cm of i2] (a2);
    \vertex[right=3cm of a2] (o2) ;

    %\vertex[below=2cm of i3] (t1);
    \vertex[above=1cm of a1] (t2);
    \vertex[above=2cm of a2] (t3);
    
    %\vertex[above right =1.3cm of t2] (t1);
    \vertex[above=1.5cm of b1] (i3) {$T^{\mu\nu}$};
    
    \diagram* { 
    (i1)  -- [fermion] (a1) -- [fermion] (b1)   --[fermion] (o1),
    
  (i2)  -- [anti fermion] (a2)  --[anti fermion] (o2),

  (i3)  -- [graviton] (b1),
  (a1)-- [gluon] (a2)
  };

  \node[] at (-0.35,-0.45) {$\pi\ $};
\node[] at (4.46,-0.45) {$\pi\ $};

\filldraw[fill=black, fill opacity=0.1] (-0.42,-0.5) ellipse (.45cm and 1.3cm);
\filldraw[fill=black, fill opacity=0.1] (4.42 ,-0.5) ellipse (.45cm and 1.3cm);

\draw[double, double distance=1.5,line width=1.2, postaction={decorate, decoration={
			markings,% switch on markings
			mark=at position 0.8 with  {\arrow{ latex} } } } ] (-1.8,-0.5) -- node [above] {} (-0.9,-0.5);

\draw[double, double distance=1.5,line width=1.2, postaction={decorate, decoration={
			markings,% switch on markings
			mark=at position 0.8 with  {\arrow{ latex} } } } ] (4.9,-0.5) -- node [above] {} (5.85,-0.5);
\end{feynman}
\end{tikzpicture}
\hspace{1cm}
\begin{tikzpicture}
\begin{feynman}
    \vertex (i1);
    \vertex[right=1cm of i1] (a1);
    \vertex[right=3cm of a1] (o1) ;
    \vertex[below=1cm of i1] (i2) ;
    \vertex[right=3cm of i2] (a2);
    \vertex[right=1cm of a2] (o2) ;

    %\vertex[below=2cm of i3] (t1);
    \vertex[above=1cm of a1] (t2);
    \vertex[above=2cm of a2] (t3);
    \vertex[blob] (t1) at (2,1.3) {{$TJJ$}};
    %\vertex[above right =1.3cm of t2] (t1);
    \vertex[above=1.5cm of t1] (i3) {$T^{\mu\nu}$};
    
    \diagram* { 
    (i1)  -- [fermion] (a1)   --[fermion] (o1),
    
  (i2)  -- [anti fermion] (a2)  --[anti fermion] (o2),

  (i3)  -- [graviton] (t1),
  (t1) -- [gluon] (a1) ,
  (t1) -- [gluon] (a2),
  };
  
 \node[] at (-0.35,-0.45) {$\pi\ $};
\node[] at (4.46,-0.45) {$\pi\ $};

\filldraw[fill=black, fill opacity=0.1] (-0.42,-0.5) ellipse (.45cm and 1.3cm);
\filldraw[fill=black, fill opacity=0.1] (4.42 ,-0.5) ellipse (.45cm and 1.3cm);

\draw[double, double distance=1.5,line width=1.2, postaction={decorate, decoration={
			markings,% switch on markings
			mark=at position 0.8 with  {\arrow{ latex} } } } ] (-1.8,-0.5) -- node [above] {} (-0.9,-0.5);

\draw[double, double distance=1.5,line width=1.2, postaction={decorate, decoration={
			markings,% switch on markings
			mark=at position 0.8 with  {\arrow{ latex} } } } ] (4.9,-0.5) -- node [above] {} (5.85,-0.5);
\end{feynman}
\end{tikzpicture}
    \caption{Leading (left) and NLO contributions to the GFF of the pion.}
    \label{fig:1}
\end{figure}

\subsection{The GFF-DVCS relation}
Gravitational form factors (GFFs) are related to deeply virtual Compton scattering (DVCS) of an electron (e) off a nucleon ($N$) ($eN\to e^\prime N^\prime \gamma$) with a final state photon. 
DVCS is a process where a high-energy electron scatters off a hadron (such as a proton or pion) by exchanging a virtual photon, which subsequently emits a real photon. An analysis of the process with other neutral currents is also possible \cite{Amore:2004ng}. \\
The process interpolates kinematically between the soft region, where a description in terms of QCD sum rules is possible \cite{Coriano:1993mr} and the process is dominated by the Feynman mechanism (overlap of intitial and final state hadron wavefunctions), and the inelastic region at higher energy \cite{Coriano:1998ge} \cite{Sterman:1997sx}.\\
A way to access EMT form factors is with GPDs
\cite{Ji:1996ek,Radyushkin:1996nd,Radyushkin:1996ru,
Ji:1996nm,Collins:1996fb,Radyushkin:1997ki,Vanderhaeghen:1998uc},
which describe hard-exclusive reactions, such as deeply virtual 
Compton scattering (DVCS) $eN\to e^\prime N^\prime \gamma$ sketched 
in Fig.~\ref{dvcsx} or hard exclusive meson production 
$eN\to e^\prime N^\prime M$. In the case of the nucleon,
the second Mellin moments of unpolarized GPDs yield 
the EMT form factors $A$, $B$ and $D$
\begin{align}
	\int_{-1}^1{\rm d}x\;x\, H^a(x,\xi,t) = A^a(t) + \xi^2 D^a(t) \,, \quad \quad
        \int_{-1}^1{\rm d}x\;x\, E^a(x,\xi,t) = B^a(t) - \xi^2 D^a(t) \,.
    	\label{Eq:GPD-Mellin}
\end{align}
$H$ and $E$ parameterize light-cone amplitudes with nonforward kinematics, describing the amplitude for removing from the nucleon
a parton carrying the fraction $x-\xi$ of the average momentum $P$ and 
reinserting back in the nucleon with a fraction fraction $x+\xi$ on the light-cone.
In the process, the nucleon receives the momentum transfer $\Delta$, with $\xi$ representing a second scaling variable. 
 Through their moments, this information provides insights into the GFFs.\\
The significance of GFFs in delineating hadron structure has spurred dedicated experimental endeavors, leading to the initial determinations of proton quark~\cite{Burkert:2018bqq} and gluon~\cite{Duran:2022xag} GFFs through measurements involving deeply virtual Compton scattering and $J/\psi$ photoproduction, respectively. Progress towards discerning the pion GFFs has been more constrained, with the first phenomenological constraints of the pion quark GFFs obtained from data recorded by the Belle experiment at KEKB \cite{Belle:2015oin,Savinov:2013hda,Kumano:2017lhr}. Anticipated advances in various hadron GFF determinations are expected from ongoing and forthcoming facilities such as the JLab 12 GeV program~\cite{JeffersonLabHallA:2022pnx,CLAS:2022syx} and the Electron-Ion Collider (EIC)~\cite{AbdulKhalek:2021gbh}.\\

\begin{figure}
    \centering
    \begin{tikzpicture}
\begin{feynman}
    \vertex (i1);
    \vertex[right=1cm of i1] (a1);
    \vertex[right=1cm of a1] (b1);
    \vertex[right=3cm of a1] (o1) ;

    \vertex[below=1cm of i1] (i2) ;
    \vertex[right=1cm of i2] (a2);
    \vertex[right=2cm of a2] (b2);
    \vertex[right=3cm of a2] (o2) ;

    \vertex[below=1cm of i2] (i4) ;
    \vertex[right=1cm of i4] (a4);
    \vertex[right=2cm of a4] (b4);
    \vertex[right=3cm of a4] (o4) ;

    %\vertex[below=2cm of i3] (t1);
    \vertex[above=1cm of a1] (t2);
    \vertex[above=2cm of a2] (t3);
    
    %\vertex[above right =1.3cm of t2] (t1);
    \vertex[above=1.5cm of t1] (i3) {$T^{\mu\nu}$};
    
    \diagram* { 
    (i1)  -- [fermion] (a1) -- [fermion] (b1)   --[fermion] (o1),
    
  (i2)  -- [fermion] (a2)-- [fermion] (b2)  --[fermion] (o2),
  (i4)  -- [fermion] (b4)  --[fermion] (o4),

  (i3)  -- [graviton] (b1),
  (a1)-- [gluon] (a2),
  (b2)-- [gluon] (b4)
  };
  
 \node[] at (-0.4,-1.) {$P $};
\node[] at (4.46,-1.) {$P$};

\filldraw[fill=black, fill opacity=0.1] (-0.435,-1.) ellipse (.48cm and 1.9cm);
\filldraw[fill=black, fill opacity=0.1] (4.435 ,-1.) ellipse (.48cm and 1.9cm);

\draw[double, double distance=1.5,line width=1.2, postaction={decorate, decoration={
			markings,% switch on markings
			mark=at position 0.8 with  {\arrow{ latex} } } } ] (-1.8,-1.) -- node [above] {} (-0.9,-1.);

\draw[double, double distance=1.5,line width=1.2, postaction={decorate, decoration={
			markings,% switch on markings
			mark=at position 0.8 with  {\arrow{latex} } } } ] (4.9,-1.) -- node [above] {} (5.85,-1.);
\end{feynman}
\end{tikzpicture}
\hspace{1cm}
    \begin{tikzpicture}
\begin{feynman}
    \vertex (i1);
    \vertex[right=1cm of i1] (a1);
    \vertex[right=1cm of a1] (b1);
    \vertex[right=3cm of a1] (o1) ;

    \vertex[below=1cm of i1] (i2) ;
    \vertex[right=1cm of i2] (a2);
    \vertex[right=2cm of a2] (b2);
    \vertex[right=1cm of a2] (c2);
    \vertex[right=3cm of a2] (o2) ;

    \vertex[below=1cm of i2] (i4) ;
    \vertex[right=1cm of i4] (a4);
    \vertex[right=2cm of a4] (b4);
    \vertex[right=1cm of a4] (c4);
    \vertex[right=3cm of a4] (o4) ;

    %\vertex[below=2cm of i3] (t1);
    \vertex[above=1cm of a1] (t2);
    \vertex[above=2cm of a2] (t3);
    
    %\vertex[above right =1.3cm of t2] (t1);
    \vertex[above=1.5cm of t1] (i3) {$T^{\mu\nu}$};
    \vertex[blob] (t1) at (2,1.3) {{$TJJ$}};
    
    \diagram* { 
    (i1)  -- [fermion] (a1)  --[fermion] (o1),
    
  (i2)  -- [fermion] (c2)-- [fermion] (b2)  --[fermion] (o2),
  (i4)  -- [fermion] (c4)  --[fermion] (o4),

  (i3)  -- [graviton] (t1),
  (a1)-- [gluon] (t1),
  (b2)-- [gluon] (t1),
  (c2)-- [gluon] (c4),
  };

 \node[] at (-0.4,-1.) {$P $};
\node[] at (4.46,-1.) {$P$};

\filldraw[fill=black, fill opacity=0.1] (-0.435,-1.) ellipse (.48cm and 1.9cm);
\filldraw[fill=black, fill opacity=0.1] (4.435 ,-1.) ellipse (.48cm and 1.9cm);

\draw[double, double distance=1.5,line width=1.2, postaction={decorate, decoration={
			markings,% switch on markings
			mark=at position 0.8 with  {\arrow{ latex} } } } ] (-1.8,-1.) -- node [above] {} (-0.9,-1.);

\draw[double, double distance=1.5,line width=1.2, postaction={decorate, decoration={
			markings,% switch on markings
			mark=at position 0.8 with  {\arrow{latex} } } } ] (4.9,-1.) -- node [above] {} (5.85,-1.);

\end{feynman}
\end{tikzpicture}
    \caption{Typical leading (left) and NLO contributions (right) to the GFF of the proton. }
    \label{fig:2}
\end{figure}

\subsection{The $TJJ$ from factorization and the conformal anomaly}
 At sufficiently large momentum  transfer, the GFF is described by a factorization formula, with an insertion of the $TJJ$ vertex, which is at the center of our investigation. Also in this case one can resort to an ordinary collinear factorization, using distribution amplitudes, or to a modified factorization with the inclusion of Sudakov effects, as discussed in the case of the electromagnetic form factor of the proton \cite{Li:1992nu}. \\
The light-front Fock state expansion provides a natural representation of hadronic structures~\cite{Brodsky:1997de}. This expansion formally consists of an infinite series of Fock states, each characterized by distinct partonic configurations and their associated light-front wave functions. However, for exclusive processes involving large momentum transfers, general power counting arguments indicate that the dominant contributions arise from the lowest Fock states, those with the minimal number of partons and the least orbital angular momentum \cite{Matveev:1973ra,Brodsky:1973kr} \cite{Ji:2003fw}. 
In the factorization approach, we are going to show, before our conclusions, to briefly illustrate how the leading three-valence-quark Fock state of the proton proceeds \cite{Ji:2002xn,Ji:2003yj}, following the steps indicated in \cite{Tong:2022zax}.   \\
In general, hadronic and partonic matrix elements are connected by hadronic wave functions integrated over the fractional momenta of the hard scattering amplitude. For example, in the case of the electromagnetic form factor of a pion of initial momentum $p$ and final momentum $p'$ \cite{Sterman:1997sx}
\beq
(p' + p)_\mu F_{\pi}(q^2) = \langle \pi(p') | J_\mu(0) | \pi(p) \rangle ,
\eeq
which is the simplest case, the corresponding form factor $F_{\pi}(q^2)$ is expressed at large $q^2$ in the form 
\beq
F_{\pi}(q^2) = \int_0^1 dx \, dy \, \phi_{\pi}(y, \mu^2) \, T_H(y, x, q^2, \mu^2) \, \phi_{\pi}(x, \mu^2)
\eeq
$J^\mu$ is the electromagnetic current of the valence quarks in the pion. 
In the collinear factorization limit, the pion wave function \( \phi_{\pi}(y, \mu^2) \) describes the distribution of fractional momenta \( (x, y) \) carried by the valence \( q/\bar{q} \) pair within the pion, in its lowest Fock state. Here, \( \mu \) represents the factorization scale, whose evolution is governed by the renormalization group equation, specifically the Efremov-Radyushkin-Brodsky-Lepage (ERBL) evolution equation~\cite{Lepage:1980fj, Efremov:1979qk}.  \\
For a gravitational form factor (GFF), a similar analysis can be performed. However, identifying the anomaly contribution in the associated hard scattering process, denoted as \( T_H \), is significantly more complex. A detailed discussion of this issue will be presented in Section 11.
The process is depicted in Fig. \ref{fig:1} for the pion and in Fig. \ref{fig:2} for the proton. 
 The amplitude is interpolated by the $TJJ$, due to the insertion at NLO of this correlator on the hard scattering amplitude. A more recent analysis of radiative corrections to the  the trace anomaly has been presented in \cite{Hatta:2018sqd}. \\
 The emergence of a dilaton pole in this vertex was pointed out in QED and QCD in 
\cite{Giannotti:2008cv,Armillis:2009pq} \cite{Armillis:2010qk}. 
This point will be examined in greater detail in the next section, where we first review the result for the $TJJ$ computed in QCD with on-shell gluons before discussing its sector decomposition. Our decomposition is motivated by the $CFT_p$ approach, which effectively resolves the conformal constraints. Following the method introduced in \cite{Bzowski:2013sza}, this provides a highly economical parameterization of the vertex. The anomaly contribution arises at NLO, with the two gluons in the $TGG$ (graviton/gluon/gluon) configuration attaching to the hard scattering process in all possible ways. In contrast, the non-anomalous contributions begin at $O(\alpha_s)$ and extend to $O(\alpha_s^2)$.  
\begin{figure}[t]
{\par\centering \resizebox*{12cm}{!}{\includegraphics{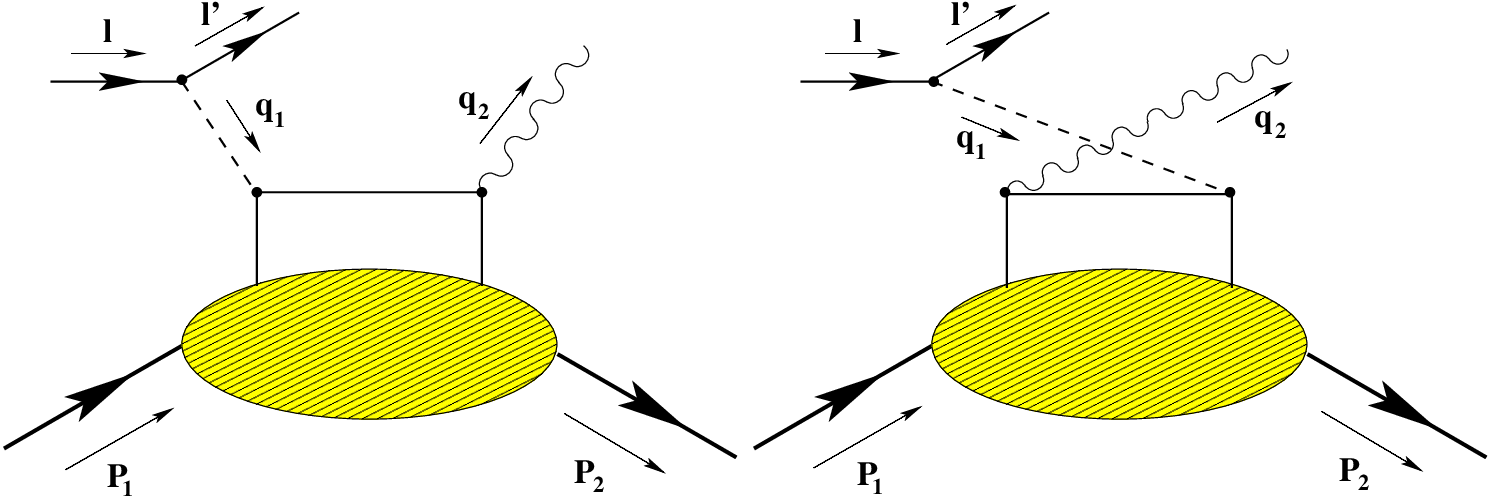}} \par}
\caption{Leading hand-bag diagrams for the DVCS process}
\label{DVCS_1.eps}
\label{dvcsx}
\end{figure}

\section{The QCD gravitational effective action and the conformal limit}
\label{three1}
To derive the conformal constraints for the $TJJ$ correlation function, we employ the general formalism of the gravitational effective action, examining the partition function within gravitational and gauge field backgrounds.\\
In QCD, the presence of the gluon sector and the gauge-fixing procedure requires treating the quark and gluon sectors separately from a CFT perspective, with each sector following its own hierarchy of conformal constraints at the one-loop level. Although this separation is no longer valid at higher orders in $\alpha_s$ within the correlator's expansion, it suffices for our current analysis.\\
The anomaly in the trace of the stress-energy tensor and the appearance of the dilaton pole stem from the renormalization of the effective action, which involves distinct counterterm contributions from the quark and gluon/ghost sectors.\\
The quark sector of the correlator behaves similarly to the Abelian case, with its hierarchical identities proportional to $n_f$, the number of fermions running in the loops. Contributions from this sector comply with standard vector Ward identities on the external gluon lines. The CWIs in this sector are analogous to those in the Abelian scenario and are only broken by the anomaly.\\
Within this sector, the CWIs can be addressed both non-perturbatively by directly solving the associated differential equations and perturbatively, as we will illustrate. Reconstructing the quark contribution to the correlator can be achieved using standard CFT methods in momentum space, employing a sector decomposition as shown in Eq. \eqref{Abar}. However, a similar analysis for the gluon sector is more challenging due to gauge-fixing conditions and the presence of Slavnov-Taylor identities (STIs), which replace the standard WIs. Consequently, the gluon sector is treated by us perturbatively, and its contributions are organized according to the same tensor decomposition of the quark sector, from which the dilaton state naturally emerges as a combination of the trace anomaly from both sectors.

\subsection{The action in a weak gravitational field }
For the derivation of all the relevant WIs we will be relying on the formalism of the gravitational effective action, that we are going to investigate in this and in the next sections. 
In the case of a non-Abelian gauge theory the action is given by
\bea
S_0[g,A,\psi]&=&-\frac{1}{4}\int d^4 x \sqrt{-g}F_{\mu\nu}^{{a}} F^{\mu\nu a} +\int d^4 x \sqrt{-g}\, i \bar{\psi}\gamma^\mu D_\mu \psi \nn\\
 S_1[g,A]&=&\int  \sqrt{-g}J_\mu^{a} A^{\mu a}
\eea 
with
\bea
F_{\mu\nu}^{a} &=&\nabla_\mu A_{\nu}^{a} - \nabla_{\nu} A_{\mu}^{ a} + g f^{a  b  c} A_\mu^{ b} A_\nu^c\nonumber \\
&=&\partial_\mu A_{\nu}^{a} - \partial_{\nu} A_{\mu}^{ a} + g f^{a  b  c} A_\mu^{ b} A_\nu^c, \nonumber \\
\nabla_\mu A^{\nu a} &=&\partial_\mu A^{\nu a} +\Gamma_{\mu\nu}^\lambda A^{\lambda a}
\eea
with $J^{\mu  a}=g_c\bar{\psi}\gamma^\mu T^{a}\psi$ denoting the fermionic current, with $T^a$ the generators of the theory and $\nabla_\mu$ denoting the covariant derivative in the curved background on a vector field. The local Lorentz and gauge covariant derivative $(D)$ on the fermions acts via the spin connection $\omega$
\be
D_\mu \psi=\left( \partial_\mu \psi  + A_\mu^a T^a +\frac{1}{4}\omega_{\mu}^{\underline{a} \underline{b}}\sigma_{
\underline a \underline b}\right)\psi
\ee
with $\sigma^{\ul{a}\ul{b}}=1/2 [\gamma^{\ul{a}},\gamma^{\ul{b}}]$,
having denoted with $\underline{a}\underline{b}$ the local Lorentz indices.  A local Lorentz covariant derivative $(D)$ can be similarly defined for a vector field, say $V^{\underline{a}}$, via the Vielbein $e^{\underline{a}}_{\, \,\mu}$ and its inverse  $e^{\,\,\mu}_{\underline a}$
\be
D_\mu V^{\underline a}= \partial_\mu V^{\underline a} +\omega^{\underline a}_{\mu \underline b} V^{\underline b}
\ee
with 
\be
\nabla_\mu V^\rho = e_{\underline a}^{\,\,\rho} D_\mu V^{\underline a}.
\ee
The Christoffel and the spin connection related via the holonomic relation 
 \be
 \Gamma^\rho_{\mu\nu}=e^{\,\,\rho}_{\ul a}\left( \partial_\mu e^{\ul a}_{\,\,\nu} +\omega^{\ul a}_{\mu \ul b} e^{\ul b}_{\,\,\nu}\right).
\ee
 with
\beq
 \omega_\mu^{\ul{a} \ul{b}}(x) =e_{\underline{a}}^{\,\nu}(x)e_{\underline{b}\nu;\mu}(x)\, ,
\eeq
where we have introduced the vielbein $e_{\underline{a}}^\mu(x)$,  
giving a Lagrangian of the form
\beqa \mathcal{L}_f& = & \sqrt{-g} \bigg\{\frac{i}{2}\bigg[\bar\psi\g^\mu(\mathcal{D}_\mu\psi)
 - (\mathcal{D}_\mu\bar\psi)\g^\mu\psi \bigg] - m\bar\psi\psi\bigg\}       \eeqa
for the fermion sector. The gauge fixing and ghost sectors are given by 
\beq
\mathcal{L}_{\text{gf}} = -\frac{1}{2\xi} \, g^{\mu\nu} (\nabla_\mu A_\nu^a)^2\,, \qquad \mathcal{L}_{\text{ghost}} = \bar{c}^a \, \left( - g^{\mu\nu} \nabla_\mu D_\nu^{ab} \right) c^b,  \qquad \eeq
with
\beq
D_\nu^{ab} = \delta^{ab} \nabla_\nu + g f^{abc} A_\nu^c
\eeq
the gauge and diffeomorphism covariant derivative. \\
 A symmetric stress energy tensor can be defined as 
\beq
T^{\mu\nu}=\frac{2}{\sqrt{-g}} \frac{\delta Z}{\delta g_{\mu\nu}}, 
\eeq
starting from the QCD partition function 
\bea
&& Z[J,\eta,\bar\eta,\chi,\bar\chi,g] =\mathcal{N}
\int \mathcal D A \, \mathcal D \psi \, \mathcal D \bar\psi \,
\mathcal D c \, \mathcal D \bar c \, \exp\bigg\{ i \int \sqrt{-g} d^4
x \left( \mathcal{L} + J_{\mu}^a A^{\mu  a} \right. \nn \\
&&\left. \hspace{6cm}   + \bar \eta \psi +
\bar \psi \eta + \bar \chi c + \bar c \chi \right)\bigg\}.
\eea
with a metric $g$ in the background. $J,\eta,\bar\eta,\chi,\bar\chi$ are the
sources of the gauge field $A$, of the fermion and antifermion fields ($\bar{\eta}, \eta$) and of the ghost and antighost fields ($\bar{\chi}$, $\chi$) respectively. \\
$\mathcal{L}$ is the standard QCD action, that a linearized level 
generates the term 
\beq
\label{Lgrav} \mathcal{L}_{grav}(x) = -\frac{\kappa}{2}T^{\mu\nu}(x)h_{\mu\nu}(x), 
\eeq 
with $h_{\mu\nu}$, describing the linear expansion of the metric around a flat spacetime.  
We use the convention $\eta_{\mu\nu}=(1,-1,-1,-1)$ for the metric in flat spacetime, parameterizing its deviations from the flat case as
\beq\label{QMM} g_{\mu\nu}(x) = \h_{\mu\nu} + \kappa \, h_{\mu\nu}(x)\,,\eeq
 with the symmetric rank-2 tensor $h_{\mu\nu}(x)$ accounting for its fluctuations. We have set $\kappa^2 =16 \pi G_N$, with $G_N$ the gravitational constant.We set $\kappa\to 1$. \\

\subsection{The stress energy tensor of the gauge-fixed action}
The stress energy tensors obtained by adding the components of the field strength $(fs)$, fermionic, gauge fixing and ghost sectors of the Lagrangian  
\bea
T_{\mu\nu} = T^{f.s.}_{\mu\nu} + T^{ferm.}_{\mu\nu} + T^{g.fix.}_{\mu\nu} + T^{ghost}_{\mu\nu},
\eea
with
\beq
T^{f.s.}_{\mu\nu}
 = 
\eta_{\mu\nu}\frac{1}{4}F^a_{\r\s}F^{a\,\r\s} -  F^a_{\mu\r}F^{a\,\r}_\nu  
\eeq

\bea
T^{g.f.}_{\mu\nu} &=& {1 \over \xi}\left[A_\nu^a \partial_\mu(\partial \cdot A^a) +A_\mu^a \partial_\nu(\partial \cdot A^a)\right] -{1 \over \xi}g_{\mu \nu}
\left[- \frac{1}{2} (\partial \cdot A)^2 + \partial^\rho (A_\rho^a \partial \cdot A^a)\right], \\
T^{gh}_{\mu\nu} &=& \partial_{\mu}\bar c^a D^{ab}_{\nu} c^b + \partial_{\nu}\bar c^a D^{ab}_{\mu}c^b - g_{\mu\nu} \partial^{\rho}\bar c^a D^{ab}_{\rho}c^b.
\label{gfghost}
\eea
After the inclusion of the fermion sector, the final expression takes the form

\bea
T_{\mu \nu} &=& -g_{\mu \nu} {\mathcal L}_{QCD}
-F_{\mu \rho}^a F_\nu^{a \rho} -{\frac{1} {\xi}}g_{\mu \nu}
\partial^\rho (A_\rho^a \partial^\sigma A_\sigma^a) +{\frac{1}{\xi}}(A_\nu^a \partial_\mu(\partial^\sigma A^a_\sigma)
  +A_\mu^a \partial_\nu(\partial^\sigma A_\sigma^a))
\nonumber\\
&+& {\frac{i}{4}} \Big[
  \overline \psi \gamma_\mu (\overrightarrow \partial_\nu
-i g T^a A_\nu^a)\psi  -\overline \psi (\overleftarrow \partial_\nu
+i g T^a A_\nu^a)\gamma_\mu\psi
 +\overline \psi \gamma_\nu (\overrightarrow \partial_\mu
-i g T^a A_\mu^a)\psi
\nonumber\\
&-& \overline \psi (\overleftarrow \partial_\mu
+i g T^a A_\mu^a)\gamma_\nu\psi \Big] +\partial_\mu \overline
c^a (\partial_\nu c^a -g f^{abc} A_\nu^c  c^b)
+\partial_\nu \overline c^a (\partial_\mu c^a -g f^{abc}
A_\mu^c c^b).
\label{EMT}
\eea
The diagrams relevant in the perturbative analysis of this correlator are shown in Fig. 4.

\subsection{The covariant expansion and the nonlocal anomaly action}
In order to derive the effective action we introduce the logarithm of the partition function that collects all the connected Green's functions 
\beq
W[J,\eta,\bar\eta,\chi,\bar\chi,h] =-i \log{Z[J,\eta,\bar\eta,\chi,\bar\chi,h]}
\eeq
(normalized to the vacuum functional) and the effective action, the generating functional $\mathcal{S}_{eff}$ of the 1-particle irreducible and truncated amplitudes. This is defined by a Legendre transform of $W$ with respect to all the non-gravitational sources
\bea
\mathcal{S}_{eff}[A_{(c)},\bar \psi_{(c)}, \psi_{(c)},
\bar c_{(c)}, c_{(c)}, h] = W[J,\eta,\bar\eta,\chi,\bar\chi,h] -
\int d^4 x \left( J_{\mu}A^{\mu}_c + \bar \eta \psi_c + \bar \psi_c \eta
+ \bar \chi c_c + \bar c_c \chi \right).
\label{1PIfunctional}
\eea
with
\bea
A^{\mu}_{(c)} = \frac{\delta W }{\delta J_{\mu}}, \qquad
\psi_{(c)} = \frac{\delta W }{\delta \bar \eta}, \qquad \bar \psi_{(c)} =
\frac{\delta W }{\delta \eta}, \qquad \omega_{(c)} = \frac{\delta W
}{\delta \bar \chi}, \qquad \bar \omega_{(c)} = \frac{\delta W }{\delta
\chi} \label{Legendre1}
\eea
being the classical external source fields. $\mathcal{S}_{eff}$ satisfies the equations

\bea
\frac{\delta\mathcal{S}_{eff}}{\delta A^{\mu}_{(c)}} = - J_{\mu}, \qquad
\frac{\delta\mathcal{S}_{eff}    }{\delta \psi_{(c)}} = - \bar \eta, \qquad
\frac{\delta\mathcal{S}_{eff}}{\delta \bar \psi_{(c)}} = - \eta, \qquad
\frac{\delta\mathcal{S}_{eff}}{\delta \omega_{(c)}} = - \bar \chi, \qquad
\frac{\delta\mathcal{S}_{eff}}{\delta \bar \omega_{(c)}} = - \chi,
 \label{Legendre2}
\eea
 
 \bea
\frac{\delta\mathcal{S}_{eff}}{\delta h_{\mu\nu}} = \frac{\delta W}{\delta h_{\mu\nu}}. \label{Legendre3}
\eea
The vev of the stress energy tensor in a gravitational background $g_{\mu\nu}$ will be simply denoted 
as the functional derivative of $\mathcal{S}_{eff}$
\be
\langle T^{\mu\nu}\rangle=\frac{2}{\sqrt{g}}\frac{\delta \mathcal{S}_{eff}}{\delta g_{\mu\nu}}. 
\ee
The covariant expansion of $\mathcal{S}_{eff}$ around a background metric $\bar{g}$ and 
	vanishing gauge field $A_{c\mu}$, with fluctuations  $\delta g$ and $\delta A_{c\,\mu}$ takes the form
	\begin{eqnarray}
	\label{exps2}
	\mathcal{S}_{eff}(g,A_c) &\equiv&\sum_{n,
	k}^\infty \frac{1}{2^{n+ k} (n+k)!} \int d^d x_1\ldots d^d x_{n+k} \sqrt{g_1}\ldots \sqrt{g_n}\,\langle T^{\mu_1\nu_1}\ldots \,T^{\mu_n\nu_n} J^{a_{n+1}}_{\mu_{n+1}}\ldots J^{a_{n+k}}_{\mu_{n + k}}\rangle_{\bar{g} }\nonumber\\
&&\times \delta g_{\mu_1\nu_1}(x_1)\ldots \delta g_{\mu_n\nu_n}(x_n)\delta A^{a_{n+1}}_{c\,\, \mu_{n+1}}(x_{n+1}) 
	\ldots \delta A^{a_{n+k}}_{c\,\, \mu_{n+k}}(x_{n+k}) +  GT \ldots
	\end{eqnarray}
	where the dots refer to contributions from the  pure gauge sector in the expansion, and $GT$ refers to the purely gravitational terms (i.e. with multiple insertions of only stress-energy tensors). In the future expressions we will remove the subscript "c" from the classical external source $A_{c \,\mu}^a$ for convenience. \\
The covariant normalization of the correlation functions is given by
	\begin{equation}
		\label{exps1}
		\langle T^{\mu_1\nu_1}(x_1)\ldots T^{\mu_n\nu_n}(x_n)\rangle \equiv\frac{2}{\sqrt{g_1}}\ldots \frac{2}{\sqrt{g_n}}\frac{\delta^n \sm_{eff}(g,A)}{\delta g_{\mu_1\nu_1}(x_1)\delta g_{\mu_2\nu_2}(x_2)\ldots \delta g_{\mu_n\nu_n}(x_n)} 
	\end{equation}
	for the $n$-graviton sector, with $\sqrt{g_1}\equiv \sqrt{|\textrm{det} \, g_{{\mu_1 \nu_1}}(x_1) |} $ and so on, while mixed correlators will be defined as 
	\begin{eqnarray}
	\label{exps1}
	\langle T^{\mu_1\nu_1}(x_1)\ldots T^{\mu_n\nu_n}(x_n)J^{a_{n+1} \ \mu_{n+1}}(x_{n+1})\ldots J^{a_{n+k} \, \mu_{n+k}}(x_{n+k})\rangle & \equiv & \nonumber\\
	\frac{2}{\sqrt{g_1}}\ldots \frac{2}{\sqrt{g_n}}\frac{\delta^n \mathcal{S}_{eff}(g,A)}{\delta g_{\mu_1\nu_1}(x_1)\ldots \delta g_{\mu_n\nu_n}(x_n) \, \delta  A^{a_{n+1}}_{\mu_{n+1}} (x_{n+1}) \ldots \,  \delta A^{a_{n+k}}_{\mu_{n+k}} (x_{n+k}) } && \nonumber \\
	\end{eqnarray}
	for the graviton/gauge sector. 
In the renormalization at one-loop of correlators like $\langle TTT\rangle_{even}$, two counterterms are required: $V_E/\epsilon$ and $V_{C^2}/\epsilon$. These are defined as 

\begin{align}
\label{ffr}
V_{E}(g,d) \equiv & \mu^{\epsilon} \int d^dx \, \sqrt{-g} \, E, \quad \quad \epsilon = d-4, \notag \\ 
V_{C^2}(g,d) \equiv & \mu^{\epsilon} \int d^dx \, \sqrt{-g} \, C^2.
\end{align}
Here, $E$ is an evanescent term, the Gauss-Bonnet term, being topological at $d=4$, yet its inclusion is necessary in order for $\mathcal{S}_{eff}$  to satisfy the Wess-Zumino consistency condition. 
$C^2$ denotes the square of the Weyl tensor.\\
The anomaly arises in dimensional regularization (DR) due to the failure of these two counterterms to remain invariant under Weyl transformations  in $d$ dimensions 

\begin{equation} \label{ren}
	\begin{aligned}
		&2 g_{\mu\nu}\frac{\delta}{\delta g_{\mu\nu}}V_{E}=\frac{\delta}{\delta \phi}V_{E}=\epsilon \sqrt{g} E,\\
		&2 g_{\mu\nu}\frac{\delta}{\delta g_{\mu\nu}}V_{C^2}=\frac{\delta}{\delta \phi}V_{C^2}=\epsilon \sqrt{g} \left[ C^2 + \frac{2}{3} \square R \right]
	\end{aligned}
\end{equation}
$(\epsilon=d-4)$.
We have defined the square of the Weyl tensor at $d=4$
\beq
C^2= C_{\lambda\mu\nu\rho}C^{\lambda\mu\nu\rho} = R_{\lambda\mu\nu\rho}R^{\lambda\mu\nu\rho}
-2 R_{\mu\nu}R^{\mu\nu}  + \frac{R^2}{3},
\eeq
while the Euler (Gauss-Bonnet) term is given by
\beq
E = ^*\hskip-.2cmR_{\lambda\mu\nu\rho}\,^*\hskip-.1cm R^{\lambda\mu\nu\rho} =
R_{\lambda\mu\nu\rho}R^{\lambda\mu\nu\rho} - 4R_{\mu\nu}R^{\mu\nu}+ R^2.
\eeq
Defining, alternatively 
\begin{align}
(C^{(d)})^2&\equiv R^{\mu \nu \rho \sigma}R_{\mu \nu \rho \sigma}-\frac{4}{d-2}R^{\mu \nu }R_{\mu \nu}+\frac{2}{(d-2)(d-1)}R^2\notag \\
&=(C^{(4)})^2+\frac{d-4}{d-2}\left(2R^{\mu \nu }R_{\mu \nu}-\frac{d+1}{3(d-1)}R^2\right)\notag 
\end{align}
to be the Weyl tensor in $d$ dimensions, the second variation in \eqref{ren} takes the form 
\begin{equation}
\label{ren1}
2g_{\mu \nu}\frac{\delta}{\sqrt{-g}\delta g_{\mu \nu}}\int d^d x  \sqrt{-g} (C^{(d)})^2=(d-4)(C^{(d)})^2,
\end{equation}
showing that $\Box R$ is prescription dependent, after a comparison of the second of \eqref{ren} with \eqref{ren1}.\\
In the renormalization of the $TJJ$ we will only need the gauge counterterm 
	\begin{align}
	\label{ct}
		S_{count}\equiv V_{F^2}\equiv -\frac{1}{\varepsilon}\int d^dx\,\sqrt{-g}\left(\beta\,F^2\right),
	\end{align}
	corresponding to the field strength $F^2$, where the coefficient $\beta$ is the ordinary QCD $\beta$ function.\\
	 In the $TJJ$ the naive trace of $T^{\mu\nu}$  is rather simple, since naive scale invariance gives the traceless condition
\begin{equation}
g_{\mu\nu}T^{\m\n}=0\label{trace},	
\eeq	
which at linear order in the gravitational fluctuations, after renormalization,  is modified at quantum level in the form
\begin{equation}
	g_{\mu\nu} \langle T^{\mu\nu}\rangle=\beta F^{a\, \mu\nu}F^a_{\mu\nu}.
\label{ann}
\end{equation}
Additional contributions to the trace anomaly, associated with $E$ and $C^2$, are captured only by extra insertions of stress energy tensors. 
The $TJJ$ is part of the hierarchies of correlators containing such extra insertions (see for instance \cite{Coriano:2022jkn}).
 For correlators with multiple insertions of the stress energy tensor, the renormalization of the effective action leads to the inclusion of all the three counterterms $V_E,V_{C^2}$ and $V_{F^2}$, with the generation of a parity-even trace anomaly
\bea
g_{\mu\nu}\langle T^{\mu\nu}\rangle =   b \, C^2 + b^{\prime} \, \left( E - \frac{2}{3} \, \square \, R\right) + b'' \, \square \, R +  c\, F^{a \, \mu \nu} F^a_{\mu \nu},
\label{var2}
\eea
where  $b$ and $b'$ are related to the massless content of the virtual contributions. 
In the non-Abelian $TJJ$ case, the expansion of the counterterm \eqref{ct} is the only one needed in order to renormalize the  the 2- 3- and 4- gluons vertices which are part of the gauge invariant contributions to the non-Abelian trace anomaly. In the analysis of the hierarchy of the quark sector, only the $n_f$ dependence of \eqref{ct} will be relevant, while its $C_A=N_c=3$ (colour) dependence will affect the gluon sector. \\
The anomaly content is gauge invariant and
can be identified by the sector decomposition that  we are going to discuss in the next sections. \\
The anomaly part of $\mathcal{S}_{eff}$, also termed "the anomaly induced action" and denoted as $\mathcal{S}_{anom}$, can be derived by solving the functional constraint in \eqref{var2} in $d=4$ dimensions in the nonlocal form 
\bea
&& \hspace{-.6cm}S_{anom}[g,A] = \label{Tnonl}\\
&&\frac {1}{8}\int d^4x\sqrt{-g}\int d^4x'\sqrt{-g'} \left(E - \frac{2}{3} \square R\right)_x
 \Delta_4^{-1} (x,x')\left[ 2b\,F
 + b' \left(E - \frac{2}{3} \square R\right) + 2\, c\, F_{\mu\nu}F^{\mu\nu}\right]_{x'} \nonumber
 \label{var1}
\eea
for a general field content of massless fields in the anomaly functional, parameterized by constants $b,b'$ and $c$. We have introduced the Green's functions of the quartic (Paneitz) operator
\beq
\Delta_4 \equiv  \nabla_\mu\left(\nabla^\mu\nabla^\nu + 2 R^{\mu\nu} - \frac{2}{3} R g^{\mu\nu}\right)
\nabla_\nu = \square^2 + 2 R^{\mu\nu}\nabla_\mu\nabla_\nu +\frac{1}{3} (\nabla^\mu R)
\nabla_\mu - \frac{2}{3} R \square\,
\label{Deldef}
\eeq
which is conformally covariant.  
Around flat space, \eqref{var1}  \cite{Riegert:1984kt} also knonw as the Riegert action, can be reformulated in the form
\beq
S_{anom}[g,A]  \rightarrow  -\frac{c}{6}\int d^4x\sqrt{-g}\int d^4x'\sqrt{-g'}\, R^{(1)}_x
\, \square^{-1}_{x,x'}\, [F_{\alpha\beta}^aF^{a\, \alpha\beta}]_{x'}\,,
\label{SSimple}
 \eeq
which holds true to the first order in the metric fluctuations around a flat background. This form is expected  to emerge in QCD and is reproduced in the perturbative picture, as shown in explicit QED \cite{Giannotti:2008cv,Armillis:2009pq}  and QCD \cite{Armillis:2010qk} computations at lowest order. \\
Indeed, in the case of the $TJJ$ in QED, which is the lowest order,  \eqref{SSimple} acquires the specific form 
\beq
 \label{pole}
\mathcal{S}_{pole}= - \frac{e^2}{ 36 \pi^2}\int d^4 x \, d^4 y \left(\square h(x) - \partial_\mu\partial_\nu h^{\mu\nu}(x)\right)  \square^{-1}_{x\, y} F_{\alpha\beta}(x)F^{\alpha\beta}(y),
\eeq
where the Ricci scalar is expanded at linear order. In the QCD case a similar relation is valid for on-shell gluons. The $\Box^{-1}$ pole contribution to the anomaly is correctly described by such nonlocal action up to 3-point functions. The consistency of \eqref{SSimple} in the generation of the correct hierarchy of correlators, 
constrained by the corresponding CWIs up to 3-point functions, has been verified in free field theory realizations with scalars, fermions and Abelian spin-1 fields running in the loops. This check has been performed for the $TTT$ correlator \cite{Coriano:2017mux,Coriano:2018bsy}. \\
However, an analysis of 4-point functions reveals that, in the Abelian case, a correlator like $TTJJ$, derived from \eqref{var1}, lacks certain Weyl-invariant terms. These terms are essential for maintaining the correct hierarchical structure to which the $TTJJ$ correlator belongs. Such additional terms have been identified in the $4T$ correlator, as shown in \cite{Coriano:2021nvn}. These findings were deduced by examining the one-loop free field theory realization of the same correlator.\\
The analysis of \cite{Armillis:2010qk}, however shows that in the case of a 3-point function such as the $TJJ$, for on-shell gluons in QCD, the nonlocal action reproduces the perturbative expansion of the anomaly form factor. \\
In the off-shell case, we are going to show, such previous analysis can be extended, and allow to completely characterize the same anomaly form factor using a sector decomposition. This allows to separate the anomaly contribution from those terms which are proportional to the equations of motion for the external gluons.  
The two sectors of the perturbative expansion are given in \secref{expansion}.
\begin{figure}[t]
\begin{center}
\includegraphics[scale=0.15]{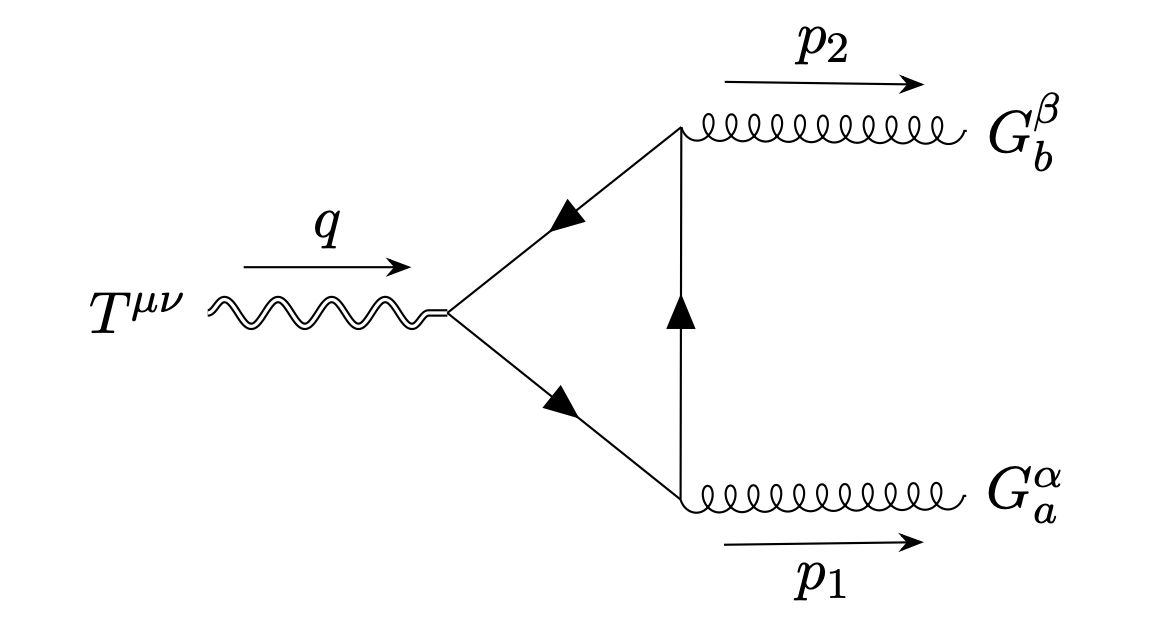}
\includegraphics[scale=0.15]{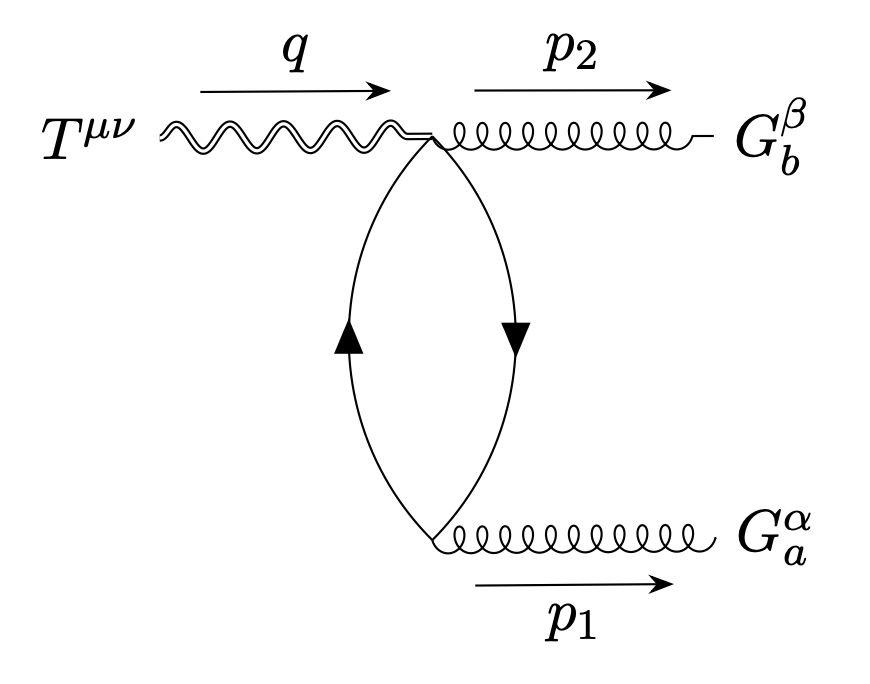}
\includegraphics[scale=0.15]{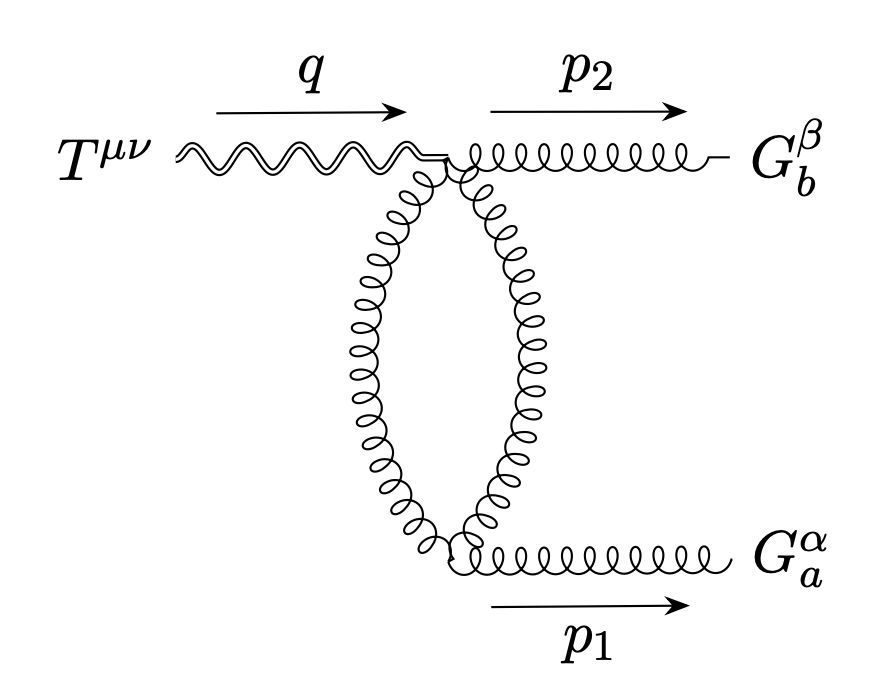}
\includegraphics[scale=0.15]{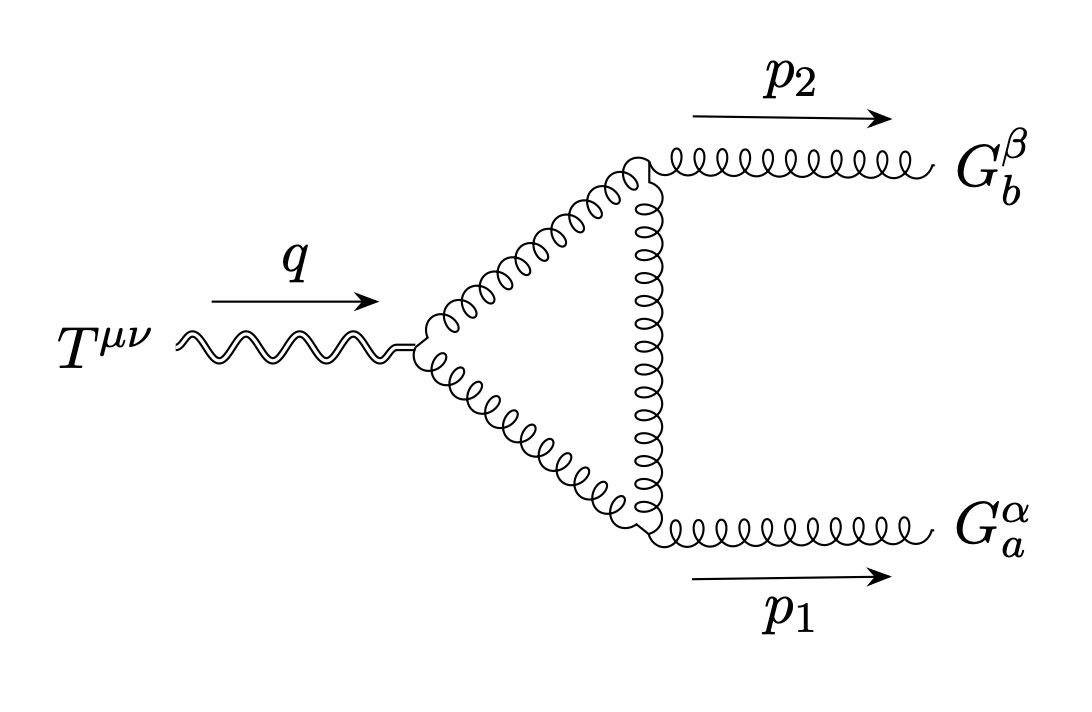}
\includegraphics[scale=0.15]{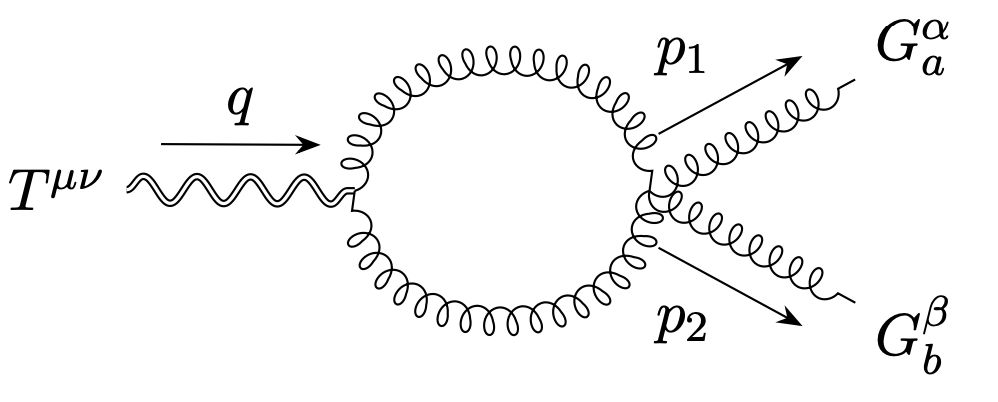}
\includegraphics[scale=0.15]{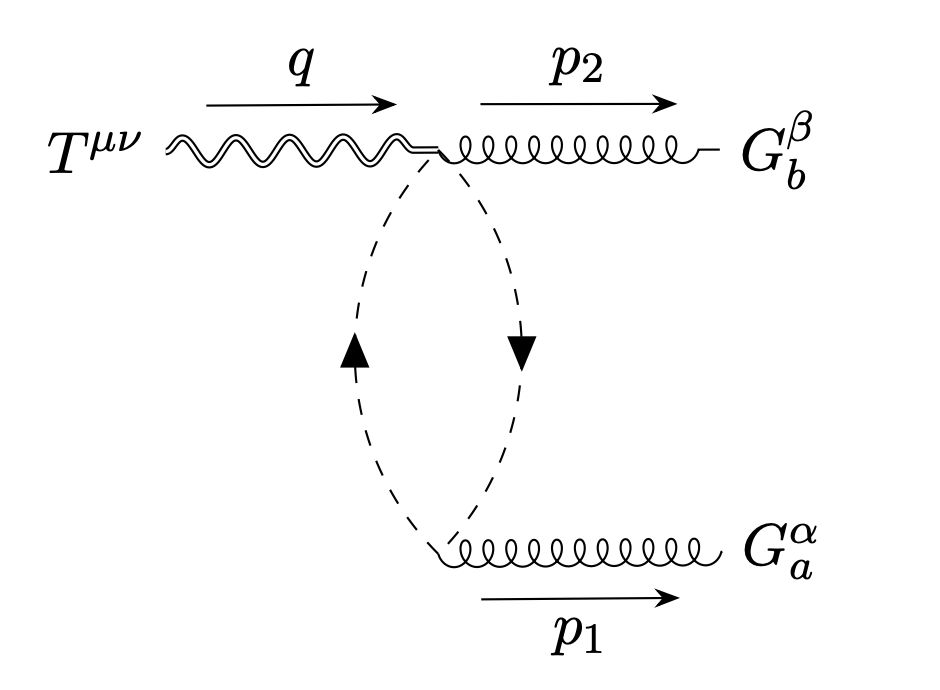}
\includegraphics[scale=0.15]{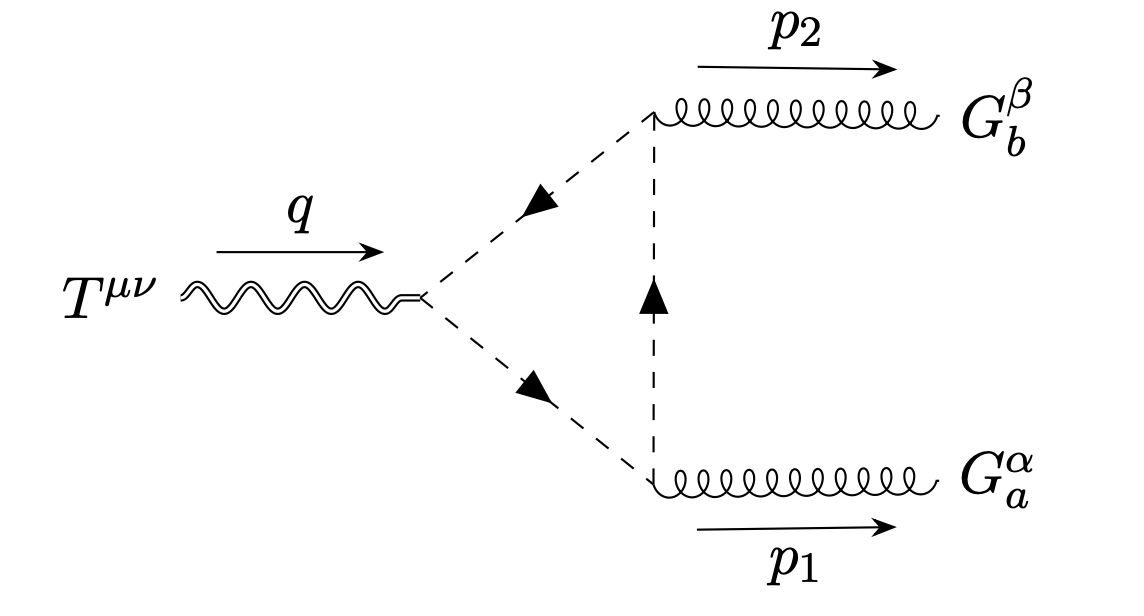}
\caption{List of the perturbative quark, gluon, and ghost (dashed lines) sectors in the non-Abelian $TJJ$. }
\label{expansion}
\end{center}
\end{figure}

\subsection{The anomaly sector and gauge invariance}
In the non-Abelian case, as is customary, the anomaly computation is primarily conducted at the level of 3-point functions. The full gauge-invariant contribution is then inferred by extending the result of the 3-point function in a gauge-invariant manner. Specifically, the perturbative computation's outcome can be made gauge-covariant, at least regarding the anomaly pole part. In the non-Abelian scenario, additional contributions arise from other sectors, corresponding to extra diagrams with three and four external gluons.\\
The $TJJ$ correlator is part of a broader set of correlation functions, including the $TJJJ$ and $TJJJJ$ correlators, all interconnected through conservation Ward identities (WIs), gauge WIs, and broken conformal WIs (CWIs) in the gluon sector. Gauge invariance permits the extraction of the gauge-invariant form of the nonlocal anomaly action from the perturbative computation of the $TJJ$. This is achieved through a standard covariantization of the results obtained from investigating this correlator. The anomaly's structure can then be recovered from the nonlocal action as in \eqref{pole}, modified by the QCD $\beta$ function 
\bea
S_{pole} &=&  \beta(g)  \, \int d^4 x \, d^4 y \,R^{(1)}(x)\, \square^{-1}(x,y) \, F^a_{\alpha \beta}F^{a \alpha \beta}
\eea
\beq
\beta(g) = \frac{dg}{d\ln(\mu^2)} = -\beta_0 \frac{g^3}{16\pi^2}, \qquad \beta_0 = \frac{11}{3} C_A - \frac{2}{3} n_f
\eeq
with $n_f$ the number of massless quark flavours.
One may define the quantum averaged stress energy tensor, with just the pole-term included  
 \be
T^{\mu\nu}_{anom}(z) =
\beta(g) \left(g^{\mu\nu}\sq - \partial^{\mu}\partial^{\nu}\right)_z \int\,d^4x'\, \sq_{z,x'}^{-1}
\left[F^a_{a\alpha\beta}F^{a \alpha\beta}\right]_{x'}\,,
\label{Tanom}
\ee
from which one may extract the anomaly-induced vertex at trilinear level
 
 \be
 \Gamma_{{anom}}^{\mu\nu\alpha\beta\, a b}(p_1,p_2) = \int\,d^4x\,\int\,d^4y\, e^{ip_1\cdot x + i p_2\cdot y}\,\frac{\delta^2 T^{\mu\nu}_{anom}(0)}
{\delta A^a_{\alpha}(x) A^b_{\beta}(y)} = \beta(g)\frac{1}{3\, q^2} \left(g^{\mu\nu}q^2 - q^{\mu}q^{\nu}\right)u^{\alpha\beta}(p_1,p_2)\delta^{a b},\,
\label{Gamanom}
\ee
 with a trace anomaly 
\be
g_{\mu\nu}\Gamma^{\mu\nu\alpha\beta a b }(p,q)\big\vert_{A=0} =\beta(g)
\,u^{\alpha\beta}(p_1,p_2)\delta^{a b},\,
\label{Gamanomtr}
\ee
where 
\beq
u^{\a \b} (p_1,p_2)\equiv (p_1\cdot p_2) g^{\alpha\beta} - p_2^\alpha p_1^\beta
\eeq
This tensor structure summarizes the conformal anomaly contribution, being the Fourier transform of the anomaly term at $O(g^2)$
\bea
&&u^{\alpha\beta}(p_1,p_2) = -\frac{1}{4}\int\,d^4x\,\int\,d^4y\ e^{ip_1\cdot x + i p_2\cdot y}\ 
\frac{\delta^2 \{F^a_{\mu\nu}F^{a\mu\nu}(0)\}} {\delta A_{\alpha}(x) A_{\beta}(y)}\vline_{A=0} \,.
\label{locvar}\\
\eea
By differentiating $T^{\mu\nu}_{\text{anom}}(0)$ to higher orders with respect to the external classical gluon field, we derive analogous relations for the 4-point function ($TJJJ$)

\beqa
 \Gamma_{anom}^{\mu\nu\alpha\beta\rho\, a_1 a_2 a_3}(p_1,p_2,p_3) &=& \int\,d^4x d^4y\, d^4 z \, e^{ip_1\cdot x + i p_2\cdot y + i p_3\cdot z}\,\frac{\delta^2 T^{\mu\nu}_{anom}(0)}
{\delta A^{a_1}_{\alpha}(x) \delta A^{a_2}_{\beta}(y)\delta A^{a_3}_{\rho}(z)}\vert_{A=0} \nn\\
&=&  \beta(g)\frac{1}{3\, q^2} \left(g^{\mu\nu}q^2 - q^{\mu}q^{\nu}\right)u^{\alpha\beta\rho\, a_1 a_2 a_3}(p_1,p_2,p_3)\,,
\label{Gamanom}
\eeqa

\begin{equation}
 u^{\mu_1\mu_2\mu_3\,\, a_1 a_2 a_3}= g\left(f^{a_1 a_2 a_3} \left(g^{\mu_1 \mu_2} \left(p_1^{\mu_3}-p_2^{\mu_3}\right)+g^{\mu_1 \mu_3} \left(p_3^{\mu_2}-p_1^{\mu_2}\right)+g^{\mu_2 \mu_3} \left(p_2^{\mu_1}-p_3^{\mu_1}\right)\right)\right)
\end{equation}

and 5-point function $(TJJJJ)$

\beqa
 \Gamma_{anom}^{\mu\nu\alpha\beta\rho\, a_1 a_2 a_3}(p_1,p_2,p_3) &=& \int\,d^4x\,d^4y\, d^4 z d^4 w \, e^{ip_1\cdot x + i p_2\cdot y + i p_3\cdot z + i p_4\cdot w}\,\frac{\delta^2 T^{\mu\nu}_{anom}(0)}
{\delta A^{a_1}_{\alpha}(x) \delta A^{a_2}_{\beta}(y) \delta A^{a_3}_{\rho}(z) \delta A^{a_4}_{\sigma}(w)} \nn\\
&=&  \beta(g)\frac{1}{3\, q^2} \left(g^{\mu\nu}q^2 - q^{\mu}q^{\nu}\right)u^{\alpha\beta\rho\sigma\, a_1 a_2 a_3 a_4}(p_1,p_2,p_3)\,,
\label{Gamanom}
\eeqa
where 
\begin{align}
u^{\mu_1\mu_2\mu_3\mu_4 \,\, a_1 a_2 a_3 a_4}=g^2 \Bigl(\Bigl(&f^{a_1 a_2 b} f^{a_3 a_4 b} \left(g^{\mu_1 \mu_3} g^{\mu_2 \mu_4}-g^{\mu_1 \mu_4} g^{\mu_2 \mu_3}\right)+\nonumber\\&+f^{a_1 a_3 b} f^{a_2 a_4 b} \left(g^{\mu_1 \mu_2} g^{\mu_3 \mu_4}-g^{\mu_1 \mu_4} g^{\mu_2 \mu_3}\right)
    +f^{a_1 a_4 b} f^{a_2 a_3 b} \left(g^{\mu_1 \mu_2} g^{\mu_3 \mu_4}-g^{\mu_1 \mu_3} g^{\mu_2 \mu_4}\right)\Bigl)\Bigl).
\end{align}

\section{Symmetries of the quark partition function }
\label{four1}
The quark sector at one-loop satisfies ordinary anomalous CWIs, together with ordinary gauge Ward identities on the gluon vector current. In this case the treatment follows the usual approach similarly to the case of QED or any scale invariant Abelian theory, with the due generalization. Formal derivation of the equations are based on the functional integral 
\be
\tilde{W}[g,A] =-i \log \tilde{Z} 
\label{GF}
\ee

where
\be
\tilde{Z} =\int {D\bar\psi}D\psi e^{-(S_0[g,\psi] + S_1[A,\psi ])}
\ee
is integrated only over the quark field $\psi$, in the background of the metric $g_{\mu\nu}$ and of the gauge field $A_\mu^a$. To fix the form of the correlator we need to impose the transverse WI on the vector lines and the conservation WI for $T^{\mu\nu}$. 
Indeed, diffeomorphism invariance of the generating functional of the connected vertices (\ref{GF}) gives 
 \be
 \int d^d x \left(\frac{\delta\tilde{ W}}{\delta g_{\mu\nu}}\delta g_{\mu\nu}(x) +  \frac{\delta\tilde{ W}}{\delta A_{\mu}^a}\delta A_{\mu}^a(x)\right)=0,
\label{aver}
\ee
where the variation of the metric and the gauge fields are the corresponding Lie derivatives, for a change of variables 
$x^\mu\to x^\mu + \epsilon^\mu(x)$
\bea
\label{lie}
\delta A_\mu^a(x) &=& -\nabla_\alpha A_\mu^a \epsilon^\alpha - A^a_\alpha \nabla_\mu \epsilon^\alpha \nonumber \\
g_{\mu\nu}&=&-\nabla_\mu \epsilon_\nu  - \nabla_\nu \epsilon_\mu 
\eea
while for a gauge transformation with a parameter $\theta^a(x)$
\be
\delta A_\mu^a=\ul{D}_\mu \theta^a \equiv\partial_\mu \theta^a + g_c f^{a b c} A_\mu^b \theta^c.
\ee
The absence of propagating gluons guarantees the preservation of all the fundamental symmetries needed for the derivaton of the hierarchical WIs of this sector.\\ CWIs, conservation WIs for the stress energy tensor as well as ordinary gauge WIs can be derived by simply requiring the invariance of this functional with respect to conformal transformations, diffeomorphisms and gauge transformations.

\begin{itemize}
\item{\bf Diffeomorphism invariance}
\end{itemize}
Using (\ref{lie}), Eq. (\ref{aver}) becomes 
\begin{align}
\label{interx1}
0=&\,\left\langle\, \int d^4 x \left( \frac{\delta (S_0 + S_1)}{\delta g_{\mu\nu}} \delta g_{\mu\nu} +\frac{\delta S_1}{\delta A_\mu^a} \delta A_{\mu}^a\right)
\right\rangle_q\nn \\
=&\,\left\langle \int d^4 x \sqrt{-g_x}\left[\nabla_\mu T^{\mu\nu} -(\nabla_\mu A_\nu^a -\nabla_\nu A_\mu^a) J^{\mu a} - \nabla_\mu J^{\mu a} A_\nu^a\right]\epsilon^\nu(x)\right\rangle_q
\end{align}
while the condition of gauge invariance gives the 
\begin{itemize}
\item{\bf gauge WI }
\end{itemize}
\be
 \int d^d x \frac{\delta \tilde{W}}{\delta A_{\mu}^a}\delta A_{\mu}^a=\left\langle \int d^4 x \sqrt{-g_x}J^\mu_a\ul{D}_\mu \theta^a \right\rangle_q=0
\ee
which, in turn, after an integration by parts, generates the gauge WI
\be
\langle \nabla_\mu J^{\mu a} \rangle_q = g_c f^{a b c} \langle J_\mu^b\rangle A^{\mu c}. 
\ee
Inserting this relation into \eqref{interx1} we obtain the conservation WI
\be
\langle \nabla^\mu T_{\mu\nu}\rangle_q -F_{\mu\nu}^a \langle J^{\mu a}\rangle_q =0.
\ee
In the quark sector, diffeomorphism and gauge invariance  then give the relations
\begin{equation}
\begin{split}
0&=\langle \nabla^\mu T_{\mu\nu}\rangle_q -F_{\mu\nu} \langle J^{\mu}\rangle_q \\
0&=\nabla_\n\braket{J^\nu}_q .
\end{split}\label{transverse}
\end{equation}

\begin{itemize}
\item{\bf Conformal invariance and differential equations}
\end{itemize}
We recall that a special conformal transformation in flat space is characterised by
\beq
x'^\mu=\frac{(x^\mu - b^\mu x^2)}{\Omega(x)}\qquad \textrm{with}\qquad \Omega(x)=1 - 2 b\cdot x + b^2 x^2\qquad \textrm{and}\qquad J_c\equiv \vline \frac{\partial x'}{\partial x}\vline=\Omega^{-d}.
\label{special}
\eeq
On the metric and on primary scalar field $\phi(x)$ of scaling  dimension $\Delta$ it will induce the transformations
\beq
g'_{\mu\nu}(x')=\Omega^2 g_{\mu\nu}(x) \qquad \phi'(x')=J_c^{-\Delta/d}\phi(x)=\Omega^{\Delta}\phi(x).\qquad 
\eeq
and for a spin-1 field 
\beq
J'^\mu(x')=\Omega^{\Delta_J}\frac{\partial x'^\mu}{\partial x^\nu}J^\nu(x).
\eeq

Differential equations can be derived for all of the transformations above. Expanding these relations for $b\ll 1$
and taking the finite part one obtains
\beq
 \mathcal{K}^k \phi(x)=\left(- x^2\frac{\partial}{\partial x^\kappa}+ 2 x^\kappa x^\alpha \frac{\partial}
{\partial x^\alpha} +2 \Delta_\phi x^\kappa\right) \phi(x)
\eeq

\beq
\mathcal{K}^k J^\mu(x)= \left(- x^2\frac{\partial}{\partial x^\kappa}+ 2 x^\kappa x^\alpha \frac{\partial}
{\partial x^\alpha} +2 \Delta_J x^\kappa\right) J^\mu(x) +2 \left(  \delta^{\mu\kappa} x_{\rho} - \delta_{\rho}^{\kappa }x^\mu  \right) J^\rho(x)
\eeq
for the special conformal tranformations.
A similar transformation holds for the spinor 
\beq
\psi'(x')=\Omega^{\Delta_\psi} S(\Lambda(x))\psi(x)\qquad S(\Lambda(x))=e^{-\frac{i}{4}\omega_{\mu\nu}(x)\Sigma^{\mu\nu}}\qquad \Sigma^{\mu\nu}=-\frac{i}{2}[\gamma^\mu,\gamma^\nu]
\eeq
where $\omega^{\mu\nu}(x)$ is the infinitesimal Lorentz tranformation induced by the conformal transformation. For a special conformal transformation as in \eqref{special}
\beq
\delta x^\mu=\omega^{\mu}_\nu x^\nu= 1-2 b\cdot x \, x^\mu + b^\mu x^2, \qquad \omega_{\mu\nu}=-2 (b_\mu x_\nu - b_\nu x_\mu),
\eeq
giving 
\beq
\mathcal{K}^k \psi=\left(- x^2\frac{\partial}{\partial x^\kappa}+ 2 x^\kappa x^\alpha \frac{\partial}
{\partial x^\alpha} +2 \Delta_\psi x^\kappa\right)\psi +\frac{1}{2}[\gamma^k,\gamma\cdot x] \,\psi.
\eeq
Due to the conformal invariance of the quark sector in the massless limit, these identities at operatorial level  act on the $TJJ$ as a derivative on each of the coordinates 
$x_i, i=1,2,3 $ giving 
\beq
\mathcal{K}^\k \langle TJJ\rangle_q=0. 
\eeq
Defining 
\be
\Gamma^{\mu\nu\alpha\beta }_q(x_1,x_2,x_3)=\langle T^{\mu\nu}(x_1) J^\alpha(x_2)J^\beta(x_3)\rangle_q
\ee
(we omit the colour indices) their explicit expression is 
\beqa
\mathcal{K}^\kappa \Gamma^{\mu\nu\a\b}_q(x_1,x_2,x_3) 
&=& \sum_{i=1}^{3} {K_i}^{ \kappa}_{scalar}(x_i) \Gamma^{\mu\nu\a\b}_q(x_1,x_2,x_3) \nonumber \\
&&\qquad + 2 \left(  \delta^{\mu\kappa} x_{1\rho} - \delta_{\rho}^{\kappa }x_1^\mu  \right)\Gamma^{\rho \nu\alpha\beta}_q
 + 2 \left(  \delta^{\nu\kappa} x_{1\rho} - \delta_{\rho}^{\kappa }x_1^\nu  \right)\Gamma_q^{\mu\rho \alpha\beta}\nonumber \\ 
 &&\qquad  2 \left(  \delta^{\a\kappa} x_{2\rho} - \delta_{\rho}^{\kappa }x_2^\a  \right)\Gamma_q^{\mu\nu \rho\beta}
 +  2 \left(  \delta^{\beta\kappa} x_{3\rho} - \delta_{\rho}^{\kappa }x_3^\beta  \right)\Gamma_q^{\mu\nu \alpha\rho}=0,
 \eeqa

where 
\be
\label{ki}
{\mathcal{K}_i}^{\kappa}_{scalar}=-x_i^2 \frac{\partial }{\partial x_\kappa} + 2 x_i^\kappa x_i^\tau\frac{\partial}{\partial x_i^\tau} + 2 \Delta_i x_i^\kappa
\ee
is the scalar part of the special conformal operator acting on the $i_{\textrm{th}}$ coordinate. $\Delta_i\equiv(\Delta_T,\Delta_J,\Delta_J)$ are the scaling dimensions of the operators in the correlation function ($\Delta_T=3, \Delta_J=3$ at $d=4$). \\
Similarly, in the case of dilatations,where $\Omega(x)=\lambda^{-1}$ and 
$x'^\mu=\lambda x^\mu$, the condition of scale invariance gives for a correlator of $n$ primary fields, scalars or tensors, 

\be
 \label{scale1}
\Phi (\lambda x_1,\lambda x_2,\ldots,\lambda x_n)=\lambda^{-\Delta} \Phi(x_1,x_2,\ldots, x_n), \qquad 
 \Delta=\Delta_1 +\Delta_2 +\ldots \Delta_n,
 \ee
 which generates the Euler equation. 
Also in this case, the operator acts on $n-1$ of the $n$ momenta.

\subsection{The transition to momentum space in the quark sector}
It is convenient to resort to a symmetric relabeling of the three momenta in order to define the action of $\mathcal{K}$. The functional differentiation of \eqref{transverse} and \eqref{trace} allows to derive ordinary Ward identities for the various correlators. In the $TJJ$ case we obtain, after a Fourier transformation, the  
\begin{itemize}
\item{\bf conservation WIs}
\end{itemize}
\begin{align}
\label{tr}
p_{1\n_1}\braket{T^{\mu_1\nu_1}(p_1)\,J^{\m_2}(p_2)\,J^{\m_3}(p_3)}_q&=4\,\big[\d^{\m_1\m_2}p_{2\l}\braket{J^\l(p_1+p_2)\,J^{\m_3}(p_3)}_q-p_2^{\m_1}\braket{J^{\m_2}(p_1+p_2)\,J^{\m_3}(p_3)}_q\big]\notag\\
&+4\,\big[\d^{\m_1\m_3}p_{3\l}\braket{J^\l(p_1+p_3)\,J^{\m_2}(p_2)}_q-p_3^{\m_1}\braket{J^{\m_3}_q(p_1+p_3)\,J^{\m_2}(p_2)}_q\big].
\end{align}
We recall that the 2-point function of two conserved vector currents $J_i$ $(i=2,3)$ \cite{Coriano:2013jba} in any conformal field theory in $d$ dimension is given by 
\beqa
\langle J_2^\alpha(p)J_3^\beta(-p) \rangle =\delta_{\Delta_2\, \Delta_3}\left(c_{123} \Gamma_J \right)\pi^{\alpha\beta}(p) (p^2)^{\Delta_2-d/2},
\qquad \Gamma_J=\frac{\pi^{d/2}}{ 4^{\Delta_2 -d/2}}\frac{\Gamma(d/2-\Delta_2)}{\Gamma(\Delta_2)},
\eeqa
with $c_{123}$ an overall constant and $\Delta_2=d-1$. Notice that the divergence of this 2-point function is removed by the counterterm 
\beq
\mathcal{L}_{\textrm{q\, count}}=-\frac{1}{(d-4)}\frac{g_s^2}{16 \pi^2}\frac{2}{3} n_f
F^{a\, \mu\nu}F^{a}_{\mu\nu}
\eeq
where $n_f$ denotes the number of flavours. In the gluon sector a similar counterterm will be introduced 
\beq
\mathcal{L}_{\textrm{count}}=\frac{1}{(d-4)}\frac{g_s^2}{16 \pi^2}(\frac{5}{3}C_A-\frac{2}{3} n_f)
F^{a\, \mu\nu}F^{a}_{\mu\nu}.
\eeq

In our case $\D_2=\D_3=d-1$ and Eq. (\ref{tr}) then takes the form 
\begin{align}
\label{2point}
p_{1\mu_1}\braket{T^{\mu_1\nu_1}(p_1)\,J^{\m_2}(p_2)\,J^{\m_3}(p_3)}_q&=4 c_{123} \Gamma_J
\left( \delta^{\nu_1\mu_2}\frac{p_{2\lambda}}{(p_3^2)^{d/2 -\Delta_2}} {\pi^{\lambda \mu_3}}(p_3) -
\frac{p_2^{\nu_1}}{(p_3^2)^{d/2 -\Delta_2}}\pi^{\mu_2\mu_3}(p_3) \right.\nn\\
& \left. + \delta^{\nu_1\mu_3}\frac{p_{3\lambda}}{(p_2^2)^{d/2-\Delta_2}}\pi^{\lambda \mu_2}(p_2) -\frac{p_3^{\nu_1}}{(p_2^2)^{d/2-\Delta_2}}\pi^{\mu_3\mu_2}(p_2)\right).
\end{align}
The equation can be checked perturbatively in the quark sector by the ordinary free field theory representation. In this case the constant $c_{123}$ is simply proportional to $n_f$, the number of quark flavours running in the quark loops. \\
One can generalize this equation in the quark case to higher orders in the number of external gluons, by the inclusion of additional gluon currents on the external lines.  
Specifically, at higher order in the hierarchy of the $TJJ$ we have the conservation WI
\begin{equation}
	\begin{aligned}
		0=&q_{\mu } \langle T^{\mu\nu}(q)J^{\alpha a_1}(p_1)J^{\beta a_2}(p_2)J^{\gamma a_3}(p_3)\rangle_q+2
		\delta^{\alpha}_{[\mu}{p_1}_{\nu]}\langle J^{\mu a_1}(p_1-q)J^{\beta a_2}(p_2)J^{\gamma a_3}(p_3) \rangle_q\\&
		+2
		\delta^{\beta}_{[\mu}{p_2}_{\nu]}\langle J^{\alpha a_1}(p_1)J^{\mu a_2}(p_2-q)J^{\gamma a_3}(p_3) \rangle_q
		+2
		\delta^{\gamma}_{[\mu}{p_3}_{\nu]}\langle J^{\alpha a_1}(p_1)J^{\beta a_2}(p_2)J^{\mu a_3}(p_3-q) \rangle_q
		\\&-2i gf^{a_1 a_2 c} \delta^{[\alpha}_\nu \langle J^{\beta] c}(p_1+p_2-q)J^{\gamma a_3}(p_3)\rangle_q
		-2i gf^{a_1 a_3 c} \delta^{[\alpha}_\nu \langle J^{\gamma] c}(p_1+p_3-q)J^{\beta a_2}(p_2)\rangle_q
				\\&-2i gf^{a_3 a_2 c} \delta^{[\gamma}_\nu \langle J^{\beta] c}(p_3+p_2-q)J^{\alpha a_1}(p_1)\rangle_q,
	\end{aligned}
\end{equation}
derived from diffeomorphism invariance.
The hierarchical equation of the $TJJJJ$ can be found in Appendix \ref{t4j}.
\\

 \begin{itemize}
 \item{\bf gauge WIs}
 \end{itemize}
 These identities take the form
\begin{align}
\label{x1}
p_{2\m_2}\braket{T^{\mu_1\nu_1}(p_1)\,J^{\m_2}(p_2)\,J^{\m_3}(p_3)}_q&=0\\
p_{3\m_3}\braket{T^{\mu_1\nu_1}(p_1)\,J^{\m_2}(p_2)\,J^{\m_3}(p_3)}_q&=0,
\end{align}
These two equations are homogeneous because the $TJ$ two-point function, which is expected on their right-hand sides, is trivially zero. \\
We also have
\begin{equation}
	\begin{aligned}
			0=&p_{1 \alpha } \langle T^{\mu\nu}(q)J^{\alpha a_1}(p_1)J^{\beta a_2}(p_2)J^{\gamma a_3}(p_3)\rangle_q\\&+i gf^{a_1 a_2 c}  \langle T^{\mu\nu}(q)J^{\beta c}(p_2+p_1)J^{\gamma a_3}(p_3)\rangle_q+i gf^{a_1 a_3 c}  \langle T^{\mu\nu}(q)J^{\beta a_2}(p_2)J^{\gamma c}(p_3+p_1)\rangle_q
	\end{aligned}
\end{equation}
and 
\begin{equation}
	\begin{aligned}
		0=&p_{1 \alpha } \langle T^{\mu\nu}(q)J^{\alpha a_1}(p_1)J^{\beta a_2}(p_2)J^{\gamma a_3}(p_3)J^{\delta a_4}(p_4)\rangle_q+i gf^{a_1 a_2 c}  \langle T^{\mu\nu}(q)J^{\beta c}(p_2+p_1)J^{\gamma a_3}(p_3)J^{\delta a_4}(p_4)\rangle_q\\ &+i gf^{a_1 a_3 c}  \langle T^{\mu\nu}(q)J^{\beta a_2}(p_2)J^{\gamma c}(p_3+p_1)J^{\delta a_4}(p_4)\rangle_q
		+i gf^{a_1 a_4 c}  \langle T^{\mu\nu}(q)J^{\beta a_2}(p_2)J^{\gamma a_3}(p_3)J^{\delta c}(p_4+p_1)\rangle_q
	\end{aligned}
\end{equation}
Applying functional derivatives to \eqref{ann} with respect to the gauge fields and Fourier transforming, one obtains similarly to 
\eqref{Gamanomtr} but to higher orders in the coupling constant $g$ 
\begin{itemize}
\item{\bf trace WIs}
\end{itemize}

\begin{equation}
\label{x2}
g_{\mu\nu} \langle T^{\mu\nu}(q)J^{\alpha a_1}(p_1)J^{\beta a_2}(p_2)\rangle_q=\beta u^{\alpha \beta  \, a_1a_2}
\end{equation}
and those of higher order in the hierarchy
\begin{equation}
	g_{\mu\nu} \langle T^{\mu\nu}(q)J^{\alpha a_1}(p_1)J^{\beta a_2}(p_2)J^{\gamma a_3}(p_3)\rangle_q=\beta u^{\alpha \beta \gamma \, a_1a_2a_3}
\end{equation}

\begin{equation}
	g_{\mu\nu} \langle T^{\mu\nu}(q)J^{\alpha a_1}(p_1)J^{\beta a_2}(p_2)J^{\gamma a_3}(p_3)J^{\delta a_4}(p_4)\rangle_q=\beta u^{\alpha \beta \gamma \delta \, a_1a_2a_3 a_4}.
\end{equation}
Finally, considering the 
 
\begin{itemize}
\item{\bf conformal Ward identities}
\end{itemize}
 in the quark sector, the non-anomalous special conformal equation takes the form 
\begin{equation}
	\begin{aligned}
			0=&\mathcal{K}^k\langle T^{\mu\nu}(q)J^{\mu_1 a_1}(p_1)\dots J^{\mu_{n-1} a_{n-1}}(p_{n-1})\rangle_q=\\&\sum_{j=1}^{n-1}\left(p_j^\kappa \frac{\partial^2}{\partial {p_j}_\alpha \partial p_j^\alpha}+2\left(\Delta_j-d\right) \frac{\partial}{\partial p_j^\kappa}-2 p_j^\alpha \frac{\partial^2}{\partial p_j^\kappa \partial p_j^\alpha}\right)\langle T^{\mu\nu}(q)J^{\mu_1 a_1}(p_1)\dots J^{\mu_{n-1} a_{n-1}}(p_{n-1})\rangle_q\\&
		+2 \sum_{j=1}^{n-1} \left[\delta^{\kappa \mu_j} \frac{\partial}{\partial p_j^{\alpha_{j}}}-\delta_{\alpha_{j}}^\kappa \frac{\partial}{\partial {p_j}_{\mu_j}}\right]\langle T^{\mu\nu}(q)J^{\mu_1 a_1}(p_1)\dots J^{\alpha_j a_j}(p_j)\dots J^{\mu_{n-1} a_{n-1}}(p_{n-1})\rangle_q
	\end{aligned}
\end{equation}
while the dilatation WIs are given by
\begin{equation}
	0=\left[\sum_{j=1}^n \Delta_j-(n-1) d-\sum_{j=1}^{n-1} p_j^\lambda \frac{\partial}{\partial p_j^\lambda}\right]\langle T^{\mu\nu}(q)J^{\mu_1 a_1}(p_1)\dots J^{\mu_{n-1} a_{n-1}}(p_{n-1})\rangle_q.
\end{equation}

\section{BRST symmetry of the gauge-fixed action and the complete partition function}
\label{BRSTX}
As we include the gluon sector in the path integral representation, special conformal symmetry is broken by the gauge-fixing term in the QCD Lagrangean. For this reason, the natural approach in the investigaton of the conformal behaviour of the theory is to turn to the effective action with a Legrendre transform of the complete partition function.
The condition of diffeomorphism invariance of the generating functional $Z$ gives
\beq\label{covW}
Z[V,J,J^\dag,\c,\bar{\c},g] = Z[V',J',J^{'\,\dag},\c',\bar{\c}',J^{'\,\mu}, g']\, ,\eeq
where we have allowed an arbitrary change of coordinates $x^{'\,\mu} = F^\mu(x)$ on the spacetime manifold, which can be
parameterized locally as $x^{'\,\mu} = x^\mu + \e^\mu(x)$.
 The measure of integration is invariant under general coordinate transformations under such changes $( \mD\Phi' = \mD\Phi)$ as far as we stay in $d$ dimensions
 and we obtain to first order in $\e^\mu(x)$
\beqa\label{preWard}
\int \mD\Phi \, e^{i \tilde{S}}
& = & \int \mD\Phi \quad e^{i \tilde{S}}\,\bigg( 1  + i\int d^4xd^4y\, \bigg\{- V\Theta^\mu_{\,\,\,\underline{a}}\bigg[-\d^{(4)}(x-y)\pd_\nu V^{\underline{a}}_\mu(x)
- [\pd_\mu\d^{(4)}(x-y)]V^{\underline{a}}_\nu\bigg]\nn\\
&& - \pd_\nu[\d^{(4)}(x-y)\bar{\c}(x)]\psi(x) - \bar{\psi}(x)\pd_\nu[\d^{(4)}(x-y)\c(x)] \nn\\
&& - [\pd_\nu[J^\mu(x)\d^{(4)}(x-y)] + [\pd_\r\d^{(4)}(x-y)]\d^\mu_\nu J^\r(x)]A_\mu(x)\bigg\}\e^\nu(y)\bigg),
\eeqa

\beq \Theta^{\mu\nu} = - \frac{1}{V}\frac{\d S}{\d e^{{\underline{a}}}_\mu}e^{{\underline{a}}\nu}, \qquad 
\Theta^{\mu\ul{a}}= \Theta^{\mu\nu} e^{\ul{a}}_{\nu} \eeq
in terms of the determinant of the vielbein $V(x){\equiv} \left|V^{\underline{a}}_\mu(x)\right|$.
Notice that this expression of the EMT is non-symmetric. The symmetric expression can be easily defined by the relation
 \beq T^{\mu\nu} = \frac{1}{2}(\Theta^{\mu\nu} + \Theta^{\nu\mu})\,
 \label{tmunu} \eeq
that will be used below.\\
\eqref{preWard}  can be brought into the form 
\beqa\label{Ward}
 &&  \int\mD\Phi\, e^{i\tilde{S}}\, \bigg[\pd_\a T^{\a\b}(y)  -  J^\a(y)\pd^\b A_\a(y) + \pd_\a[J^\a(y) A^\b(y)]\nn\\
  && -  \pd^\b\bar{\psi}(y)\c(y) - \bar{\c}(y)\pd^\b\psi(y)  -      \frac{1}{2}\pd_\a\bigg(\frac{\d S}{\d\psi(y)}\s^{\a\b}\psi(y)
            - \bar{\psi}(y)\s^{\a\b}\frac{\d S}{\d\bar{\psi}(y)}\bigg)\bigg] = 0\,.
\eeqa
where 
\beqa
\label{ActionToyModel}
\tilde{S}&            =          & S + i\int d^4x\,\left( J^{\mu a} A^a_\mu  + \bar{\c}(x)\psi(x) + \textrm{h.c.}\right).\,
\eeqa
The conservation equation of the energy-momentum tensor takes the following form off-shell 
\bea
\partial^{\mu}T_{\mu\nu} &=& -\frac{\delta S}{\delta \psi} \partial_{\nu}\psi - \partial_{\nu}\bar \psi \frac{\delta S}{\delta \bar \psi} + \frac{1}{2}\partial^{\mu}\left( \frac{\delta S}{\delta \psi}\sigma_{\mu\nu}\psi - \bar \psi \sigma_{\mu\nu}\frac{\delta S}{\delta \bar \psi} \right) - \partial_{\nu}A_{\mu}^a \frac{\delta S}{\delta A_{\mu}^a} \nn\\
&+& \partial_{\mu}\left( A_{\nu}^a \frac{\delta S}{\delta A_{\mu}^a} \right) - \frac{\delta S}{\delta c^a} \partial_{\nu}c^a - \partial_{\nu}\bar c^a\frac{\delta S}{\delta \bar c^a} \,  
\label{EMTdivergence}
\eea
where $\sigma_{\mu\nu}= \frac{1}{4}[\gamma_{\mu},\gamma_{\nu}]$, where ${S}$ is the QCD action.\\
The conservation of the energy-momentum tensor summarized in Eq.~(\ref{EMTdivergence}) in terms of classical fields, can be re-expressed
in a functional form by a differentiation of $W$ with respect to $h_{\mu\nu}$ and the use of Eq.~\eqref{EMTdivergence}
under the functional integral. One derives the relation \cite{Armillis:2010qk}
\bea
\partial_{\mu} \frac{\delta \mathcal{S}_{eff}}{\delta h_{\mu\nu}} &=& - \frac{\delta \mathcal{S}_{eff}}{\delta \psi_c}\partial^{\nu}\psi_c
- \partial^{\nu}\bar\psi_c \frac{\delta\mathcal{S}_{eff}}{\delta \bar \psi_c} + \frac{1}{2}\partial_{\mu}\left( \frac{\delta \mathcal{S}_{eff}}{\delta \psi_c} \sigma^{\mu\nu} \psi_c
- \bar \psi_c \sigma^{\mu\nu}\frac{\delta \mathcal{S}_{eff}}{\delta \bar \psi_c}  \right) - \partial^{\nu}A^{\mu}_c\frac{\delta \mathcal{S}_{eff}}{\delta A^{\mu}_c}
+\partial^{\mu}\left(A^{\nu}_c\frac{\delta \mathcal{S}_{eff}}{\delta A^{\mu}_c} \right) \nn \\
&-& \frac{\delta \mathcal{S}_{eff}}{\delta c_c}\partial^{\nu}c_c - \partial^{\nu}\bar c_c \frac{\delta \mathcal{S}_{eff}}{\delta \bar c_c}
\label{FuncWI},
\eea
obtained from Eq.~\eqref{EMTdivergence} with the help of Eqs.~(\ref{Legendre1}), (\ref{Legendre2}), (\ref{Legendre3}). Some additional details on the derivation of these identities can be found in Appendix \ref{STTI} and in \cite{Coriano:2011zk}, where the analysis is extended to all the gauge sectors of the Standard Model. The relevant 
WIs and STIs that can be used in order to fix the expression of the correlator in terms of truncated graphs are given by
\bea
\partial^{\mu}\langle T_{\mu\nu}(x) A_\alpha^a (x_1)A_{\beta}^b (x_2)\rangle_{trunc}  &=& - \partial_{\nu}\delta^4(x_1-x) D^{-1}_{\alpha\beta}(x_2,x)
- \partial_{\nu}\delta^4(x_2-x) D^{-1}_{\alpha\beta}(x_1,x) \nn \\
&+& \partial^{\mu}\left( g_{\alpha \nu} \delta^4(x_1-x) D^{-1}_{\beta\mu}(x_2,x) + g_{\beta \nu} \delta^4(x_2-x) D^{-1}_{\alpha\mu}(x_1,x)\right) \label{WIcoord} \nonumber \\
\eea
where $D^{-1}_{\alpha\beta}(x_1,x_2)$ is the inverse gluon propagator defined as
\bea
D^{-1}_{\alpha\beta }(x_1,x_2) = \langle A_{\alpha}(x_1) A_{\beta}(x_2) \rangle_{trunc}  =  \frac{\delta^2 \mathcal{S}_{eff}}{\delta A^{\alpha }_c(x_1) \delta A^{\beta }_c(x_2) }
\eea
and where we have omitted, for simplicity, both the colour indices and the symbol of the $T$-product.
The first Ward identity (\ref{WIcoord}) becomes
\bea
k^{\mu}\langle T_{\mu\nu}(k) A_{\alpha}(p) A_{\beta}(q) \rangle_{trunc} =
q_{\mu} D^{-1}_{\alpha\mu}(p) g_{\beta\nu} + p_{\mu} D^{-1}_{\beta\mu}(q) g_{\alpha\nu}  - q_{\nu} D^{-1}_{\alpha\beta}(p) - p_{\nu} D^{-1}_{\alpha\beta}(q) \, .
\label{WImom}
\eea
with 
\bea
D^{-1}_{\alpha\beta}(p) = (p^2 g_{\alpha \beta} - p_{\alpha} q_{\beta}) \Pi (p^2)
\eea
being the gluon inverse 2-point function in momentum space and $\Pi(p^2)$ its scalar form factor. 
A second WI can be derived using the BRST symmetry of $\mathcal{S}_{eff}$ involving two derivative. 
For this purpose, we choose an appropriate  Green's function,  $\langle T_{\mu\nu} \partial^{\alpha} A_{\alpha}^a \bar c^b \rangle$, and then use its BRST invariance to obtain

\bea
\delta \langle T_{\mu\nu} \partial^{\alpha} A_{\alpha}^a \bar c^b \rangle = \langle \delta T_{\mu\nu} \partial^{\alpha} A_{\alpha}^a \bar c^b \rangle + \lambda \langle T_{\mu\nu} \partial^{\alpha} D^{ac}_{\alpha} c^c \bar c^b \rangle - \frac{\lambda}{\xi} \langle T_{\mu\nu} \partial^{\alpha} A_{\alpha}^a \partial^{\beta} A_{\beta}^b \rangle = 0,
\label{BRSTward}
\eea
where the first two correlators, built with operators proportional to the equations of motion, contribute only with disconnected amplitudes, that are not part of the one-particle irreducible vertex function.
 From Eq.~(\ref{BRSTward}) we obtain the identity
 \bea
\partial_{x_1}^{\alpha}\partial_{x_2}^{\beta}\langle T_{\mu\nu}(x) A_{\alpha}^a(x_1)A_{\beta}^b(x_2)\rangle_{trunc} = 0,
\eea
which in momentum space becomes
\bea
p_1^{\alpha} p_2^{\beta} \langle T_{\mu\nu}(k) A_{\alpha}^a(p_1)A_{\beta}^b(p_2)\rangle_{trunc} = 0. \label{brstwi}
\label{pract1}
\eea
The broken conformal WIs of QCD in $d$ dimensions can be obtained by imposing the invariance of the generating functional 
$W[J,\eta,\bar\eta,\chi,\bar\chi,h] $ under a conformal change of coordinates in the integrand and then rewriting the 
constraint in the terms of $\mathcal{S}_{eff}$ using \eqref{Legendre1} and \eqref{Legendre2} \cite{Nielsen:1975ph}. They can be recast as  constraints on the 1PI vertices and truncated 2-point functions such as the $TJJ$ and the truncated $JJ$. As already mentioned in the Introduction, in the gluon sector we will not solve these equations directly using CFT$_p$ as for the quark sector, but rely on a direct one-loop computations of the gluon contibutions, that will be added to the former. This is sufficient in order to obtain a consistent decomposition of the correlator in the LT-basis and proceed with an analysis of all of its sector i  the off-shell gluon case.

 \section{The off-shell form factors from the LT decomposition of CFT$_p$}
\label{seven}
	%%%%%%%%%%%%%%%%%%%%%%%%%%%%%%%			%%%%%%%%%%%%%%%%%%%%%%%%%%%%%%%
In this section, we begin by discussing the sector decomposition of the $TJJ$ to facilitate a direct analysis of its form factors. This decomposition applies to both the quark and gluon sectors, with their form factors initially treated separately before being combined.\\
The general form of the non-Abelian $\langle TJJ \rangle$ correlator can be constructed through a decomposition into transverse,  longitudinal and trace terms \cite{Bzowski:2013sza}, exploiting its symmetries. The analysis includes additional contributions not found in the general expression for the $TJJ$ vertex derived in \cite{Bzowski:2018fql} using the CFT approach. These extra terms manifest as longitudinal components, which are naturally present in perturbative QCD (pQCD) but absent in the abstract conformal treatment of non-Abelian correlators with gauge currents.\\
We consider the decomposition of the operators $T$ and $J$ in terms of their transverse traceless part and longitudinal (local) one, separating the quark and gluon parts, that, as we are going to see, behave differently under the application of the conformal constraints. Following \cite{Bzowski:2013sza} we define
	\begin{align}
		T^{\mu_i\nu_i}(p_i)&\equiv t^{\mu_i\nu_i}(p_i)+t_{loc}^{\mu_i\nu_i}(p_i),\label{decT}\\
		J^{a_i \, \mu_i}(p_i)&\equiv j^{a_i \, \mu_i}(p_i)+j_{loc}^{a_i \, \mu_i}(p_i),\label{decJ}
	\end{align}
	where
	\begin{align}
		\label{loct}
		t^{\mu_i\nu_i}(p_i)&=\Pi^{\mu_i\nu_i}_{\alpha_i\beta_i}(p_i)\,T^{\alpha_i \beta_i}(p_i), &&t_{loc}^{\mu_i\nu_i}(p_i)=\Sigma^{\mu_i\nu_i}_{\alpha_i\beta_i}(p)\,T^{\alpha_i \beta_i}(p_i),\\
		j^{a_i \, \mu_i}(p_i)&=\pi^{\mu_i}_{\alpha_i}(p_i)\,J^{a_i \, \alpha_i }(p_i), &&\hspace{1ex}j_{loc}^{a_i \, \mu_i}(p_i)=\frac{p_i^{\mu_i}\,p_{i\,\alpha_i}}{p_i^2}\,J^{a_i \, \alpha_i}(p_i).
	\end{align}
	having introduced the transverse-traceless ($\Pi$), transverse $(\pi)$, longitudinal ($\Sigma$) projectors, given respectively by 
	\begin{align}
		\label{prozero}
		\pi^{\mu}_{\alpha} & = \delta^{\mu}_{\alpha} - \frac{p^{\mu} p_{\alpha}}{p^2}, \\
		%%%%%%%%%%%%%%%%%%%%%%%%%%%%%%%
		\Pi^{\mu \nu}_{\alpha \beta} & = \frac{1}{2} \left( \pi^{\mu}_{\alpha} \pi^{\nu}_{\beta} + \pi^{\mu}_{\beta} \pi^{\nu}_{\alpha} \right) - \frac{1}{d - 1} \pi^{\mu \nu}\pi_{\alpha \beta}\label{TTproj}, \\
		%%%%%%%%%%%%%%%%%%%%%%%%%%%%%%%
		\Sigma^{\mu_i\nu_i}_{\alpha_i\beta_i}&=\frac{p_{i\,\beta_i}}{p_i^2}\Big[2\delta^{(\nu_i}_{\alpha_i)}p_i^{\mu_i)}-\frac{p_{i\alpha_i}}{(d-1)}\left(\delta^{\mu_i\nu_i}+(d-2)\frac{p_i^{\mu_i}p_i^{\nu_i}}{p_i^2}\right)\Big]+\frac{\pi^{\mu_i\nu_i}(p_i)}{(d-1)}\delta_{\alpha_i\beta_i}\equiv\mathcal{I}^{\mu_i\nu_i}_{\alpha_i}p_{i\,\beta_i} +\frac{\pi^{\mu_i\nu_i}(p_i)}{(d-1)}\delta_{\alpha_i\beta_i}\label{Lproj}.
		%%%%%%%%%%%%%%%%%%%%%%%%%%%%%%%
	\end{align}
	
with
\begin{equation} \label{a:T}
\mathscr{T}_{\mu\nu \alpha} (\bs{p}) = \frac{1}{p^2} \left[ 2 p_{(\mu} \delta_{\nu)\alpha} - \frac{p_\alpha}{d-1} \left( \delta_{\mu\nu} + (d-2) \frac{p_\mu p_\nu}{p^2} \right) \right]
\end{equation}	
and
\[\label{Idecomp}
\delta_{\mu(\alpha}\delta_{\beta)\nu}=\Pi_{\mu\nu\alpha\beta}(\bs{p})+\mathscr{T}_{\mu\nu (\alpha}(\bs{p})\,p_{\beta)}+\frac{1}{d-1}\pi_{\mu\nu}(\bs{p})\delta_{\alpha\beta}.
\]
	Turning to the $TJJ$ case, we can divide the 3-point function into two parts: the \emph{transverse-traceless} part and the \emph{semi-local} part (indicated by subscript $loc$) expressible through the transverse and trace Ward Identities. These parts are obtained by using the projectors $\Pi$ and $\Sigma$, previously defined. 
	We can then decompose the full 3-point function as 
	
	\begin{align}
	\label{dec1}
		\braket{T^{\m \n}\,J^{a \, \a}\,J^{b \, \b}}&=
		\braket{t^{\m \n }\,j^{a \, \a}\,j^{b \, \b}}+\braket{T^{\m \n}\,J^{a \, \a}\,j_{loc}^{b \, \b}}+\braket{T^{\m \n}\,j_{loc}^{a \, \a}\,J^{b \, \b}}+\braket{t_{loc}^{\m \n}\,J^{a \, \a}\,J^{b \, \b}}\notag\\
		&\quad-\braket{T^{\m \n }\,j_{loc}^{a \, \a}\,j_{loc}^{b \, \b}}-\braket{t_{loc}^{\m \n}\,j_{loc}^{a \, \a}\,J^{b \, \b}}-\braket{t_{loc}^{\m \n}\,J^{a \, \a}\,j_{loc}^{b \, \b}}+\braket{t_{loc}^{\m \n }\,j_{loc}^{a \, \a}\,j_{loc}^{b \, \b}}.
	\end{align}
In a CFT approach, all the terms on the right-hand side of the decomposition, apart from the first one, may be computed by means of transverse and trace Ward Identities, with the anomaly induced by the renormalization of the hierachy by a single counterterm, proportional to the square of the field strength $(F^2)$. 

 \section{The sectors decomposition and the solution of the CWIs in the quark sector}
 \label{eight}
Using the projectors $\Pi$ and $\pi$ one can write the most general form of the transverse-traceless part as
	\begin{equation}
		{\braket{t^{\m \n}(p_1)\,j^{a \, \a}(p_2)\,j^{b \, b }(p_3)}}_q =\Pi^{\m \n}_{\m_1 \n_1}(p_1)\pi^{\a}_{\a_1 }(p_2)\pi^{\b}_{\b_1}(p_3)\,\,X^{a b \, \m_1 \n_1 \a_1 \b_1}_q,
	\end{equation}
	where $X^{a b \, \m_1 \n_1 \a_1 \b_1}_q$ is a general tensor of rank four built out of the metric and momenta. We can enumerate all possible tensor that can appear in $X^{a b \, \m_1 \n_1 \a_1 \b_1}$ preserving the symmetry of the 
	correlator 
	\begin{align}
		\langle t^{\mu \nu }(p_1)j^{a \, \a}(p_2)j^{b \, \b}(p_3)\rangle_q & =
		{\Pi_1}^{\mu \nu}_{\m_1 \n_1}{\pi_2}^{\a}_{\a_1}{\pi_3}^{\b}_{\b_1}
		\left( A_{1}^{(q) a b} \ p_2^{\m_1 }p_2^{\n_1}p_3^{\a_1}p_1^{\b_1} + 
		A_{2}^{(q) a b}\ \delta^{\a_1 \b_1} p_2^{\m_1}p_2^{\n_1} + 
		A_{3}^{(q) a b}\ \delta^{\m_1\a_1}p_2^{\n_1}p_1^{\b_1}\right. \notag\\
		& \left. + 
		A_{3}^{(q) a b}(p_2\leftrightarrow p_3)\delta^{\m_1\b_1}p_2^{\n_1}p_3^{\a_1}
		+ A_{4}^{(q) a b}\  \delta^{\m_1\b_1}\delta^{\a_1\n_1}\right)\label{DecompTJJ}
	\end{align}
	with the reconstruction taking the form \cite{Bzowski:2013sza}
\begin{framed}
	\begin{align}
& \la T_{\mu_1 \nu_1}(\bs{p}_1) J^{\mu_2 a_2}(\bs{p}_2) J^{\mu_3 a_3}(\bs{p}_3) \ra_q = \la t_{\mu_1 \nu_1}(\bs{p}_1) j^{\mu_2 a_2}(\bs{p}_2) j^{\mu_3 a_3}(\bs{p}_3) \ra_q \nn\\[0.5ex]
& \qquad +  2 \mathscr{T}_{\mu_1 \nu_1}^{\quad\,\,\alpha}(\bs{p}_1)\Big[\delta_{[\alpha}^{\mu_3}p_{3\beta]}  \la J^{\mu_2 a_2}(\bs{p}_2) J^{\beta a_3}(-\bs{p}_2) \ra_q
+\delta_{[\alpha}^{\mu_2}p_{2\beta]}  \la J^{\mu_3 a_3}(\bs{p}_3) J^{\beta a_2}(-\bs{p}_3) \ra_q\Big]
 \nn\\
& \qquad + \frac{1}{d - 1}\,\pi_{\mu_1 \nu_1}(\bs{p}_1) \mathcal{A}^{\mu_2 \mu_3 a_2 a_3}_q, \label{tjjdec}
\end{align}
\end{framed}
where $\mathscr{T}_{\mu_1 \nu_1\alpha}$ is defined in \eqref{a:T} and $\langle JJ\rangle_q$ is the 2-point function of the gluon with a virtual quark. 

The transverse Ward identities are 
\begin{align}
& p_1^{\nu_1} \la T_{\mu_1 \nu_1}({p}_1) J^{\mu_2 a_2}({p}_2) J^{\mu_3 a_3}({p}_3) \ra_q  \nn\\[0.5ex]
& \qquad = \: 2 \delta^{\mu_3}_{[\mu_1} p_{3\alpha]} \la J^{\mu_2 a_2}({p}_2) J^{\alpha a_3}(-{p}_2) \ra_q + 2 \delta^{\mu_2}_{[\mu_1} p_{2\alpha]} \la J^{\alpha a_2}({p}_3) J^{\mu_3 a_3}(-{p}_3) \ra_q, \\[1ex]\label{TWI_TJJ_2}
& p_{2 \mu_2} \la T_{\mu_1 \nu_1}({p}_1) J^{\mu_2 a_2}({p}_2) J^{\mu_3 a_3}({p}_3) \ra_q = 0, \\[1ex]
\end{align}
while the anomalous trace WI is given by
\begin{align}
& \la T({p}_1) J^{\mu_2 a_2}({p}_2) J^{\mu_3 a_3}({p}_3) \ra_q = \mathcal{A}^{\mu_2 \mu_3 a_2 a_3}_q, 
\end{align}
where 
\beq
\la J^{\alpha a}({p}_1) J^{\beta b}(-{p}_1) \ra_q=
 \frac{2}{3}n_f  \,  \frac{g_s^2}{16\pi ^2} {B}_0\left({{p_1}}^2\right) {p_1}^2   \delta ^{a b}\pi^{\alpha\beta}(p_1).\eeq 

They satisy the renormalized dilatation equations (diagonal in colour space) 
\cite{Bzowski:2017poo}\cite{Coriano:2018bbe}

\begin{align}
\left(\sum_{i}^{3}p_i\sdfrac{\partial}{\partial p_i}+2\right)\,A^{(q)}_1&=0=-\m\sdfrac{\partial}{\partial \m}A^{(q)}_1\\
\left(\sum_{i}^{3}p_i\sdfrac{\partial}{\partial p_i}\right)\,A^{(q)}_2&=\sdfrac{8\p^2\,g^2}{3}=-\m\sdfrac{\partial}{\partial \m}A^{(q)}_2\\
\left(\sum_{i}^{3}p_i\sdfrac{\partial}{\partial p_i}\right)\,A^{(q)}_3(p_2\leftrightarrow p_3)&=\sdfrac{8\p^2\,g^2}{3}=-\m\sdfrac{\partial}{\partial \m}A^{(q)}_3\\
\left(\sum_{i}^{3}p_i\sdfrac{\partial}{\partial p_i}-2\right)\,A^{(q)}_4&=-\sdfrac{4}{3}\p^2\,g^2(s-s_1-s_2)=-\m\sdfrac{\partial}{\partial \m}A^{(q)}_4.
\end{align}
 while for the primary CWI's take the form 
 \begin{equation}
\label{fst1}
\begin{split}
&K_{13}A^{(q)}_1=0\\
&K_{13}A^{(q)}_2=-2A^{(q)}_1\\
&K_{13}A^{(q)}_3=4A^{(q)}_1\\
&K_{13}A^{(q)}_3(p_2\leftrightarrow p_3)=0\\
&K_{13}A^{(q)}_4=2A^{(q)}_3(p_2\leftrightarrow p_3)-\sdfrac{16\,\pi^2 g^2}{3}
\end{split}
\hspace{1.5cm}
\begin{split}
&K_{23}A^{(q)}_1=0\\
&K_{23}A^{(q)}_2=0\\
&K_{23}A^{(q)}_3=4A^{(q)}_1\\
&K_{23}A^{(q)}_3(q_2\leftrightarrow q_3)=-4A^{(q)}_1\\
&K_{23}A^{(q)}_4=-2A^{(q)}_3+2A^{(q)}_3(p_2\leftrightarrow p_3).
\end{split}
\end{equation}
similar to the Abelian case, discussed in \cite{Bzowski:2013sza}\cite{Coriano:2018bbe}

\be
{ K}_i\equiv \frac{\partial^2}{\partial    p_i \partial    p_i} 
+\frac{d + 1 - 2 \Delta_i}{   p_i}\frac{\partial}{\partial   p_i} \qquad  K_{l n}\equiv K_l -K_n \qquad i,l,n=1,2,3
\ee
In $d=4$ the operator $K_i$ take the forms
\begin{equation}
K_1\equiv 4s\sdfrac{\partial^2}{\partial s^2}-4\sdfrac{\partial}{\partial s},\quad K_2\equiv 4s_1\sdfrac{\partial^2}{\partial s_1^2},\quad K_3\equiv 4s_2\sdfrac{\partial^2}{\partial s_2^2}
\end{equation}
The solutions of these equations in terms of ordinay master integrals $B_0$ and $C_0$, is completely equivalent to the solutions expressed in terms of 3K integrals, i.e. integrals of the product of three Bessel functions, as discussed in the appendix \ref{dd}. The differential equations for master integrals allow to reformulate the 3K solutions in terms of the ordinary free-field theory realization. \\
It is possible, in the quark sector, to check the general CFT solution against the perturbative one. 
For this purpose one can use the identities \cite{Coriano:2018bbe}

\begin{align}
\sdfrac{\partial}{\partial s}\,C_0&= \sdfrac{1}{s\,\s}\,\big[s(s_1+s_2-s)C_0+B_{0,R}(s_1)(s+s_1-s_2)+B_{0,R}(s_2)(s-s_1+s_2)-2s\,B_{0,R}(s)\big]\\
\sdfrac{\partial}{\partial s_1}\,C_0&= \sdfrac{1}{s_1\,\s}\,\big[s_1(s+s_2-s_1)C_0+B_{0,R}(s)(s+s_1-s_2)+B_{0,R}(s_2)(s_1-s+s_2)-2s_1\,B_{0,R}(s_1)\big]\\
\sdfrac{\partial}{\partial s_2}\,C_0&= \sdfrac{1}{s_2\,\s}\,\big[s_2(s+s_1-s_2)C_0+B_{0,R}(s_1)(s_2+s_1-s)+B_{0,R}(s)(s-s_1+s_2)-2s_2\,B_{0,R}(s_2)\big].
\end{align}
We have defined $\s=s^2-2s(s_1+s_2)+(s_1-s_2)^2$,  $B_{0,R}(s_i)\equiv B_{0,R}(s_i,0,0)=2-\log\left(-\sdfrac{s_i}{\mu^2}\right)$ and, for simplicity, $C_0\equiv C_0(s,s_1,s_2)$. \\
The tensor nature of the correlator necessitates the imposition of additional first-order differential constraints, referred to as secondary conformal Ward identities (CWIs) in \cite{Bzowski:2013sza}. 
These constraints can be solved at specific kinematic points as when the invariant masses of the two photons are equal (\(p_2^2 = p_3^2\)) or in the massless limit of the graviton line. We have left to appendix \ref{secc} a more detailed discussion of the procedure.
At these points, the undetermined constants in the general solutions of the primary CWIs are constrained. The secondary CWIs are associated with the longitudinal and trace components of the correlators, and consequently, with contact terms.  Defining

 \begin{equation}
A_{j }^{(q) a b}= -\,  n_f\,   \frac{g_s^2}{16\pi^2}\, \bar{A}_j^{{(q)}a b}, \qquad j=1,2\ldots 4
 \end{equation}
 where the label $(q)$ refers to the contribution from a single quark in the first diagram 
 of Fig. 4, a direct computation gives
 
\begin{eqnarray}
\label{Abar}
		\bar{A}_1^{(q) ab}  &=& \frac{\delta^{ab}}{48 \left(p_1^2 p_2^2-(p_1 \cdot p_2)^2 \right)^4} \Big[ A_{10} + A_{11}  B_0 (p_1^2) + A_{12} B_0(p_2^2) + A_{13} B_0(q^2) + A_{14} C_0 (p_1^2 , p_2^2 , q^2 ) \Big] \notag \\
		\bar{A}_2^{(q) ab}  &=& -\frac{\delta ^{ab}}{144 \left(p_1^2 p_2^2-(p_1 \cdot p_2)^2 \right)^3} \Big[ A^{(q)}_{20} + A^{(q)}_{21}  B_0 (p_1^2) + A^{(q)}_{22} B_0(p_2^2)  + A^{(q)}_{23} B_0(q^2) + A^{(q)}_{24} C_0 (p_1^2 , p_2^2 , q^2 ) \Big] \notag \\
		\bar{A}_3^{(q) ab}  &=& \frac{\delta ^{ab}}{72 \left(p_1^2 p_2^2-(p_1 \cdot p_2)^2 \right)^3} \Big[A^{(q)}_{31}  B_0 (p_1^2) + A^{(q)}_{32} B_0(p_2^2)  + A^{(q)}_{33} B_0(q^2) + A^{(q)}_{34} C_0 (p_1^2 , p_2^2 , q^2 ) \Big] \notag \\
		\bar{A}_4^{(q) ab}  &=& -\frac{ \delta^{ab}}{72 \left(p_1^2 p_2^2-(p_1 \cdot p_2)^2 \right)^2} \Big[A^{(q)}_{40} + A^{(q)}_{41}  B_0 (p_1^2) + A^{(q)}_{42} B_0(p_2^2) + A^{(q)}_{43} B_0(q^2) + A^{(q)}_{44} C_0 (p_1^2 , p_2^2 , q^2 ) \Big]. \notag \\
\end{eqnarray}
The expressions of the $A^{(q)}_i$,\, $ i=2,3,4$, are given in the Appendix \ref{ff}.  They can be isolated from the general expressions in \secref{ff} by selecting the $n_f$ dependent terms of the $A_i^{a b}$. For $i=1$ they are identical in the quark and gluon sectors, where their contributions to the final expression of the vertex differ just by $SU(3)$ colour factors.  \\
In the explicit evaluation we can use the relations  

\bea
 &  C_0 ( p_1^2,p_2^2,q^2) = \frac{ 1}{q^2} \Phi (x,y),
 \eea
where the function $\Phi (x, y)$  is defined as
\cite{Usyukina:1993ch}
\bea
\Phi( x, y) &=& \frac{1}{\lambda} \biggl\{ 2 [Li_2(-\rho  x) + Li_2(- \rho y)]  +
\ln \frac{y}{ x}\ln \frac{1+ \rho y }{1 + \rho x}+ \ln (\rho x) \ln (\rho  y) + \frac{\pi^2}{3} \biggr\},
\label{Phi}
\eea
with
\bea
 \lambda(x,y) = \sqrt {\Delta},
 \qquad  \qquad \Delta=(1-  x- y)^2 - 4  x  y,
\label{lambda} \\
\rho( x,y) = 2 (1-  x-  y+\lambda)^{-1},
  \qquad  \qquad x=\frac{p_1^2}{q^2} \, ,\qquad \qquad y= \frac {p_2^2}{q^2}\, .
\eea
and 
\begin{equation}
{B}_0(p^2)=\sdfrac{1}{i\p^\frac{d}{2}}\int\,d^d l\ \frac{1}{l^2(l-p_1)^2}=\frac{ \left[\G\left(\frac{d}{2}-1\right)\right]^2\G\left(2-\frac{d}{2}\right)}{\G\left(d-2\right)(p^2)^{2-\frac{d}{2}}}\label{B0ex}.
\end{equation}
with 
\beq
B_0(p^2)=\frac{1}{\varepsilon}+\bar{B}_0(p^2)
\eeq
and
\beq
B_0^R(p^2,0,0)\equiv \bar{B}_0(p^2)= 2 + \log(\mu^2/p^2) 
\label{BB}
\eeq
is the finite part in $d=4$ of the scalar integral in the $\overline{MS}$ scheme. One can check the direct cancellation of the $1/(d-4)$ poles after renormalization using the counterterm \eqref{ct}, and the $B_0$'s can be taken to be of the form $\bar{B}_0(p^2)$ in \eqref{BB}. We have left to Appendix \ref{secc} a discussion of the secondary (first order) equations for the same quark sector.

\section{The gluon sector and the complete correlator at one-loop}
\label{nine}
The decomposition at one-loop of the gluon sector follows \eqref{dec1}, 
but its final expression is modified compared to 
\eqref{tjjdec}, which is affected by new trace contributions not present in the quark sector. This sector provides a contribution of the form
\begin{framed}
\begin{equation}
\begin{aligned}
	& \langle T^{\mu \nu}(q)  J^{ a\alpha}(p_1) J^{ b\beta}(p_2)\rangle_g=\langle t^{\mu \nu}(q)  j^{ a\alpha}(p_1) j^{ b\beta}(p_2)\rangle_g +\langle t^{\mu \nu }(q)j_{loc}^{a  \a}(p_1)j^{b  \b}(p_2)\rangle_g +\langle t^{\mu \nu }(q)j^{a  \a}(p_1)j_{loc}^{b  \b}(p_2)\rangle_g\\
	& \qquad+2 \mathcal{I}^{\mu \nu \rho }(q)\left[\delta_{[\rho}^{\beta} p_{2 \sigma ]}\langle J^{ a\alpha}({p}_1) J^{ b\sigma}(-{p}_1)\rangle_g +\delta_{[\rho}^{\alpha} p_{1 \sigma]}\langle J^{ b\beta}({p}_2) J^{ a\sigma}(-p_2)\rangle\right]_g  
	+\frac{1}{d-1} \pi^{\mu \nu}(q) \left[\mathcal{A}^{\alpha \beta a b}_g+\mathcal{B}^{\alpha \beta a b}_g\right]
\end{aligned}
\label{res}
\end{equation}
\end{framed} 
There are additional local terms of the form 
\beq
\langle t^{\mu \nu }(q)j^{a \, \a}(p_1)j_{loc}^{b \, \b}(p_2)\rangle_g 
\label{adds}
\eeq
which are not part of the transverse traceless sector, but appear in the longitudinal sector of the decomposition. As we are going to discuss below, these terms are not set to zero by the Slavnov-Taylor identities, as in the quark sector, or in the general conformal solution.  
At the same time, the trace sector is modified by the presence of extra terms which are proportional to the equations of motion of the gluons, here indicated as $\mathcal{B}^{\alpha \beta a b}_g, $ which are absent in the on-shell decomposition. Defining  
\begin{allowdisplaybreaks}
\begin{equation}
A_{i }^{(g) a b}= -\,  C_A\,   \frac{g_s^2}{16\pi^2}\, \bar{A}_i^{{(g)}a b}, \qquad i=1,2\ldots 4
 \end{equation}
 $(C_A=N_c=3)$, where the functions $\bar{A}_i^{{(g)}a b}$ are extracted from  \secref{ff} by selecting the part of the 
 $A_i$ proportional to the Casimir $C_A$, their explicit expressions are given by 
\begin{eqnarray}
\label{Abar}
		\bar{A}_1^{(g) ab}  &=& -\frac{\delta^{ab}}{48 \left(p_1^2 p_2^2-(p_1 \cdot p_2)^2 \right)^4} \Big[ A_{10} + A_{11}  B_0 (p_1^2) + A_{12} B_0(p_2^2) + A_{13} B_0(q^2) + A_{14} C_0 (p_1^2 , p_2^2 , q^2 ) \Big] \notag \\
		\bar{A}_2^{(g) ab}  &=& -\frac{\delta ^{ab}}{144 \left(p_1^2 p_2^2-(p_1 \cdot p_2)^2 \right)^3} \Big[ A^{(g)}_{20} + A^{(g)}_{21}  B_0 (p_1^2) + A^{(g)}_{22} B_0(p_2^2)  + A^{(g)}_{23} B_0(q^2) + A^{(g)}_{24} C_0 (p_1^2 , p_2^2 , q^2 ) \Big] \notag \\
		\bar{A}_3^{(g) ab}  &=& \frac{\delta ^{ab}}{72 \left(p_1^2 p_2^2-(p_1 \cdot p_2)^2 \right)^3} \Big[A^{(g)}_{31}  B_0 (p_1^2) + A^{(g)}_{32} B_0(p_2^2)  + A^{(g)}_{33} B_0(q^2) + A^{(g)}_{34} C_0 (p_1^2 , p_2^2 , q^2 ) \Big] \notag \\
		\bar{A}_4^{(g) ab}  &=& -\frac{ \delta^{ab}}{72 \left(p_1^2 p_2^2-(p_1 \cdot p_2)^2 \right)^2} \Big[A^{(g)}_{40} + A^{(g)}_{41}  B_0 (p_1^2) + A^{(g)}_{42} B_0(p_2^2) + A^{(g)}_{43} B_0(q^2) + A^{(g)}_{44} C_0 (p_1^2 , p_2^2 , q^2 ) \Big]. \notag \\ 
\end{eqnarray}
\end{allowdisplaybreaks}
 The new longitudinal terms \ref{adds} take the form 
  \begin{framed}
 \begin{align}
	\langle t^{\mu \nu }(q)j^{a \, \a}(p_1)j_{loc}^{b \, \b}(p_2)\rangle_g & =
	{\Pi}^{\mu \nu}_{\m_1 \n_1}(q) \, \pi^\a _{\a_1} (p_1)\, {{p_2}_{\beta_1} p_2^\beta \over p_2^2}
	\left( 	B^{ab}_1 \, p_1^{\m_1} \, p_1^{\nu_1} \, p_2^{\a_1} \, p_2^{\b_1} + B_2^{ab} \, p_1^{\m_1} \, p_2^{\b_1} \, \delta^{\a_1 \n_1}  \right) 
\end{align}
\end{framed}
which is orthogonal to the trace sector. Notice that these local contributions vanish when the gluons are on-shell. The $B_i, \,\, i=1,2$ are given by  

\begin{eqnarray}
	B^{ab}_1 & = &  \frac{ \,  \,  \, C_A \,  g_s^2 \,  \delta^{ab} \, p_1^2}{64\pi^2 \, p_2^2 \, \big(p_1^2 p_2^2 - (p_1 \cdot p_2)^2 \big)^2} \, \Big( -\Big[2 \, (p_1 \cdot p_2)^2 + p_1^2 p_2^2 + 3 \, p_1^2 p_1 \cdot p_2 \Big]\, B_0(p_1^2)  \notag \\
	&& -\Big[2\, (p_1 \cdot p_2)^2 + p_1^2 p_2^2 + 3 \, p_2^2 p_1 \cdot p_2 \Big] \, B_0(p_2^2) +  \Big[ 4 \, (p_1 \cdot p_2)^2 + 2 \, p_1^2 p_2^2 + 3\, (p_1^2 + p_2^2) p_1 \cdot p_2 \Big]\, B_0(q^2)  \notag \\
	&&  +\Big[  q^2 \, (p_1 \cdot p_2)^2 + 2 \, q^2 \, (p_1 \cdot p_2)^2  \Big] \,C_0(p_1^2, p_2^2, q^2) - 2 \, (p_1 \cdot p_2)^2 - 2 \, p_1^2 p_2^2 \Big) \\
	B^{ab}_2 & = & \frac{ \,  \,  \, C_A \,  g_s^2 \,  \delta^{ab} \, p_1^2}{32\pi^2 \, p_2^2 \, \big(p_1^2 p_2^2 - (p_1 \cdot p_2)^2 \big)^2} \Big( - p_1 \cdot p_2 \, B_0(p_1^2)  - p_2^2 \cdot p_2  \, B_0(p_2^2) + (p_2^2+p_1 \cdot p_2) \, B_0(q^2) \notag \\ &&
	+ \Big[p_1^2 p_2^2 + p_2^2 \, (p_1 \cdot p_2) \Big] \, C_0(p_1^2, p_2^2 , q^2) \Big).
\end{eqnarray}

The trace sector is also affected by terms that vanish for 
on-shell gluons, indicated as $\mathcal{B}_g^{\alpha \beta a b}$, as well as the genuine anomaly term 
$\mathcal{A}_g^{\alpha \beta a b}$. Defining 
\begin{framed}
\beq
\mathcal{B}_g^{\alpha \beta a b}=\left[ C_1^{ab} \, p_1^\alpha \, p_1 ^\beta + C_2^{ab} \, p_1^\alpha \, p_2^\beta + C_3^{ab}  \, p_1^\beta \, p_2^\alpha + C_4^{ab} \, p_2^\alpha \, p_2^\beta + C_5^{ab} \, \delta^{\alpha \beta} \right]  
\label{bb}
\eeq
\end{framed}
the trace sector is characterised by the anomaly contributions from the gluon sector $\mathcal{A}_g$, plus the 
$\mathcal{B}_g$ terms proportional to the equations of motion of the gluons 
\begin{eqnarray}
	&\mathcal{A}_g^{\alpha \beta a b}+\mathcal{B}_g^{\alpha \beta a b}=\left. g_{\mu \nu} \langle T^{\mu \nu} (q) \, J^{a \a} (p_1) \, J^{b \b} (p_2)\, \rangle_g \right| \\&
	\end{eqnarray}
with the gluon contribution to the anomaly given by

\begin{eqnarray}
 \mathcal{A}_g^{\alpha\beta ab} &=& {11 \over 3} \,  \,  \,  \, \frac{g_s^2}{16\pi^2} \, C_A  \delta^{ab}u^{\alpha\beta}(p_1,p_2)
\end{eqnarray}
and $u^{\alpha\beta}$ defined in \eqref{locvar}.
The form factors proportional to the equations of motion in \ref{bb} take the form

\begin{eqnarray}
	C_1^{ab} &=& -\frac{ \,  \,  \, C_A \, g_s^2 \, \delta^{ab} }{ 32\pi^2 \big( (p_1 \cdot p_2)^2 - p_1^2 \, p_2^2 \big)} \, \Big(
	2 \, (p_1 \cdot p_2)^2 - 2 \, p_1^2 \, p_2^2 
	 - p_1^2 \, (p_1 \cdot p_2 + p_2^2 ) B_0 (p_1^2)  \notag \\ &&
	 + \big[ p_2^2 \, (p_1^2 - p_1 \cdot p_2) - 2 \, (p_1 \cdot p_2)^2 \big]  B_0(p_2^2)
	 + (p_1 \cdot p_2) \, (2 \, p_1 \cdot p_2 + p_1^2 + p_2^2) \, B_0 (q^2)  \notag \\ &&
	 + \big[ 2\, (p_1 \cdot p_2)^2 \, (p_1 \cdot p_2)^2 - p_2^2 \, \big(4 (p_1 \cdot p_2)^2   + p_1^4\big) + 5 \, p_1^2 \, p_4^4  \big] C_0(p_1^2, p_2^2 , q^2)  \Big) \\
	 C_2^{ab} &=& -2  \,  \,  \, C_A \, \frac{g_s^2}{16\pi^2} \, \delta^{ab} \, (p_1 \cdot p_2) \, C_0 (p_1^2 , p_2^2 , q^2) \\
	 C_3^{ab} &=& -\frac{ \,  \,  \, g_s^2 \, C_A \, (p_1^2 + p_2^2) \, \delta^{ab}}{32\pi^2 \big(p_1^2 \, p_2^2 - (p_1 \cdot p_2)^2 \big)} \Big( 
	 (-p_1\cdot p_2-p_1^2)  B_0(p_1^2)+(-(p_1\cdot p_2)-p_2^2) \, B_0(p_2^2) \notag \\ &&
	 +q^2 \, B_0(q^2)+\big(p_1^2 \, (p_1\cdot p_2-2 p_2^2) +(p_1\cdot p_2) \, \big(4 \, (p_1\cdot p_2)+p_2^2) \big) \, C_0(p_1^2,p_2^2, q^2)  \Big)  \\
	 C_4^{ab} &=&  -\frac{ \, \,  \, g_s^2 \, C_A \, g_s^2 \,  \delta^{ab}}{32\pi^2\big((p_1 \cdot p_2)^2 - p_1^2 \, p_2^2 \big)} \Big( 
	 (p_1^2 \, (p_2^2-p_1\cdot p_2)\notag \\ &&  -2 (p_1\cdot p_2)^2) \, B_0(p_1^2) -p_2^2 \, (p_1\cdot p_2+p_1^2) \, B_0(p_2^2)+(p_1\cdot p_2) \, q^2 \, B_0(q^2) \notag \\ &&  
	 + \big[-p_1^2 \, (4 \, (p_1\cdot p_2)^2+p_2^4)+2 \, (p_1\cdot p_2+p_2^2) \, (p_1\cdot p_2)^2+5 \, p_1^4 \, p_2^2 \, \big] \, C_0(p_1^2,p_2^2,q^2) \, \Big) \\
	 C_5^{ab} &=& \, \,  \, \frac{g_s^2}{32\pi^2} \, C_A \,  \delta^{ab} \Big( (p_1^2 - p_2^2) \, \big[ B_0 (p_1^2) - B_0(p_2^2) \big] 
	 + \big[ p_1^4 + p_2^4 - 2 \, (p_1^2 + p_2^2) \, p_1 \cdot p_2 - 6 \, p_1^2 \, p_2^2 \big] \, C_0(p_1^2, p_2^2, q^2) \Big). \nonumber \\
\end{eqnarray}

\section{The general structure of the TJJ from the CFT decomposition}
\label{ten}
We may combine the parameterization of both sectors in order to derive the general expression of the hard scattering vertex. It is given by the expression 

\begin{framed}
\begin{equation}
\begin{aligned}
	& \langle T^{\mu \nu}(q)  J^{ a\alpha}(p_1) J^{ b\beta}(p_2)\rangle=\langle t^{\mu \nu}(q)  j^{ a\alpha}(p_1) j^{ b\beta}(p_2)\rangle+\langle t^{\mu \nu }(q)j_{loc}^{a  \a}(p_1)j^{b  \b}(p_2)\rangle_g +\langle t^{\mu \nu }(q)j^{a  \a}(p_1)j_{loc}^{b  \b}(p_2)\rangle_g\\
	& \qquad+2 \mathcal{I}^{\mu \nu \rho }(q)\left[\delta_{[\rho}^{\beta} p_{2 \sigma ]}\langle J^{ a\alpha}({p}_1) J^{ b\sigma}(-{p}_1)\rangle+\delta_{[\rho}^{\alpha} p_{1 \sigma]}\langle J^{ b\beta}({p}_2) J^{ a\sigma}(-p_2)\rangle\right] 
	+\frac{1}{ 3 \, q^2} \hat\pi^{\mu \nu}(q) \left[\mathcal{A}^{\alpha \beta a b}+\mathcal{B}^{\alpha \beta a b}_g\right]
\end{aligned}
\label{res}
\end{equation}
\end{framed}

where the traceless sector is
\begin{equation}
\begin{aligned}
 \langle T^{\mu \nu}(q)  J^{ a\alpha}(p_1) J^{ b\beta}(p_2)\rangle_{tls}=&
\langle t^{\mu \nu}(q)  j^{ a\alpha}(p_1) j^{ b\beta}(p_2)\rangle+\langle t^{\mu \nu }(q)j_{loc}^{a  \a}(p_1)j^{b  \b}(p_2)\rangle_g +\langle t^{\mu \nu }(q)j^{a  \a}(p_1)j_{loc}^{b  \b}(p_2)\rangle_g\\
	& \qquad+2 \mathcal{I}^{\mu \nu \rho }(q)\left[\delta_{[\rho}^{\beta} p_{2 \sigma ]}\langle J^{ a\alpha}({p}_1) J^{ b\sigma}(-{p}_1)\rangle+\delta_{[\rho}^{\alpha} p_{1 \sigma]}\langle J^{ b\beta}({p}_2) J^{ a\sigma}(-p_2)\rangle\right] 
\end{aligned}
\label{ali}
\end{equation}

and the trace part is 

\beq
\label{trc}
\langle T^{\mu \nu}(q)  J^{ a\alpha}(p_1) J^{ b\beta}(p_2)\rangle_{tr}=\frac{1}{ 3 \, q^2} \hat\pi^{\mu \nu}(q) \left[\mathcal{A}^{\alpha \beta a b}+\mathcal{B}^{\alpha \beta a b}_g\right].
\eeq
%Both $\mathcal{A}$ and $\mathcal{B}$
The trace part contains the anomaly contribution 
$\mathcal{A}^{\alpha \beta a b}=\mathcal{A}^{\alpha \beta}\delta^{ab} $and a second term proportional to the gluon equations of motion $\mathcal{B}^{\alpha \beta a b}_g$. This is not to be considered part of the trace anomaly although it is part of the trace of the correlator since 
\beq
g_{\mu\nu}\langle T^{\mu \nu}(q)  J^{ a\alpha}(p_1) J^{ b\beta}(p_2)\rangle_{tr}= \left[\mathcal{A}^{\alpha \beta a b}+\mathcal{B}^{\alpha \beta a b}_g\right].
\eeq
The first tensor term $\mathcal{A}$ "projects into $FF$" :
\beq
\mathcal{A}^{\alpha\beta a b}(p_1,p_2)= A_n \delta^{a b} u^{\alpha\beta}(p_1,p_2)
\eeq
and it is explicitly given by  
\beq
\mathcal{A}^{\alpha\beta ab} = {1 \over 3} \, \,  \, \, \frac{g_s^2}{16\pi^2} \, (11C_A - 2 n_f) \delta^{ab}u^{\alpha \beta}(p_1,p_2),
\eeq
defining the conformal anomaly term coming from the quark and gluon sectors. $\mathcal{B}^{\alpha \beta a b}_g$ is given in \eqref{bb}. The anomaly form factor characterising the exchange of a 
dilaton pole in the $TJJ$ is then distilled in the form 
\be
  \frac{\beta}{q^2} \delta^{ab} \subset \langle T^{\mu \nu}(q)  J^{ a\alpha}(p_1) J^{ b\beta}(p_2)\rangle
\ee
shown in Fig.~\ref{scvertex},
where the $1/q^2$ anomaly pole has been explicitly extracted from the longitudinal projector $\pi^{\mu\nu}$ of the last term on the right-hand side of \ref{res} by defining 
 \beq
 \hat\pi^{\mu\nu}\equiv q^2 g^{\mu\nu} - q^\mu q^\nu.
 \eeq
Eq. \eqref{res} shows that the structure of the effective vertex corresponding to the $TJJ$ 
correlator is modified compared to the ordinary CFT case, encountered in \eqref{tjjdec},
with modifications affecting both the longitudinal and the trace sectors. \\
As we have demonstrated, the decomposition is constructed using appropriate projectors, enabling a separate analysis of the different sectors of the correlator. Consequently, it remains valid at all orders in perturbation theory.

\subsection{The off-shell decompositon and the anomaly form factor for massive quarks}
 Let's now consider massive quarks. The decomposition given in \eqref{ali} remains valid, but the trace sector acquires a different form, that is naturally decomposed into two contributions: a trace part which is directly projected on the tensor structure $u^{\alpha\beta}$, therefore projected onto $FF$, indicated as $\phi_1$, and a second form factor 
generalizing the $\mathcal{B}^{\alpha \beta a b}_g$ contributions given above, once we allow for massive quarks. This second term is denoted as $\phi_2$. The modifications are only present in the quark sector. This sector gives

\begin{equation}
    g_{\m\n}\braket{T^{\m\n}J^{\a a}J^{\b b}}_q=\left(\phi_1^{\a\b}(p_1,p_2,q,m)+\phi_2^{\a\b}(p_1,p_2,q,m)\right)\d^{ab}
\end{equation}

where

\begin{equation}
   \phi_1^{\alpha\beta a b}=\left(-\frac{2}{3}   n_f  \frac{g_s^2}{16\pi^2}+ \chi_0 (p_1,p_2,q,m)\right)\d^{ab} u^{\a\b}(p_1,p_2)
\end{equation}
contains the $n_f$ contribution to the anomaly pole, now with an extra term generated by the mass dependence. The $\chi_0$ function is defined as follows
\beq
\chi_0(p_1,p_2,q,m)\equiv \frac{B_1 - B_4 }{2\, p_1\cdot p_2}+\frac{2}{3}   n_f  \frac{g_s^2}{16\pi^2}\,,
\eeq
in order to make the anomaly contribution explicit.
The trace contains extra terms that vanish upon use of the equations of motion, related to the tensor structure 
${\phi_2}$, given by 

\begin{equation}
    \phi_2^{\a\b}(p_1,p_2,q,m)= \chi_1(p_1,p_2,q,m) v^{\a\b}+ B_2 p_1^{\a}p_1^{\b}+B_2(p_1\leftrightarrow p_2) p_2^{\a}p_2^{\b}+B_3 p_1^{\a}p_2^{\b}.
\end{equation}
where
\beq
\chi_1(p_1,p_2,q,m)\equiv \frac{B_1 + B_4  }{2\, p_1\cdot p_2 }
\eeq
with
\begin{equation}
    v^{\a\b}(p_1,p_2)=(p_1\cdot p_2) g^{\a\b}+p_2^\a p_1^\b
\end{equation}

and with $B_1\ldots B_4$ given below. Explicitly

	{\smaller 
	\begin{align}
		\chi_0=&\frac{3 m^4 \left(p_1^4+p_2^4+q^4-2 \left(p_1^2+p_2^2\right) q^2\right) \text{C}_0\left(p_1^2,p_2^2,q^2,m^2,m^2,m^2\right) g_s^2 }{2 \pi ^2 \left(-p_1^2-p_2^2+q^2\right) \left(q^4-2 \left(p_1^2+p_2^2\right) q^2+\left(p_1^2-p_2^2\right)^2\right)}\nonumber\\&+m^2 g_s^2 \Biggl(\frac{3 \left(p_1^4+p_2^4+q^4-2 \left(p_1^2+p_2^2\right) q^2\right) }{4 \pi ^2 \left(-p_1^2-p_2^2+q^2\right) \left(q^4-2 \left(p_1^2+p_2^2\right) q^2+\left(p_1^2-p_2^2\right)^2\right)}  \nonumber\\&\quad\quad\quad\quad +\frac{3\, p_2^2\left(p_2^6+\left(p_1^2-3 q^2\right) p_2^4+\left(p_1^4+3 q^4\right) p_2^2-\left(q^2-p_1^2\right)^2 \left(3 p_1^2+q^2\right)\right)\bar B_0(p_2^2,m^2)}{4 \pi ^2 \left(-p_1^2-p_2^2+q^2\right) \left(q^4-2 \left(p_1^2+p_2^2\right) q^2+\left(p_1^2-p_2^2\right)^2\right)^2}  \nonumber\\&\quad\quad\quad\quad -\frac{3\, p_1^2\left(-3 p_2^6+\left(p_1^2+5 q^2\right) p_2^4+\left(p_1^4-q^4\right) p_2^2-\left(q^2-p_1^2\right)^3\right)\bar B_0(p_1^2,m^2)) }{4 \pi ^2 \left(p_1^2+p_2^2-q^2\right) \left(q^4-2 \left(p_1^2+p_2^2\right) q^2+\left(p_1^2-p_2^2\right)^2\right)^2}  \nonumber\\&\quad\quad\quad\quad +\frac{3 \,q^2 \left(-p_2^8+\left(2 p_1^2+3 q^2\right) p_2^6-\left(p_1^2+q^2\right) \left(2 p_1^2+3 q^2\right) p_2^4+\left(q^2-p_1^2\right) \left(-2 p_1^4+3 q^2 p_1^2+q^4\right) p_2^2+p_1^2 \left(q^2-p_1^2\right)^3\right)  \bar B_0(q^2,m^2)}{4 \pi ^2 q^2 \left(-p_1^2-p_2^2+q^2\right) \left(q^4-2 \left(p_1^2+p_2^2\right) q^2+\left(p_1^2-p_2^2\right)^2\right)^2}   \nonumber\\&\quad\quad\quad\quad -\frac{3 P_0 C_0\left(p_1^2,p_2^2,q^2,m^2,m^2,m^2\right)}{8 \pi ^2 \left(-p_1^2-p_2^2+q^2\right) \left(q^4-2 \left(p_1^2+p_2^2\right) q^2+\left(p_1^2-p_2^2\right)^2\right)^2}\Biggl)
	\end{align}
} 
where we have defined the function

	\begin{align}
	P_0= &-p_2^{10}+\left(p_1^2+5 q^2\right) p_2^8-2 q^2 \left(3 p_1^2+5 q^2\right) p_2^6+2 \left(5 q^2-3 p_1^2\right) q^2 \left(p_1^2+q^2\right) p_2^4\nonumber\\&-\left(q^2-p_1^2\right)^2 \left(p_1^2+q^2\right) \left(5 q^2-p_1^2\right) p_2^2+\left(q^2-p_1^2\right)^5
\end{align}
$\bar B_0(p^2,m^2,m^2)$ stands for the renormalized 2-point function while $C_0(p_1^2,p_2^2,q^2,m^2,m^2,m^2) $ denotes the ordinary scalar off-shell 3-point function given in Appendix \ref{scalars}.

The scalar form factors related to terms in the trace which are proportional to the equations of motion are given by

{\smaller
	\begin{align}
		B_1=& - \frac{g_s^2}{16\p^2} m^2    \Biggl(\frac{3     \left(p_1^6-p_1^4 \left(p_2^2+3 q^2\right)-p_1^2 \left(p_2^2-3 q^2\right) \left(p_2^2+q^2\right)+\left(p_2^2-q^2\right)^3\right) C_0\left(p_1^2,p_2^2,q^2,m^2,m^2,m^2\right)}{p_1^4-2 p_1^2 \left(p_2^2+q^2\right)+\left(p_2^2-q^2\right)^2}
		\nn\\
		&\qquad \qquad \qquad +\frac{6   p_1^2 \bar B_0(p_1^2,m^2) \left(p_1^2-p_2^2-q^2\right)}{p_1^4-2 p_1^2 \left(p_2^2+q^2\right)+\left(p_2^2-q^2\right)^2} 
		+\frac{6 q^2\bar  B_0(q^2,m^2)   \left(q^2 \left(p_1^2+p_2^2\right)-\left(p_1^2-p_2^2\right)^2\right)}{q^2 \left(p_1^4-2 p_1^2 \left(p_2^2+q^2\right)+\left(p_2^2-q^2\right)^2\right)} \nn\\
		&\qquad \qquad \qquad  -\frac{6 p_2^2
			\bar B_0(p_2^2,m^2)\left(p_1^2-p_2^2+q^2\right)}{p_1^4-2 p_1^2 \left(p_2^2+q^2\right)+\left(p_2^2-q^2\right)^2} +6 \Biggl)-12   m^4 \frac{g_s^2}{16\p^2} C_0\left(p_1^2,p_2^2,q^2,m^2,m^2,m^2\right)
	\end{align}

	\begin{align}
		B_2= & - \frac{g_s^2}{16\p^2} m^2 \Biggl(\frac{12    p_2^2  \left(p_1^6-p_1^4 \left(p_2^2+q^2\right)-p_1^2 \left(p_2^4-6 p_2^2 q^2+q^4\right)+\left(p_2^2-q^2\right)^2 \left(p_2^2+q^2\right)\right) C_0\left(p_1^2,p_2^2,q^2,m^2,m^2,m^2\right)}{\left(p_1^4-2 p_1^2 \left(p_2^2+q^2\right)+\left(p_2^2-q^2\right)^2\right)^2}\nn\\ &\qquad \qquad \qquad+\frac{24    p_2^2 p_1^2\bar B_0(p_1^2,m^2)\left(2 p_1^4-p_1^2 \left(p_2^2+q^2\right)-\left(p_2^2-q^2\right)^2\right)}{p_1^2 \left(p_1^4-2 p_1^2 \left(p_2^2+q^2\right)+\left(p_2^2-q^2\right)^2\right)^2}\nn\\ &\qquad \qquad \qquad-\frac{24    p_2^2q^2\bar B_0(q^2,m^2)\left(q^2 \left(p_1^2+p_2^2\right)+\left(p_1^2-p_2^2\right)^2-2 q^4\right)}{q^2 \left(p_1^4-2 p_1^2 \left(p_2^2+q^2\right)+\left(p_2^2-q^2\right)^2\right)^2}\nn\\ &\qquad \qquad \qquad-\frac{24    p_2^2\bar B_0(p_2^2,m^2) \left(p_1^4+p_1^2 \left(p_2^2-2 q^2\right)-2 p_2^4+p_2^2 q^2+q^4\right)}{\left(p_1^4-2 p_1^2 \left(p_2^2+q^2\right)+\left(p_2^2-q^2\right)^2\right)^2}\nn\\ &\qquad \qquad \qquad+\frac{24    p_2^2  }{p_1^4-2 p_1^2 \left(p_2^2+q^2\right)+\left(p_2^2-q^2\right)^2}\Biggl)-\frac{3   m^4 p_2^2 g_s^2 C_0\left(p_1^2,p_2^2,q^2,m^2,m^2,m^2\right)}{\p^2\left(p_1^4-2 p_1^2 \left(p_2^2+q^2\right)+\left(p_2^2-q^2\right)^2\right)}
	\end{align}

	%\begin{align}
	%    A_3=& m^2 \Biggl(\frac{12 i \pi ^2 \kappa  p_1^2 g_s^2 \delta ^{a b} \left(p_1^6-p_1^4 \left(p_2^2+q^2\right)-p_1^2 \left(p_2^4-6 p_2^2 q^2+q^4\right)+\left(p_2^2-q^2\right)^2 \left(p_2^2+q^2\right)\right) C_0}{\left(p_1^4-2 p_1^2 \left(p_2^2+q^2\right)+\left(p_2^2-q^2\right)^2\right)^2}\nn\\ &+\frac{24 i \pi ^2 \kappa  g_s^2 \sqrt{p_1^2 \left(p_1^2-4 m^2\right)} \delta ^{a b} \log \left(\s(p_1,m)\right) \left(2 p_1^4-p_1^2 \left(p_2^2+q^2\right)-\left(p_2^2-q^2\right)^2\right)}{\left(p_1^4-2 p_1^2 \left(p_2^2+q^2\right)+\left(p_2^2-q^2\right)^2\right)^2}\nn\\ &-\frac{24 i \pi ^2 \kappa  p_1^2 g_s^2 \sqrt{q^2 \left(q^2-4 m^2\right)} \delta ^{a b} \log \left(\s(q,m)\right) \left(q^2 \left(p_1^2+p_2^2\right)+\left(p_1^2-p_2^2\right)^2-2 q^4\right)}{q^2 \left(p_1^4-2 p_1^2 \left(p_2^2+q^2\right)+\left(p_2^2-q^2\right)^2\right)^2}\nn\\ &-\frac{24 i \pi ^2 \kappa  p_1^2 g_s^2 \sqrt{p_2^2 \left(p_2^2-4 m^2\right)} \delta ^{a b} \log \s(p_2,m) \left(p_1^4+p_1^2 \left(p_2^2-2 q^2\right)-2 p_2^4+p_2^2 q^2+q^4\right)}{p_2^2 \left(p_1^4-2 p_1^2 \left(p_2^2+q^2\right)+\left(p_2^2-q^2\right)^2\right)^2}\nn\\ &+\frac{24 i \pi ^2 \kappa  p_1^2 g_s^2 \delta ^{a b}}{p_1^4-2 p_1^2 \left(p_2^2+q^2\right)+\left(p_2^2-q^2\right)^2}\Biggl)+\frac{48 i \pi ^2 \kappa  m^4 p_1^2 g_s^2 \delta ^{a b} C_0}{p_1^4-2 p_1^2 \left(p_2^2+q^2\right)+\left(p_2^2-q^2\right)^2}
	%\end{align}
	
	\begin{align}
		B_3=&- \frac{g_s^2}{16\p^2} m^2 \Biggl(\frac{6   \left(p_1^2+p_2^2-q^2\right) \left(p_1^6-p_1^4 \left(p_2^2+q^2\right)-p_1^2 \left(p_2^4-6 p_2^2 q^2+q^4\right)+\left(p_2^2-q^2\right)^2 \left(p_2^2+q^2\right)\right) C_0\left(p_1^2,p_2^2,q^2,m^2,m^2,m^2\right)}{\left(p_1^4-2 p_1^2 \left(p_2^2+q^2\right)+\left(p_2^2-q^2\right)^2\right)^2} \nn\\ &\qquad \qquad \qquad+\frac{12  p_1^2\bar B_0(p_1^2,m^2)\left(p_1^2+p_2^2-q^2\right) \left(2 p_1^4-p_1^2 \left(p_2^2+q^2\right)-\left(p_2^2-q^2\right)^2\right)}{p_1^2 \left(p_1^4-2 p_1^2 \left(p_2^2+q^2\right)+\left(p_2^2-q^2\right)^2\right)^2}\nn\\ &\qquad \qquad \qquad-\frac{12   q^2\bar B_0(q^2,m^2) \left(-3 q^4 \left(p_1^2+p_2^2\right)+4 p_1^2 p_2^2 q^2+\left(p_1^2-p_2^2\right)^2 \left(p_1^2+p_2^2\right)+2 q^6\right)}{q^2 \left(p_1^4-2 p_1^2 \left(p_2^2+q^2\right)+\left(p_2^2-q^2\right)^2\right)^2}\nn\\ &\qquad \qquad \qquad-\frac{12  p_2^2\bar B_0(p_2^2,m^2) \left(p_1^2+p_2^2-q^2\right) \left(p_1^4+p_1^2 \left(p_2^2-2 q^2\right)-2 p_2^4+p_2^2 q^2+q^4\right)}{p_2^2 \left(p_1^4-2 p_1^2 \left(p_2^2+q^2\right)+\left(p_2^2-q^2\right)^2\right)^2}\nn\\ &\qquad \qquad \qquad+\frac{12    \left(p_1^2+p_2^2-q^2\right)}{p_1^4-2 p_1^2 \left(p_2^2+q^2\right)+\left(p_2^2-q^2\right)^2}\Biggl)-\frac{3   m^4 g_s^2  \left(p_1^2+p_2^2\right) C_0\left(p_1^2,p_2^2,q^2,m^2,m^2,m^2\right)}{2\p^2\left(p_1^4-2 p_1^2 \left(p_2^2+q^2\right)+\left(p_2^2-q^2\right)^2\right)}
	\end{align}
	
	\begin{align}
		B_4=&- \frac{g_s^2}{16\p^2} m^2 \, p_1\cdot p_2 \Biggl(\frac{6   \left(p_1^2+p_2^2-q^2\right) \left(p_1^6-p_1^4 \left(p_2^2+3 q^2\right)+p_1^2 \left(-p_2^4+10 p_2^2 q^2+3 q^4\right)+\left(p_2^2-q^2\right)^3\right) C_0\left(p_1^2,p_2^2,q^2,m^2,m^2,m^2\right)}{\left(p_1^4-2 p_1^2 \left(p_2^2+q^2\right)+\left(p_2^2-q^2\right)^2\right)^2}\nn\\ &\qquad \qquad \qquad\qquad+\frac{12  p_2^2\bar B_0(p_2^2m^2) \left(-5 p_1^4+4 p_1^2 \left(p_2^2+q^2\right)+\left(p_2^2-q^2\right)^2\right)}{\left(p_1^4-2 p_1^2 \left(p_2^2+q^2\right)+\left(p_2^2-q^2\right)^2\right)^2}\nn\\ &\qquad \qquad \qquad\qquad+\frac{12   p_1^2\bar B_0(p_1^2,m^2)\left(p_1^4+p_1^2 \left(4 p_2^2-2 q^2\right)-5 p_2^4+4 p_2^2 q^2+q^4\right)}{\left(p_1^4-2 p_1^2 \left(p_2^2+q^2\right)+\left(p_2^2-q^2\right)^2\right)^2}\nn\\ &\qquad \qquad \qquad\qquad-\frac{12   q^2\bar B_0(q^2,m^2)\left(q^4 \left(p_1^2+p_2^2\right)+\left(p_1^2-p_2^2\right)^2 \left(p_1^2+p_2^2\right)-2 q^2 \left(p_1^4-4 p_1^2 p_2^2+p_2^4\right)\right)}{q^2 \left(p_1^4-2 p_1^2 \left(p_2^2+q^2\right)+\left(p_2^2-q^2\right)^2\right)^2}\nn\\ &\qquad \qquad \qquad\qquad+\frac{12   \left(p_1^2+p_2^2-q^2\right)}{p_1^4-2 p_1^2 \left(p_2^2+q^2\right)+\left(p_2^2-q^2\right)^2}\Biggl)-  \frac{ 3\, m^4 \,  g_s^2 \left(p_1^2+p_2^2-q^2\right)^2 C_0\left(p_1^2,p_2^2,q^2,m^2,m^2,m^2\right)}{2\left(p_1^4-2 p_1^2 \left(p_2^2+q^2\right)+\left(p_2^2-q^2\right)^2\right)}
	\end{align}
}

The corresponding form factors $A^{(q)}_i$ for the transverse traceless part of the massive case can be found in Appendix \ref{aiq}.

\subsection{Comments on the structure of the decomposition}
At this stage, we pause for some comments concerning the structure of this result also with respect to the  prediction for this correlator coming from CFT, in a non-Lagrangian formulation, as discussed in \cite{Bzowski:2013sza} both in the Abelian and non-Abelian cases. \\
Since the gauge fixing condition breaks conformal symmetry in the non-Abelian case for a Lagrangian theory, the reconstruction of  the 
correlator  allows extra non-conformal terms. 
As we have discussed in the previous ections, in  the conformal case, the Ward identity on the two gauge currents are imposed by contracting the correlator with momenta either $p_1$ or $p_2$. In the case of a gauge-fixed theory, instead, the relevant STI requires a quadratic contraction with both momenta $p_1$ and $p_2$ in the form given by \eqref{brstwi} that replaces the ordinary single derivative WIs. 
Therefore \eqref{res} differs in its structure compared to \eqref{tjjdec}, which is typical of a conformal theory. \\
The first term, $\langle t^{\mu \nu}(q)  j^{ a\alpha}(p_1) j^{ b\beta}(p_2)\rangle$, satisfies both Ward identities (WIs) independently. In contrast, the second and third terms, such as $\langle t^{\mu \nu }(q)j_{loc}^{a \alpha}(p_1)j^{b \beta}(p_2)\rangle_g$, are constrained to vanish under the ordinary (single derivative) WI but are not constrained by the second derivative Slavnov-Taylor identity (STI).\\
Coming to the the contribution
\beq
\delta_{[\rho}^{\beta} p_{2 \sigma ]}\langle J^{ a\alpha}(p_1) J^{ b\sigma}(-p_1)\rangle + \delta_{[\rho}^{\alpha} p_{1 \sigma ]}\langle J^{ b\beta}(p_2) J^{ a\sigma}(-p_2)\rangle,
\eeq
the single derivative WIs, both in $p_1$ and $p_2$, set both terms to zero due to the transversality condition of the gauge currents in both the Abelian and non-Abelian cases. The STI also enforces this condition.\\
Finally, the new term, $\mathcal{B}^{\alpha \beta a b}_g$, is not permitted by the single derivative WI but is allowed by the STI. The anomaly contribution and the dilaton pole, for off-shell gluons, can be extracted from the trace sector by the steps that we have outlined, taking CFT$_p$ in the quark contributions as a 
guideline.  

\begin{figure}[t]
\begin{center}
\includegraphics[scale=1]{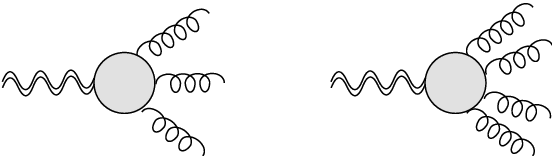}
\caption{The $TJJJ$ and $TJJJJ$ contributions in the gauge covariant expansion of the QCD trace anomaly vertex.}
\end{center}
\end{figure}

\begin{figure}[t]
\begin{center}
\includegraphics[scale=1]{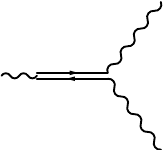}
\caption{The dilaton intermediate state in the QCD trace anomaly vertex.}
\label{scvertex}
\end{center}
\end{figure}

\subsection{The anomaly in Duff's definition and the pole in DR}
If an anomaly is understood as the failure of the trace operation to commute with a quantum average, the trace anomaly can also be defined as the difference between two trace operations on the stress-energy tensor: one performed before the quantum average and the other after. This definition was proposed by Duff \cite{Duff:1993wm,Duff:2020dqb}
\begin{equation} \label{eq:defanomduff}
	\mathcal{A}=g^{\mu \nu}(x)\left\langle T_{\mu \nu}(x)\right\rangle-\left\langle T_\mu^\mu(x)\right\rangle,
\end{equation}
and is the definition considered in \cite{Bonora:2014qla} \cite{Ferrero:2023unz}, \\
Previous computations of the $TJJ$ correlator have demonstrated the emergence of a pole, relying on a secondary decomposition first introduced in \cite{Giannotti:2008cv}, which in the case of QCD can be immeditely implemented in the quark sector
\begin{equation}
\Gamma_q^{\m_1\n_1\m_2\m_3 a b }(p_2,p_3)=\sum_{i=1}^{13}\,F_i(s;s_1,s_2,0)\,t_i^{\m_1\n_1\m_2\m_3}(p_2,p_3)\delta^{a b},
\label{fs}
\end{equation} 
where the invariant amplitudes $F_i$ are functions of the kinematic invariants $s=p_1^2=(p_2+p_3)^2$, $s_1=p_2^2$, $s_2=p_3^2$, and the $t_i^{\m_1\n_1\m_2\m_3}$ define the basis of the independent tensor structures reproduced in Appendix \ref{genbasis1} in Table \ref{genbasis}.
This decomposition can be directly mapped onto the current approach, closely paralleling the analysis presented in \cite{Coriano:2018zdo}. The method is readily applicable to the quark sector, which we will illustrate here, as it is sufficient to reveal the underlying pattern. This pattern naturally extends to the gluon sector as well.\\
 \eqref{fs} is built by imposing on the $TJJ$ vertex all the Ward identities derived from diffeomorphism invariance and gauge invariance, together with Bose symmetry and conservation WIs. Additional details have been left to appendix \ref{ffdc}.
 Using the completeness of the basis in \eqref{fs} and by a direct analysis of the CWIs, we can identify the mapping between the form factors of such basis and those of the $A$-basis. In \( d=4 \), the presence of two tensor structures 
\beq
t_1=\left(k^2 g^{\mu\nu} - k^{\mu } k^{\nu}\right) u^{\alpha\beta}(p.q)\qquad
t_2 =\left(k^2g^{\mu\nu} - k^{\mu} k^{\nu}\right) w^{\alpha\beta}(p.q)
\eeq
with nonzero trace in the \( F \)-basis initially raises questions, particularly regarding the unique relationship between anomaly poles (and associated traces) and the renormalization process. The remaining tensor structures $t_i$ are traceless. 
The expectation is to identify a single anomaly pole originating from renormalization, while any additional poles introduced by expansions should not be linked to this process. \\
The sets \( A_j \) presented above and \( F_i \) differ significantly, each highlighting distinct aspects of the same \( TJJ \) correlator. The \( F \)-basis, as we will show, is particularly effective in tracing the origin of the anomaly pole to a single form factor, \( F_{13} \), which diverges and thus requires renormalization.\\
Previous analyses, such as those in \cite{Bzowski:2017poo}, indicate that the singularities in the \( A_i \)'s, specifically \( A_2 \), \( A_3 \), and \( A_4 \), align with the mapping \eqref{mapping1}, which precisely identifies the combinations involving the divergent form factor \( F_{13} \).\\
This clear identification of the singularity's origin within the \( F \)-basis contrasts with the less direct approach in the \( A \)-basis. While the \( A_i \)'s constitute a minimal set of form factors for resolving the conformal Ward identities (CWI's) of the correlators, they obscure the origin of the singular behavior, as three out of four of these form factors exhibit UV singularities and necessitate renormalization. In contrast, the \( F \)-basis provides a straightforward method to pinpoint singularities, specifically in \( F_{13} \), which can be shown to be singular by dimensional counting. The correspondence is givel in \eqref{mapping1}.\\
To investigate the origin of the anomaly pole in the $TJJ$ correlator, we begin by considering the correlator in $d$ dimensions using the $F$-basis. Our goal is to impose that this correlator remains traceless, thereby anomaly-free, in the higher-dimensional theory. However, as we approach the physical limit $d \rightarrow 4$ using dimensional regularization, the anomaly manifests. The Ward identities associated with the trace provide crucial constraints. Specifically, imposing that the trace WI is satisfied, we derive the following conditions

\begin{equation}
\label{eq:F1}
F_1 = \frac{(d-4)}{p_1^2(d-1)} \left[F_{13} - p_2^2 F_3 - p_3^2 F_5 - p_2 \cdot p_3 F_7\right],
\end{equation}
\begin{equation}
\label{eq:F2}
F_2 = \frac{(d-4)}{p_1^2(d-1)} \left[p_2^2 F_4 + p_3^2 F_6 + p_2 \cdot p_3 F_8\right].
\end{equation}

These equations are pivotal for understanding the renormalization process of the correlator. As $d \rightarrow 4$, it is evident from Eq. (\ref{eq:F2}) that $F_2$ vanishes:
\begin{equation}
F_2 = \frac{\epsilon}{(d-1) p_1^2} \left[p_2^2 F_4 + p_3^2 F_6 + p_2 \cdot p_3 F_8\right] \rightarrow 0,
\end{equation}
where $\epsilon \equiv d-4$. Since the form factors $F_4$, $F_6$, and $F_8$ are finite due to their dimensional scaling, $F_2$ indeed approaches zero as $d \rightarrow 4$. Consequently, in the limit $d = 4$, the $F$-basis reduces to four independent combinations of the original seven form factors (as shown in \eqref{mapping1}), which fully describe the transverse traceless sector of the theory. Additionally, one extra form factor, $F_1$, remains, which corresponds to a nonzero trace and accounts for the anomaly in four dimensions. \\
Importantly, $F_{13}$ is the only form factor within the $F$-basis that requires renormalization. It exhibits a simple pole in $1/\epsilon$ under dimensional regularization. The fact that the singularity remains of the form $1/\epsilon$ at all perturbative orders, without higher-order poles, is a key feature of this construction. This behavior is consistent with conformal field theory, where the only available counterterm that regulates the theory is \eqref{ct} which renormalizes the two-point function $\langle JJ \rangle$ and thereby $F_{13}$. The quark sector gives (for a single fermion)

\begin{equation}
\label{eq:F13}
F_{13} = G_0(p_1^2, p_2^2, p_3^2) - \frac{1}{2} \left[\Pi(p_2^2) + \Pi(p_3^2)\right],
\end{equation}
where $G_0$ can be shown to be a finite function as $d \rightarrow 4$, and the singularity is traced back to the scalar form factor $\Pi(p^2)$ of the quark contribution to the gluon 2-point function $\Pi_{\mu\nu}^{ab}(p)$. 
\[
\Pi_{\mu\nu}^{ab}(p) = \delta^{ab} \Pi_{\mu\nu}(p),
\]
where the gluon polarization tensor $\Pi_{\mu\nu}(p)$ is
\[
\Pi_{\mu\nu}(p) = \left(p_\mu p_\nu - g_{\mu\nu} p^2\right) \Pi(p^2),
\]
where $\Pi(p^2)$ is the scalar form factor that captures the momentum dependence of the quark contribution to the gluon self-energy.
For a single quark flavor in the one-loop approximation, $\Pi(p^2)$ takes the form
\[
\Pi(p^2) = -\frac{g^2 T_F}{2\pi^2} \left(\frac{1}{\epsilon} - \gamma_E + \log(4\pi) + \log\left(\frac{\mu^2}{-p^2}\right) + \cdots \right),
\]

where $T_F = \frac{1}{2}$. The divergence in $F_{13}$ is then given by a single pole in $\epsilon$ is of the form

\be 
\label{renf}
F_{13}=\frac{1}{d-4} \bar{F}_{13} + F_{13\, f},
\eeq
where $F_{13 f}$ is finite in the limit $d\to 4$. It is then sufficient to insert this expansion into \eqref{eq:F1} to notice the emergence of a $1/p_1^2$ dilaton pole in the limit, since all the other form factors are finite and do not contribute as $d -4 \to 0$. Therefore, this secondary parameterization shows that the anomaly is related 
to a unique tensor structure which necessarily has to contain a pole. A similar analysis can be performed in the gluon sector. This analysis shows that the anomaly pole present in the form factor $F_1$ is the only result of the breaking of conformal symmetry as $d\to 4$ in DR. 
\section{Connecting our parameterization of the $TJJ$ with the hadronic GFF at large $-t$}

\begin{figure}
	\centering
	\begin{tikzpicture}
		\begin{feynman}
			\vertex (i1);
			\vertex[right=1cm of i1] (a1);
			\vertex[right=1cm of a1] (b1);
			\vertex[right=3cm of a1] (o1) ;
			
			\vertex[below=1cm of i1] (i2) ;
			\vertex[right=1cm of i2] (a2);
			\vertex[right=2cm of a2] (b2);
			\vertex[right=3cm of a2] (o2) ;
			
			\vertex[below=1cm of i2] (i4) ;
			\vertex[right=1cm of i4] (a4);
			\vertex[right=2cm of a4] (b4);
			\vertex[right=3cm of a4] (o4) ;
			
			%\vertex[below=2cm of i3] (t1);
			\vertex[above=1cm of a1] (t2);
			\vertex[above=2cm of a2] (t3);
			
			%\vertex[above right =1.3cm of t2] (t1);
			\vertex[above=0.6 cm of t1] (i3) {$T^{\mu\nu}_q$};
			
			\diagram* { 
				(i1)  -- [fermion] (a1) -- [fermion] (b1)   --[fermion] (o1),
				
				(i2)  -- [fermion] (a2)-- [fermion] (b2)  --[fermion] (o2),
				(i4)  -- [fermion] (b4)  --[fermion] (o4),

				(i3)  -- [graviton] (b1),
				(a1)-- [gluon] (a2),
				(b2)-- [gluon] (b4)
			};
	
		\end{feynman}
	\end{tikzpicture}
	\hspace{1cm}
		\begin{tikzpicture}
		\begin{feynman}
			\vertex (i1);
			\vertex[right=1cm of i1] (a1);
			
			\vertex[right=3cm of a1] (o1) ;
			
			\vertex[below=1cm of i1] (i2) ;
			\vertex[right=1cm of i2] (a2);
			\vertex[right=2cm of a2] (b2);
			\vertex[right=3cm of a2] (o2) ;
			
			\vertex[below=1cm of i2] (i4) ;
			\vertex[right=1cm of i4] (a4);
			\vertex[right=2cm of a4] (b4);
			\vertex[right=3cm of a4] (o4) ;
			
			%\vertex[below=2cm of i3] (t1);
			\vertex[above=1cm of a1] (t2);
			\vertex[above=2cm of a2] (t3);
			
			%\vertex[above right =1.3cm of t2] (t1);
			%\vertex[right=0.6 cm of t1] (i3);
			\vertex[above=0.5cm of a2] (b1);
			\vertex[right=2cm of b1] (i3) {$T^{\mu\nu}_g$};
			
			\diagram* { 
				(i1)  -- [fermion] (a1)   --[fermion] (o1),
				
				(i2)  -- [fermion] (a2)-- [fermion] (b2)  --[fermion] (o2),
				(i4)  -- [fermion] (b4)  --[fermion] (o4),

				(i3)  -- [graviton] (b1),
				(a1)-- [gluon] (a2),
				(b2)-- [gluon] (b4)
			};
			
		\end{feynman}
	\end{tikzpicture}
	\caption{Typical leading $O(\alpha_s^2)$ contributions to the GFF of the proton. Shown are insertions of the graviton/$f \bar{f}$ vertex contained in $T_q$ (left) and the graviton/gg vertex contained in $T_g$ (right).}
	\label{figX:2}
\end{figure}

Before presenting our conclusions, we briefly comment on the connection between our parametrization of hard scattering and \eqref{fund1} at the hadron level, focusing on the proton case. A detailed phenomenological analysis will be provided elsewhere. 
We recall that the analysis of the form factors \( A(t) \), \( B(t) \), and \( C(t) \) at large momentum transfers, where \( t \equiv q^2 \), can be performed using a standard factorization formula. \\
As shown in \cite{Tong:2022zax}, a convenient approach to studying the gravitational form factor (GFF) matrix element at the hadron level for large momentum transfer is to use a helicity basis. In this basis, the hadron helicities \( \Lambda \) and \( \Lambda' \) label the initial and final states, along with their respective momenta \( p \) and \( p' \)
\begin{align}
 \langle p', \Lambda' | T^{\mu\nu}_a|p,\Lambda\rangle.
 \end{align}
For the nucleon, the leading Fock state consists of three valence quarks, specifically \( u, u, d \), where each quark carries helicity \( \lambda_i = \pm 1/2 \). The analysis is most conveniently performed in the Breit frame, where the momenta are given by  

\begin{align}
&p^\mu = (p^+, p^-, \boldsymbol{p}_\perp) = \frac{1}{\sqrt{2}} \left(\sqrt{-t}, 0, \boldsymbol{0}_\perp \right), \\
&p'^\mu = (p'^+, p'^-, \boldsymbol{p}'_\perp) = \frac{1}{\sqrt{2}} \left(0, \sqrt{-t}, \boldsymbol{0}_\perp \right), \\
&\Delta^\mu = (\Delta^+, \Delta^-, \boldsymbol{\Delta}_\perp) = \left(-\sqrt{-t}, \sqrt{-t}, \boldsymbol{0}_\perp \right),
\end{align}
where \( t = q^2 \) is the squared momentum transfer. \\
The total helicity of the valence quarks can take the values \( 3/2, 1/2, -1/2, -3/2 \). The relation between the hadron helicity \( \Lambda \) and the individual quark helicities \( \lambda_i \) is determined by the component of the hadron's angular momentum \( l_z \) along the scattering direction \( z \)

\begin{figure}
	\centering
	\begin{tikzpicture}
	\begin{feynman}
		\vertex (i1);
		\vertex[right=1cm of i1] (a1);
		\vertex[right=1cm of a1] (b1);
		\vertex[right=3cm of a1] (o1) ;
		
		\vertex[below=1cm of i1] (i2) ;
		\vertex[right=1cm of i2] (a2);
		\vertex[right=2cm of a2] (b2);
		\vertex[right=1cm of a2] (c2);
		\vertex[right=3cm of a2] (o2) ;
		
		\vertex[below=1cm of i2] (i4) ;
		\vertex[right=1cm of i4] (a4);
		\vertex[right=2cm of a4] (b4);
		\vertex[right=1cm of a4] (c4);
		\vertex[right=3cm of a4] (o4) ;
		
		%\vertex[below=2cm of i3] (t1);
		\vertex[above=1cm of a1] (t2);
		\vertex[above=2cm of a2] (t3);
		
		%\vertex[above right =1.3cm of t2] (t1);
		\vertex[above=1.5cm of t1] (i3) {$T^{\mu\nu}$};
		\vertex[blob] (t1) at (2,1.3) {{$TJJ$}};
		
		\diagram* { 
			(i1)  -- [fermion] (a1)  --[fermion] (o1),
			
			(i2)  -- [fermion] (c2)-- [fermion] (b2)  --[fermion] (o2),
			(i4)  -- [fermion] (c4)  --[fermion] (o4),

			(i3)  -- [graviton] (t1),
			(a1)-- [gluon] (t1),
			(b2)-- [gluon] (t1),
			(c2)-- [gluon] (c4),
		};

	\end{feynman}
\end{tikzpicture}
\hspace{2cm}
\begin{tikzpicture}
	\begin{feynman}
		\vertex (i1);
		\vertex[right=1cm of i1] (a1);
		\vertex[right=1cm of a1] (b1);
		\vertex[right=3cm of a1] (o1) ;
		
		\vertex[below=2cm of i1] (i2) ;
		\vertex[right=1cm of i2] (a2);
		\vertex[right=2cm of a2] (b2);
		\vertex[right=1cm of a2] (c2);
		\vertex[right=3cm of a2] (o2) ;
		
		\vertex[below=1cm of i1] (i4) ;
		\vertex[right=1cm of i4] (a4);
		\vertex[right=2cm of a4] (b4);
		\vertex[right=1cm of a4] (c4);
		\vertex[right=3cm of a4] (o4) ;
		
		%\vertex[below=2cm of i3] (t1);
		\vertex[above=1cm of a1] (t2);
		\vertex[above=2cm of a2] (t3);
		
		%\vertex[above right =1.3cm of t2] (t1);
		\vertex[above=1.5cm of t1] (i3) {$T^{\mu\nu}$};
		\vertex[blob] (t1) at (2,1.3) {{$TJJ$}};
		
		\diagram* { 
			(i1)  -- [fermion] (a1)  --[fermion] (o1),
			
			(i2)  -- [fermion] (c2)-- [fermion] (b2)  --[fermion] (o2),
			(i4)  -- [fermion] (c4)  --[fermion] (o4),

			(i3)  -- [graviton] (t1),
			(a1)-- [gluon] (t1),
			(b2)-- [gluon] (t1),
			(c2)-- [gluon] (c4),
		};

	\end{feynman}
\end{tikzpicture}
\caption{ Typical $O(\alpha_s^3)$ corrections appearing in the expansion of the hard scattering associated with the conformal anomaly.}
\label{figX:3}
\end{figure}
\beq
\Lambda=l_z - \sum_i \lambda_i.
\eeq
The hadronic matrix elements of definite helicities in terms of the angular momentum of the hadron in the $z$ direction
\begin{align}
  \big| P_\uparrow  \big\rangle=\big| P_\uparrow \big\rangle^{l_z=2}_{-3/2}+\big| P_\uparrow \big\rangle^{l_z=1}_{-1/2}+\big| P_\uparrow \big\rangle^{l_z=0}_{1/2}+\big| P_\uparrow \big\rangle^{l_z=-1}_{3/2},
 \end{align}

\begin{align}
  \big| P_\downarrow  \big\rangle=\big| P_\downarrow \big\rangle^{l_z=1}_{-3/2}+\big| P_\downarrow \big\rangle^{l_z=0}_{-1/2}+\big| P_\downarrow \big\rangle^{l_z=-1}_{1/2}+\big| P_\downarrow \big\rangle^{l_z=2}_{3/2}\ .
 \end{align}
 The hadronic matrix elements with definite helicities can be expressed in terms of the hadron's angular momentum along the \( z \)-direction as  

\begin{align}
  \big| P_\uparrow  \big\rangle = \big| P_\uparrow \big\rangle^{l_z=2}_{-3/2} + \big| P_\uparrow \big\rangle^{l_z=1}_{-1/2} + \big| P_\uparrow \big\rangle^{l_z=0}_{1/2} + \big| P_\uparrow \big\rangle^{l_z=-1}_{3/2},
\end{align}

\begin{align}
  \big| P_\downarrow  \big\rangle = \big| P_\downarrow \big\rangle^{l_z=1}_{-3/2} + \big| P_\downarrow \big\rangle^{l_z=0}_{-1/2} + \big| P_\downarrow \big\rangle^{l_z=-1}_{1/2} + \big| P_\downarrow \big\rangle^{l_z=2}_{3/2}.
\end{align}
Here, the hadronic states with definite helicities are defined through either integrated or unintegrated wave functions. The integrated wave functions describe the interaction in the collinear limit, where the transverse momenta of the partons are neglected, while the unintegrated wave functions account for their transverse momentum dependence. \\
The kinematic variables are parameterized by \( \kappa_i = (x_i, k_i) \), where \( x_i \) denotes the longitudinal momentum fraction of the parton, and \( k_i \) represents its intrinsic transverse momentum. In the unintegrated case, the wave function explicitly depends on the transverse momenta, which serve as integration variables. The hadronic initial and final states are then defined in terms of the integrated wave function \( \phi(x_1, x_2, x_3) \) and the unintegrated wave function \( \psi(\kappa_1, \kappa_2, \kappa_3) \), both of which are functions of the momentum fractions \( x_i \) and \( y_i \) corresponding to the initial and final hadron momenta \( p \) and \( p' \), respectively

 \begin{align}
 \big| p_\uparrow \big\rangle_{1/2}^{l_z=0}= & \int \
    \frac{[d x][d^2 \bs k]}{\sqrt{24 x_1 x_2 x_3}}
 \psi_1
   ( \kappa_1,\kappa_2,\kappa_3) \ \big |\{x_i p+ \bs  k_i \}\big \rangle_{1/2},
 \label{eq_fock1} \\ 
 \big| p_\downarrow \big\rangle_{1/2}^{l_z=-1}= & \int \
    \frac{[d x][d^2 \bs k]}{\sqrt{24 x_1 x_2 x_3}}
   \big[  k_{1L} \psi_3( \kappa_2,\kappa_1,\kappa_3)+  k_{2L} \psi_4( \kappa_2,\kappa_1,\kappa_3) \big]
    \big |\{x_i p+ \bs  k_i \}\big \rangle_{1/2}.
       \label{eqf} 
 \end{align}

The unintegrated wave functions \( \psi_1, \psi_2, \psi_3, \psi_4 \), classified in \cite{Ji:2002xn, Ji:2003yj}, describe the proton's internal partonic structure. The inclusion of the complex combination \( k_L = k^x - i k^y \) accounts for the presence of nonzero orbital angular momentum. However, the dominant contributions originate from the three-quark light-cone wave function with zero orbital angular momentum:

\begin{align}
\langle p'_{ \uparrow }  |
  {T^{\mu\nu }_a} | p_ \uparrow  \rangle=\langle p'_{ \uparrow }  |^{l'_z=0}_{1/2} 
 {T^{\mu\nu }_a} | p_ \uparrow  \rangle^{l_z=0}_{1/2},
\end{align}

where \( a = q, g \) denotes the quark and gluon contributions from the QCD stress-energy tensor. To illustrate how a correspondence can be established, we focus on this leading contribution, whose expression is given by  

\begin{align}
\langle p'_\uparrow |_{1/2}T^{\mu\nu}_a|p_\uparrow\rangle_{1/2}
 =&  \int \
   \frac{[d x][d y][d^2 \boldsymbol{k}'][d^2 \boldsymbol{k}]}{24(x_1 x_2 x_3 y_1 y_2 y_3)^{1/2}}
 \psi_1^*
   ( \kappa'_1,\kappa'_2,\kappa'_3)\psi_1
   ( \kappa_1,\kappa_2,\kappa_3) 
 \big\langle  \{y_i p'+\boldsymbol{k}'_i\}\big|_{1/2}
  {T^{\mu\nu}_a}
\big|\{x_i p+\boldsymbol{k}_i\}\big\rangle_{1/2},
\end{align}

where the integration variables include both the momentum fractions and the intrinsic transverse momenta of the partons:  

\begin{align}
[d x ] &= d x_1 d x_2 d x_3\ \delta(1-x_1-x_2-x_3), \notag \\
[d^2  k ] &= \frac{1}{(2\pi)^6} d^2\boldsymbol{k}_{1} d^2\boldsymbol{k}_{2} d^2\boldsymbol{k}_{3} \delta^{(2)}(\boldsymbol{k}_{1}+\boldsymbol{k}_{2}+\boldsymbol{k}_{3}).
\label{eqm}
\end{align}

The hard scattering expansion in Fock space follows the usual approach, employing quark creation operators for plane waves with defined color and helicity. The operators \( \hat u^\dagger_{a,\lambda} \) and \( \hat d^\dagger_{a,\lambda} \) create \( u \) and \( d \) quarks, respectively, with helicity \( \lambda \) and color index \( a \) in the fundamental representation of \( SU(3) \), normalized as

\begin{align}
 \{\hat u_{a,\lambda}(k), \hat u^\dagger_{b,\lambda'}(k')\} = \delta_{\lambda \lambda'}\delta_{ab}2k^+\delta(k^+-k'^+)\delta^{(2)}(\boldsymbol{k}'-\boldsymbol{k}).
\end{align}

The light-front wave functions \( \psi_1, \psi_2, \psi_3, \psi_4 \) describe the proton in terms of the kinematic variables \( \kappa_i = (x_i, \boldsymbol{k}_i) \), where \( x_i \) represents the longitudinal momentum fraction of each parton, satisfying \( \sum_i x_i = 1 \), and \( \boldsymbol{k}_i \) denotes the intrinsic transverse momentum, constrained by \( \sum_i \boldsymbol{k}_i = 0 \). 

Neglecting intrinsic transverse momentum and considering only the collinear limit, we define the twist-three light-cone amplitude of the proton as~\cite{Braun:1999te}  

\begin{align}
\Phi_3(x_1,x_2,x_3)=2\int[d^2 \boldsymbol{k}] \psi_1(\kappa_1,\kappa_2,\kappa_3).
\end{align}

This expression encapsulates the leading contributions to the hadronic matrix elements in the collinear approximation, simplifying the analysis of the internal structure of the proton.
 
The leading $A(t)$-form factor can be extracted form the amplitude $(P=(p + p')/2)$
\begin{align}
&\langle p'_\uparrow | T^{\mu\nu}_a|p_\uparrow\rangle
=A_a(t)\bar u_\uparrow(p') \gamma^{(\mu} P^{\nu)} 
u_\uparrow(p)\ ,
\label{eq_consAmp}
\end{align}
while $B(t)$ is power-suppressed. One obtains 

\begin{align}
&A(t)\bar u_\uparrow(p') \gamma^{(\mu}  P^{\nu)} 
u_\uparrow(p)
 \notag\\
 =&  \int \
   \frac{[d x][d y]}{96(x_1 x_2 x_3 y_1 y_2 y_3)^{1/2}}
\Phi_3^*(y_1,y_2,y_3)
  \Phi_3(x_1,x_2,x_3) \big\langle  uud-udu,\{y_i p'\}\big|
 {T^{\mu\nu}}
\big| uud-udu,\{x_i p\}\big\rangle\ ,
\label{xs1}
 \end{align}
 where the partonic matrix element 
 \beq
  \mathcal{T}^{\mu\nu}_{1/2\,1/2}\equiv\langle  uud-udu,\{y_i p'\}\big|
  {T^{\mu\nu}}
\big| uud-udu,\{x_i p\}\big\rangle\
\label{tau}
\eeq
defines the hard scattering contribution, and we have summed over the quark and gluon contributions ($a=q,g$). 
The dominant partonic configurations in the hard scattering process are determined by extracting the leading quark contributions from the proton's partonic wave function within the Fock space representation
\begin{figure}
	\centering
	\begin{tikzpicture}
		\begin{feynman}
			\vertex (i1);
			\vertex[right=1cm of i1] (a1);
			\vertex[right=1cm of a1] (b1);
			\vertex[right=3cm of a1] (o1) ;
			
			\vertex[below=1cm of i1] (i2) ;
			\vertex[right=1cm of i2] (a2);
			\vertex[right=2cm of a2] (b2);
			\vertex[right=3cm of a2] (o2) ;
			
			\vertex[below=1cm of i2] (i4) ;
			\vertex[right=1cm of i4] (a4);
			\vertex[right=2cm of a4] (b4);
			\vertex[right=3cm of a4] (o4) ;
			
			%\vertex[below=2cm of i3] (t1);
			\vertex[above=1cm of a1] (t2);
			\vertex[above=2cm of a2] (t3);
			
				\vertex[right=1cm of a2] (c2);
				\vertex[below=1cm of c2] (c4);
			%\vertex[above right =1.3cm of t2] (t1);
			\vertex[above=0.6 cm of t1] (i3) {$T^{\mu\nu}_q$};
			
			\diagram* { 
				(i1)  -- [fermion] (a1) -- [fermion] (b1)   --[fermion] (o1),
				
				(i2)  -- [fermion] (a2)-- [fermion] (c2)-- [fermion] (b2)  --[fermion] (o2),
				(i4) -- [fermion] (c4) -- [fermion] (b4)  --[fermion] (o4),

				(i3)  -- [graviton] (b1),
				(a1)-- [gluon] (a2),
				(c2)-- [gluon] (c4),
				(b2)-- [gluon] (b4)
			};
			
		\end{feynman}
	\end{tikzpicture}
	\caption{Typical $O(\alpha_s^3)$ contributions from  $T_q$ not related to the conformal anomaly.}
	\label{figX:4}
\end{figure}

\begin{align}
\big|%uud,
\{p_i\} \big\rangle_{1/2}=&
\frac{\epsilon_{abc}}{\sqrt{6}}
  \hat u_{a,\uparrow}^\dagger (p_{ 1} ) 
   \big [\hat u_{b,\downarrow}^\dagger (p_{2 })
  \hat d_{c,\uparrow}^\dagger ( p_{3 })  - \hat d_{b,\downarrow}^\dagger( p_{2 })\hat  u_{c,\uparrow}^\dagger ( p_{3 })\big]\big|0 \big\rangle
         \label{fk3}
\end{align}
where $p_i\equiv x_i p$ are the momenta of the initial collinear quarks and $p_i=y_i p'$ for the final state ones 
 \begin{align}
&|uud,\{p_i\} \rangle =\frac{\epsilon_{abc}}{\sqrt{6}}
  \hat u_{a,\uparrow}^\dagger (p_{ 1} ) 
   \hat u_{b,\downarrow}^\dagger (p_{2 })
   \hat d_{c,\uparrow}^\dagger ( p_{3 })\big|0 \big\rangle
   ~,\notag 
   \\
   &
  |udu,\{p_i\} \rangle =\frac{\epsilon_{abc}}{\sqrt{6}}
    \hat  u_{a,\uparrow}^\dagger (p_{ 1} ) \hat d_{b,\downarrow}^\dagger( p_{2 }) \hat u_{c,\uparrow}^\dagger ( p_{3 })\big|0 \big\rangle~.
     \label{eq_partonstate}
\end{align} 
 We illustrate in \figref{figX:2} the typical topologies that emerge at \( O(\alpha_s^2) \) in the perturbative expansion of the hard scattering process. At tree level, the insertions of \( T_q \) into the hard scattering involve both a graviton/fermion/antifermion (\( O(g^0) \), lowest order) vertex and a graviton/gluon/fermion/antifermion (\( O(g) \)) vertex. The left diagram represents the lowest-order contribution to the hard scattering, featuring the insertion of the leading \( T_q \) vertex, while the right diagram illustrates the lowest-order insertion of \( T_g \) via the graviton/two-gluon \( O(g^0) \) vertex. \\
Insertions of \( T_g \) include both the graviton?two-gluon and graviton?three-gluon vertices, corresponding to \( O(g) \) and \( O(g^2) \), respectively, though the latter is not explicitly shown in the same figure. According to the quark/gluon counting rules, hard rescattering via gluon insertions must be maintained at large \( -t \), requiring an additional gluon propagator to connect the remaining valence quark in order to ensure an elastic transition to the final state.\\
The anomaly contributions, depicted in \figref{figX:3}, arise from replacing each gluon propagator that connects separate quark lines with the insertion of the \( O(\alpha_s) \) corrections introduced in this work. These corrections effectively replace the tree-level gluon propagator with the modified relation derived from the decomposition discussed. We have
\beq
S^{\alpha\beta\, a b}(l)\to S^{\alpha\alpha' \, a a'}(l)\,\Gamma^{\alpha'\beta' \, a' b'}(q, l, l+q))\,
S^{\beta'\beta \, b' b}(l +q)
\eeq
where \( l^\mu \) represents a typical transverse momentum of the gluon exchanged between valence quark lines at leading order. We denote by \( \Gamma^{\alpha'\beta' \, a' b'}(q, l, l+q) \) the momentum dependence of the \( TJJ \) vertex when included in the hard scattering, where \( q \) is the incoming momentum of the graviton line. 

In the pion case, such modifications affect only a single lowest-order topology, characterized by the exchange of a single gluon between the quark-antiquark pair in the hard scattering process. In contrast, for the proton, there are three relevant topologies, two of which are depicted in \figref{figX:3}. 

The analysis in \cite{Tong:2022zax}, which examines the leading-order hard partonic amplitude at \( O(\alpha_s^2) \), can be systematically extended to the next perturbative order, \( O(\alpha_s^3) \), with only minor modifications. This computation requires incorporating all standard insertions into the hard scattering, including an additional gluon connecting pairs of valence quark lines, as well as self-energy corrections in the gluon propagator. The insertion of the \( TJJ \) vertex follows the same procedure as outlined earlier, though at this order, a large number of additional corrections arise that are unrelated to the anomaly. \\
For illustration, one such correction is shown in \figref{figX:4}, where hard rescattering between two valence quark lines is mediated by the exchange of two gluons. Other relevant corrections involve graviton?three-gluon and graviton/four-gluon vertices, originating from the tree-level insertions of \( T_g \) in the \( T_g ggg \) and \( T_g gggg \) interactions. These contribute at \( O(g^5) \) and \( O(g^6) \), respectively, in the proton case. \\
By incorporating all these corrections into a matrix element that accounts for such background processes, denoted as \( \mathcal{T}^{\mu\nu}_b \), and reinstating the dependence on the strong coupling constant, the matrix element can be expressed in the form

\beq
 \langle p'| T^{\mu\nu} | p\rangle
 =\alpha_s^2 \mathcal{T}^{\mu\nu} + \alpha_s^3\mathcal{T}_b^{\mu\nu} +\alpha_s^3\mathcal{T}_{\textrm{an}}^{\mu\nu}
\eeq
 where the leading-order contributions $\mathcal{T}^{\mu\nu}$ may be decomposed in the helicity basis, of which \eqref{tau} is one of the components. \\
 In general the structure of the hard scattering, in the collinear limit will be of the form 
 \beq
 \Gamma^{\mu\nu\alpha\beta a b}(x_i p, y_j p',  t)\mathcal{N}^{\alpha\beta a b}(x_i p, y_j p',  t)
\eeq
to be convoluted with the ordinary hadronic wave functions as in \eqref{xs1}. We have denoted with $\mathcal{N}^{\alpha\beta a b}$ the ordinary Dirac trace appearing in the hard scattering, which can be worked out as in \cite{Tong:2022zax}. 
The extraction of the contribution of the anomaly form factor can be obtained by tracing the hadronic matrix element. For example, one immediately obtains that $A(t)$, at large 
$-t$, if we neglect the hadron mass, is not affected by the anomaly since the left hand side of \eqref{xs1} vanish, by use of the equations of motion on the spinors.  \\
Notice that imposing any trace condition on the matrix element eliminates the anomaly pole from the hard scattering. This feature is explicitly revealed in the perturbative analysis of \eqref{polepole}, where the anomaly pole arises in the on-shell gluon case and for massless fermions.  \\
In a related work, we will demonstrate that the same form factor obeys a general sum rule in perturbation theory. This result follows from the specific structure of the spectral density of the anomaly form factor, as given in \eqref{trc}.
\beq
\Phi_{an}\equiv \mathcal{A}^{\alpha \beta a b}/q^2
\label{anf}
 \eeq
confirming and extending previous analysis in the QED \cite{Giannotti:2008cv} and 
QCD cases \cite{Armillis:2010qk}. The pattern, in the case of the conformal anomaly form factor \eqref{anf} is pretty similar to that discussed in the case of chiral and gravitational 
anomalies, recently addressed in \cite{Coriano:2025ceu}. 
 
\section{Comments and Conclusions}
\label{dec22} 
This work has been focused on the analysis of a specific vertex, the $TJJ$, in the non-Abelian case that, as we have illustrated, plays a key role in QCD in the GFFs of hadrons. \\
The analysis that we have presented is directly linked with the factorization picture of exclusive processes. Such picture has brought us to consider the role of the perturbative one-loop insertion of the $TJJ$ in the hard scattering. Naturally, this insertion, in the hadron case, turns relevant at order $\alpha_s^2$, and is therefore subleading compared to the leading $O(\alpha_s)$ corrections, obtained by the direct coupling of the graviton to the collinear quarks of the hard scattering. The vertex, as pointed out, is essentially described by a nonlocal interaction that has been investigated in the past in several anomalous correlators. The extraction of this 
interaction is rather nontrivial and requires the formalism presented in our work, which is the result of a long-term analysis of such matrix elements in $CFT_p$ using a combination of general CFT approaches and specific free field theory realizations.\\
The interaction is described using a longitudinal/transverse/trace (sector) decomposition of such vertices, with the pole emerging in the trace channel. This factorization of the hard scattering that we have presented can be immediately extended at hadron level, in the proton and pion cases, in order to provide a possible phenomenological basis of invariant amplitudes in which parameterize the GFF form factors.  \\
We have aimed to bridge recent advancements in CFT$_p$ and their anomalies developed over the past decade for correlators of even and odd parity with the physics of strong interactions. This effort is particularly timely as anomalies have gained renewed attention in the context of the Electron-Ion Collider (EIC) program on the proton spin \cite{Tarasov:2020cwl,Bhattacharya:2022xxw,Castelli:2024eza}. This program is poised to play a pivotal role in the scientific agenda at BNL, contributing significantly to proton tomography and the determination of the spin and partonic content of hadrons and the anomaly contribution.\\
 We have shown that conformal anomalies are intrinsically linked to the presence of effective dilaton degrees of freedom in the hard scattering. In the perturbative framework, these anomalies are associated with the emergence of a dilaton pole in the hard scattering process. This phenomenon is accompanied by a sum rule, which we have verified in perturbation theory at the one-loop level. \\
 We have illustrated how  the standard 
 $CFT_p$ approach can be modified to account for the gauge fixing sector of QCD. 
 We have shown by an explicit computation, though limited to the on-shell gluon case, that the anomaly form factor is characterised by a dispersive part that satisfies a sum rule.
 The presence of  sum rules is a hallmark of chiral and conformal anomalies, transforming the pole into a cut, in the presence of extra scales. In other words, anomaly poles are not ordinay particle poles and their manifestation is essentially associated with sum rules. Successful verification of this character would serve as a strong indication of the exchange of a dilaton state, validating the theoretical predictions.

\centerline{\bf Acknowledgements}  
We thank Giovanni Chirilli, Olindo Corradini, Hsiang-nan Li  and Yoshitaka Hatta for discussions/ correspondence. This work is partially funded by the European Union, Next Generation EU, PNRR project "National Centre for HPC, Big Data and Quantum Computing", project code CN00000013; by INFN, inziativa specifica {\em QG-sky} and by the grant PRIN 2022BP52A MUR "The Holographic Universe for all Lambdas" Lecce-Naples.

\appendix
\section{Slavnov-Taylor identities }
\label{STTI}
We denote with $S[V_{\underline{a}}^\mu,\psi,A_\mu]$ the action of the model. Its expression depends on the vielbein, the fermion field $\psi$ and the Abelian gauge field $A_\mu$. We can use this action and the vielbein to derive a useful form of the EMT
We introduce the generating functional of the model, given by
\beqa\label{Z} Z[V,J^\mu,\c,\bar\c]
&=& \int \mD\bar{\psi}\mD\psi\mD A_\mu\,\mbox{exp}\bigg\{i S[V,\psi,A_\mu]  + i\int d^4x\, \bigg[\\
&+& \bar{\c(x)}\psi(x) + \bar{\psi}(x)\c(x) + J^\mu(x)A_\mu(x)\bigg]\bigg\}\, ,\eeqa
where we have denoted with $J(x)$, $J^\mu(x)$ and $\chi(x)$ the sources for the scalar, the gauge field and the spinor field respectively.
We will exploit the invariance of $Z$ under diffeomorphisms for the derivation of the corresponding Ward identities.
For this purpose we introduce a condensed notation to denote the functional integration measure of all the fields
\beqa
\label{SourcesToymodel}
\mD \Phi {\equiv} \mD\psi\mD\bar{\psi}\mD A_\mu\,
\eeqa
and redefine the action with the external sources included
\beqa
\label{ActionToyModel}
\tilde{S}&            =          & S + i\int d^4x\,\left( J^\mu A_\mu + \bar{\c}(x)\psi(x) + \textrm{h.c.}\right).\,
\eeqa
Notice that we have absorbed a factor $\sqrt{-g}$ in the definition of the sources, which clearly affects their transformation under changes of
coordinates. The transformations of the fields are given by (we have absorbed a factor $\sqrt{-g}$ in their definitions)
\beqa
V^{'\,\underline{a}}_\mu(x)           &=& V^{\underline{a}}_\mu(x) -\int d^4y\,[\delta^{(4)}(x-y)\partial_\nu V^{\underline{a}}_\mu(x)
                               + [\partial_\mu\d^{(4)}(x-y)]V^{\underline{a}}_\nu]\e^\nu(y)\, ,\nn\\
\c'(x)                     &=& \c(x) -\int d^4y\,\partial_\nu[\d^{(4)}(x-y)\c(x)]\e^\nu(y)\, .
\eeqa
The term which appears in the first line in the integrand of Eq. (\ref{preWard}) can be re-expressed in the following form
\bea \label{WIfirstterm}
&& - \int d^4 x \, V\Theta^\mu_{\,\,\,\underline{a}}\bigg[-\d^{(4)}(x-y)\partial_\nu V^{\underline{a}}_\mu(x) - [\partial_\mu\d^{(4)}(x-y)]V^{\underline{a}}_\nu\bigg] = - V\Theta^\mu_{\,\,\,\nu\,;\mu} + V\Theta^\mu_{\,\,\,\underline{a}}V^{\underline{a}}_{\mu\, ;\nu} \nn \\
&& =  - V\bigg[\Theta^\mu_{\,\,\,\nu\,;\mu} +  V_{\underline{a}\r} V^{\underline{a}}_{\mu\,;\nu}\frac{\Theta^{\mu\r} - \Theta^{\r\mu}}{2}\bigg] \,,
\eea
where in the last expression we used the covariant conservation of the metric tensor expressed in terms of the vierbein
\bea
g_{\mu\nu\,;\r} = 0 \Rightarrow V^{\underline{a}}_{\mu\,;\r}V_{\underline{a}\nu} = - V^{\underline{a}}_\mu V_{\underline{a}\nu\,;\r} = - V_{\underline{a}\mu} V^{\underline{a}}_{\nu\,;\r}.
\eea
Other simplifications are obtained using the invariance of the action under local Lorentz transformations 
parameterized as
\beqa
\d V^{\underline{a}}_\mu = {\w^{\underline{a}}}_{\underline{b}} V^{\underline{b}}_\mu\, , \qquad
\d\psi     = \frac{1}{2}\s^{a b}\w_{{\underline{a}\underline{b}}}\psi\, , \qquad
\d\bar{\psi} = -\frac{1}{2}\bar\psi\s^{\underline{a} \underline{b}}\w_{\underline{a b}}\, ,
\eeqa
that gives, using the antisymmetry of $\w^{\underline{ a b}}$
\beq\label{vincolofermioni}
\frac{\d S}{\d\psi}\s^{\underline{a b}}\psi - \bar\psi\s^{\underline{a b}}\frac{\d S}{\d\bar{\psi}}
- \frac{\d S}{\d V^{\underline{b}}_\mu}V^{\underline{a}}_\mu + \frac{\d S}{\d V^{\underline{b}}_\mu} V^{\underline{a}}_\mu = 0\, .
\eeq
The previous equation can be reformulated in terms of the energy-momentum tensor $\Theta^{\mu\nu}$
\beq
V(\Theta^{\mu\r} - \Theta^{\r\mu}) = \bar{\psi}\s^{\mu\r}\frac{\d S}{\d\bar{\psi}}
- \frac{\d S}{\d\psi}\s^{\mu\r}\psi\,,
\eeq
which is useful to re-express Eq. (\ref{WIfirstterm}) in terms of the symmetric energy-momentum tensor $T^{\mu\nu}$ and to obtain finally, in the flat space-time limit, Eq. (\ref{Ward}).

\section{Appendix. Feynman rules}
\label{rules}
%------------------
The Feynman rules used throughout the paper are collected here
\begin{itemize}
\item {Graviton - fermion - fermion vertex}
\\
%%%%%%%%%%%
\\
%\begin{align}
\bmi{95pt}
\includegraphics[scale=1.0]{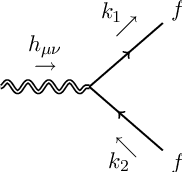}
\emi
\bmi{70pt}
\bann
&=& - i \, \frac{\kappa}{2} \, V^{\prime}_{\mu\nu}(k_1,k_2) \nn \\
&=& - i \, \frac{\kappa}{2} \, \left\{ \frac{1}{4} \left[\gamma_\mu (k_1 + k_2)_\nu
+\gamma_\nu (k_1 + k_2)_\mu \right] - \frac{1}{2} g_{\mu \nu}
[\gamma^{\lambda}(k_1 + k_2)_{\lambda} - 2 m]  \right\} \nn \\
\eann
\emi
\bea
\label{VGff}
\eea
%\end{align}
\item{Graviton - gluon - gluon vertex}
\\ \\
\bmi{110pt}
\includegraphics[scale=1.0]{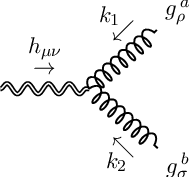}
\emi
\bmi{100pt}
\bann
&=& - i \, \frac{\kappa}{2} \, \delta_{a b} \, V^{Ggg}_{\mu\nu\rho\sigma}(k_1,k_2) \nn \\
 &= &- i \, \frac{\kappa}{2} \, \delta_{a b} \left\{ k_1\cdot k_2 \, C_{\mu\nu\rho\si} + D_{\mu\nu\rho\si}(k_1,k_2) + \frac{1}{\xi} \, E_{\mu\nu\rho\si}(k_1,k_2)  \right\}
\eann
\emi
\bea
\eea
\item{Graviton - ghost - ghost vertex}
\\ \\
\bmi{110pt}
\includegraphics[scale=1.0]{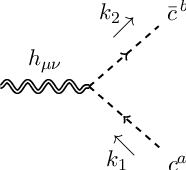}
\emi
\bmi{70pt}
\bann
& = & - i \,  \frac{\kappa}{2}\,  \delta^{a b} \, C_{\mu\nu\rho\sigma} \, k_{1\,\rho} \, k_{2\,\sigma}
\eann
\emi
\bea
\eea
%
%%%%%%%%%%%%%%%%%%%%%
%
\item{Graviton - fermion - fermion - gauge boson vertex}
\\ \\
\bmi{110pt}
\includegraphics[scale=1.0]{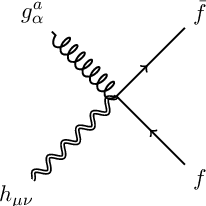}
\emi
\bmi{70pt}
\bann
&=& i g \, \frac{\kappa}{2} \,T^a\, W^{\prime}_{\mu\nu\alpha}
= i g \, \frac{\kappa}{2} \, T^a \left\{ -\frac{1}{2} (\gamma_\mu \, g_{\nu\alpha}
+\gamma_\nu \, g_{\mu\alpha}) +  g_{\mu \nu} \, \gamma_{\alpha} \right \}
\eann
\emi
\bea
\label{WGffg}
\eea
%
%%%%%%%%%%%%%%%%%%%%%%%%%%
%
%%%%%%%%%%%%%%%%%%%%%%%%%%
%
\item{Graviton - gluon - gluon - gluon  vertex}
\\ \\
\bmi{110pt}
\includegraphics[scale=1.0]{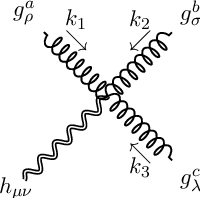}
\emi
\bmi{110pt}
\begin{eqnarray*}
&=&- g \frac{\kappa}{2} f^{a b c} V^{Gggg}_{\mu\nu\rho\sigma\lambda}(k_1,k_2,k_3) \nn \\
&=&  - g \frac{\kappa}{2} f^{a b c} \left\{ C_{\mu\nu\rho\sigma}(k_1-k_2)_{\lambda} + C_{\mu\nu\rho\lambda}(k_3-k_1)_{\sigma}   \right. \nn \\
&& \hspace{2.5cm}  + \left.  C_{\mu\nu\sigma\lambda}(k_2-k_3)_{\rho} + F_{\mu\nu\rho\sigma\lambda}(k_1,k_2,k_3)  \right\}
\hspace{1.7cm}
\end{eqnarray*}
\emi

%
%%%%%%%%%%%%%%%%%%%%%%%%%%
%%%%%%%%%%%%%%%%%%%%%%%%%%
%
\item{Graviton - ghost - ghost - gauge boson vertex}
\\ \\
\bmi{110pt}
\includegraphics[scale=1.0]{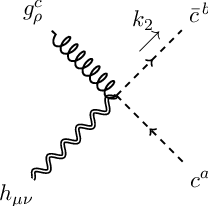}
\emi
\bmi{70pt}
\bann
&=&- \frac{\kappa}{2} \, g \, f^{a b c} \, C_{\mu\nu\rho\sigma} \, k_{2}^{\sigma}
\eann
\emi
\bea
\eea
%
%%%%%%%%%%%%%%%%%%%%%%%%%%
%
\bea
&& C_{\mu\nu\rho\sigma} = g_{\mu\rho}\, g_{\nu\sigma}
+g_{\mu\sigma} \, g_{\nu\rho}
-g_{\mu\nu} \, g_{\rho\sigma}\,
\\
&& D_{\mu\nu\rho\sigma} (k_1, k_2) =
g_{\mu\nu} \, k_{1 \, \sigma}\, k_{2 \, \rho}
- \biggl[g^{\mu\sigma} k_1^{\nu} k_2^{\rho}
  + g_{\mu\rho} \, k_{1 \, \sigma} \, k_{2 \, \nu}
  - g_{\rho\sigma} \, k_{1 \, \mu} \, k_{2 \, \nu}
  + (\mu\leftrightarrow\nu)\biggr]\, \\
&& E_{\mu\nu\rho\sigma} (k_1, k_2) = g_{\mu\nu} \, (k_{1 \, \rho} \, k_{1 \, \sigma}
+k_{2 \, \rho} \, k_{2 \, \sigma} +k_{1 \, \rho} \, k_{2 \, \sigma})
-\biggl[g_{\nu\sigma} \, k_{1 \, \mu} \, k_{1 \, \rho}
+g_{\nu\rho} \, k_{2 \, \mu} \, k_{2 \, \sigma}
+(\mu\leftrightarrow\nu)\biggr]\ , \nn  \\ \\
&& F_{\mu\nu\rho\sigma\lambda} (k_1,k_2,k_3) =
g_{\mu\rho} \,  g_{\sigma\lambda} \, (k_2-k_3)_{\nu}
+g_{\mu\sigma} \, g_{\rho\lambda} \, (k_3-k_1)_{\nu}
+g_{\mu\lambda} \, g_{\rho\sigma}(k_1-k_2)_{\nu}
+ (\mu\leftrightarrow\nu) \nn \\
\eea
%-------------------------

\section{The perturbative expansion and the $TJJ$: the quark sector}
\label{six}

We take the external momenta as incoming. We introduce the tensor components
\begin{align}
A^{\m_1\n_1\m\n}&\equiv\h^{\mu_1\nu_1}\h^{\m\n}- \big(\h^{\m\m_1}\h^{\n_1\n}+\h^{\m\n_1}\h^{\m_1\n}\big)
\end{align}
and the vertices in the fermion sector are
\begin{align}
V^{\m\, a}_{J\psi\bar\psi}(k_1,k_2) &=-ie\,\g^\m T^a\\
V^{\m_1\n_1\, a}_{T\psi\bar\psi}(p_1k_1,k_2) &=-\sdfrac{i}{4}\,A^{\m_1\n_1\m\n}\,\g_\n T^a\,(k_1+k_2)_\m\\
V^{\m_1\n_1\m_2\, a}_{TJ\psi\bar\psi}(k_1,k_2) &=\frac{i\,e}{2}A^{\m_1\n_1\m_2\n}\,\g_\m T^a.
\end{align}
giving
\begin{eqnarray}
i\Gamma^{\m_1\n_1\m_2\m_3 a b}(p_2,p_3)&\equiv&\braket{T^{\m_1\n_1}(p_1)\,J^{\m_2 a}(p_2)\,J^{\m_3 b}(p_3)}_F\nonumber \\
&&=2\,\bigg(\sum_{i=1}^2V_{F,i}^{\m_1\n_1\m_2\m_3 a b }(p_1,p_2,p_3)+\sum_{i=1}^{2}W_{F,i}^{\m_1\n_1\m_2\m_3 a b }(p_1,p_2,p_3)\bigg)
\end{eqnarray}
where the $V_{F,i}$ terms are related to the triangle topology contributions, while the $W_{F,i}$ terms denote the two bubble contributions. \\
In the quark sectors the perturbative contributions are
\begin{allowdisplaybreaks}
\begin{align}
V_{F,1}^{\m_1\n_1\m_2\m_3 a b}&=-i^3\delta^{ab}\,\int\frac{d^d\ell}{(2\p)^d}\frac{\Tr\left[V^{\m_1\n_1}_{T\psi\bar\psi}(\ell-p_2,\ell+p_3)\left(\slashed{\ell}+\slashed{p}_3\right)V^{\m_2}_{J\psi\bar\psi}(\ell,\ell-p_2)\,\slashed{\ell}\,V^{\m_3}_{J\psi\bar\psi}(\ell,\ell+p_3)\,\left(\slashed{\ell}-\slashed{p}_2\right)\right]}{\ell^2\,(\ell-p_2)^2(\ell+p_3)^2}\\
V_{F,2}^{\m_1\n_1\m_2\m_3 a b }&=-i^3\delta^{ab}\,\int\frac{d^d\ell}{(2\p)^d}\frac{\Tr\left[V^{\m_1\n_1}_{T\psi\bar\psi}(\ell-p_3,\ell+p_2)\left(\slashed{\ell}+\slashed{p}_2\right)V^{\m_2}_{J\psi\bar\psi}(\ell,\ell-p_3)\,\slashed{\ell}\,V^{\m_3}_{J\psi\bar\psi}(\ell,\ell+p_2)\,\left(\slashed{\ell}-\slashed{p}_3\right)	\right]}{\ell^2\,(\ell-p_3)^2(\ell+p_2)^2}\\
W_{F,3}^{\m_1\n_1\m_2\m_3 a b }&=-i^2\delta^{ab}\,\int\frac{d^d\ell}{(2\p)^d}\frac{\Tr\left[V^{\m_1\n_1\m_2}_{TJ\psi\bar\psi}(\ell+p_3,\ell)\left(\slashed{\ell}+\slashed{p}_3\right)V^{\m_3}_{J\psi\bar\psi}(\ell,\ell+p_3)\,\slashed{\ell}\right]}{\ell^2\,(\ell+p_3)^2}\\[1.5ex]
W_{F,2}^{\m_1\n_1\m_2\m_3 a b}&=-i^2\delta^{ab}\,\int\frac{d^d\ell}{(2\p)^d}\frac{\Tr\left[V^{\m_1\n_1\m_3}_{TJ\psi\bar\psi}(\ell+p_2,\ell)\left(\slashed{\ell}+\slashed{p}_2\right)V^{\m_2}_{J\psi\bar\psi}(\ell,\ell+p_2)\,\slashed{\ell}\right]}{\ell^2\,(\ell+p_2)^2}
\end{align}
\end{allowdisplaybreaks}
\end{itemize}
The evaluation of these one-loop integrals is rather straightforward. We use the transverse traceless $\Pi$ and the transverse ones $\pi$ in order to decompose the result of the computation in the longitudinal transverse parameterization. This procedure 
generates all the form factors appearing in Section 7. 

\section{The TJJJJ hierarchy from diffeomorphism invariance at $O(g^4)$ in the quark sector}
\label{t4j}
The conservation equation of the TJJJJ takes the form  
\begin{equation}
	\begin{aligned}
		0=&q_{\mu } \langle T^{\mu\nu}(q)J^{\alpha a_1}(p_1)J^{\beta a_2}(p_2)J^{\gamma a_3}(p_3)J^{\delta a_4}(p_4)\rangle_q+2
		\delta^{\alpha}_{[\mu}{p_1}_{\nu]}\langle J^{\mu a_1}(p_1-q)J^{\beta a_2}(p_2)J^{\gamma a_3}(p_3)J^{\delta a_4}(p_4) \rangle_q\\&
		+2
		\delta^{\beta}_{[\mu}{p_2}_{\nu]}\langle J^{\alpha a_1}(p_1)J^{\mu a_2}(p_2-q)J^{\gamma a_3}(p_3)J^{\delta a_4}(p_4) \rangle_q
		+2
		\delta^{\gamma}_{[\mu}{p_3}_{\nu]}\langle J^{\alpha a_1}(p_1)J^{\beta a_2}(p_2)J^{\mu a_3}(p_3-q) J^{\delta a_4}(p_4)\rangle_q\\&
		+2
		\delta^{\delta}_{[\mu}{p_4}_{\nu]}\langle J^{\alpha a_1}(p_1)J^{\beta a_2}(p_2)J^{\gamma a_3}(p_3) J^{\mu a_4}(p_4-q)\rangle_q
		-2i gf^{a_1 a_2 c} \delta^{[\alpha}_\nu \langle J^{\beta] c}(p_1+p_2-q)J^{\gamma a_3}(p_3)J^{\delta a_4}(p_4)\rangle_q
		\\&
		-2i gf^{a_1 a_3 c} \delta^{[\alpha}_\nu \langle J^{\gamma] c}(p_1+p_3-q)J^{\beta a_2}(p_2)J^{\delta a_4}(p_4)\rangle_q
		-2i gf^{a_3 a_2 c} \delta^{[\gamma}_\nu \langle J^{\beta] c}(p_3+p_2-q)J^{\alpha a_1}(p_1)J^{\delta a_4}(p_4)\rangle_q\\&
		-2i gf^{a_1 a_4 c} \delta^{[\alpha}_\nu \langle J^{\delta] c}(p_1+p_4-q)J^{\gamma a_3}(p_3)J^{\beta a_2}(p_2)\rangle-2i gf^{a_4 a_2 c} \delta^{[\delta}_\nu \langle J^{\beta] c}(p_4+p_2-q)J^{\gamma a_3}(p_3)J^{\alpha a_1}(p_1)\rangle_q
		\\&
		-2i gf^{a_1 a_4 c} \delta^{[\alpha}_\nu \langle J^{\delta] c}(p_1+p_4-q)J^{\beta a_2}(p_2)J^{\gamma a_3}(p_3)\rangle_q -2i gf^{a_4 a_3 c} \delta^{[\delta}_\nu \langle J^{\gamma] c}(p_4+p_3-q)J^{\beta a_2}(p_2)J^{\alpha a_1}(p_1)\rangle_q		\\&-2i gf^{a_3 a_4 c} \delta^{[\gamma}_\nu \langle J^{\delta] c}(p_3+p_4-q)J^{\alpha a_1}(p_1)J^{\beta a_2}(p_2)\rangle_q -2i gf^{a_4 a_2 c} \delta^{[\delta}_\nu \langle J^{\beta] c}(p_4+p_2-q)J^{\alpha a_1}(p_1)J^{\gamma a_3}(p_3)\rangle_q.
	\end{aligned}
\end{equation}

\section{Secondary equations in the quark sector} 
\label{secc}
The secondary conformal Ward identities are first-order partial differential equations and, in principle, involve the semi-local information contained in $j_{\text{loc}}^\mu$ and $t^{\mu\nu}_{\text{loc}}$. To write them compactly, we define two differential operators:
\begin{align}
L_N &= p_1(p_1^2 + p_2^2 - p_3^2) \frac{\partial}{\partial p_1} + 2 p_1^2\, p_2 \frac{\partial}{\partial p_2} + \big[ (2d - \Delta_1 - 2\Delta_2 +N)p_1^2 + (2\Delta_1-d)(p_3^2-p_2^2) \big], \label{Ldef} \\
R &= p_1 \frac{\partial}{\partial p_1} - (2\Delta_1-d) \label{Rdef}\,. 
\end{align}
The reason for introducing these operators arises from the special conformal constraints, once the action of $K^\kappa$ is made explicit. The separation between the two sets of constraints comes from the same equation, particularly from the terms trilinear in the momenta within the square bracket. The symmetric versions of these operators are given by:
\begin{align}
&L'_N=L_N,\quad\text{with}\ p_1\leftrightarrow p_2\ \text{and}\ \Delta_1\leftrightarrow\Delta_2,\\
&R'=R,\qquad\text{with}\ p_1\mapsto p_2\ \text{and}\ \Delta_1\mapsto\Delta_2.
\end{align}

These operators depend on the conformal dimensions of the operators involved in the 3-point function under consideration, as well as on a single parameter \(N\), which is determined by the relevant Ward identity. In the case of the \(\braket{TJJ}\) correlator, the special conformal Ward identities (CWIs) map the transverse traceless sector onto itself. Projectors within this sector can be utilized to decompose the equations into primary and secondary components

\begin{align}
&\Pi^{\rho_1\sigma_1}_{\mu_1\nu_1}(p_1)\pi^{\rho_2}_{\mu_2}(p_2)\pi^{\rho_3}_{\mu_3}(p_3)\ \bigg[\,K^\kappa\,\braket{{t^{\mu_1\nu_1}(p_1)\,j^{\mu_2}(p_2)\,j^{\mu_3}(p_3)}}\bigg]\notag\\
&=\Pi^{\rho_1\sigma_1}_{\mu_1\nu_1}(p_1)\pi^{\rho_2}_{\mu_2}(p_2)\pi^{\rho_3}_{\mu_3}(p_3)\times\notag\\
&\times\bigg[p_1^\kappa\left(C_{11}\,p_1^{\mu_3}p_2^{\mu_1}p_2^{\nu_1}p_3^{\mu_2}+C_{12}\,\delta^{\mu_2\mu_3}p_2^{\mu_1}p_2^{\nu_1}+C_{13}\delta^{\mu_1\mu_2}p_2^{\nu_1}p_1^{\mu_3}+C_{14}\delta^{\mu_1\mu_3}p_2^{\nu_1}p_3^{\mu_2}+C_{15}\delta^{\mu_1\mu_2}\delta^{\nu_1\mu_3}\right)\notag\\
&\quad+p_2^\kappa\left(C_{21}\,p_1^{\mu_3}p_2^{\mu_1}p_2^{\nu_1}p_3^{\mu_2}+C_{22}\,\delta^{\mu_2\mu_3}p_2^{\mu_1}p_2^{\nu_1}+C_{23}\delta^{\mu_1\mu_2}p_2^{\nu_1}p_1^{\mu_3}+C_{24}\delta^{\mu_1\mu_3}p_2^{\nu_1}p_3^{\mu_2}+C_{25}\delta^{\mu_1\mu_2}\delta^{\nu_1\mu_3}\right)\notag\\	
&\quad+\delta^{\mu_1\k}\left(C_{31}\,p_1^{\mu_3}p_2^{\nu_1}p_3^{\mu_2}+C_{32}\,\delta^{\mu_2\mu_3}p_2^{\nu_1}+C_{33}\,\delta^{\mu_2\nu_1}p_1^{\mu_3}+C_{34}\,\delta^{\mu_3\nu_1}p_3^{\mu_2}\right)\notag\\
&\qquad+\delta^{\mu_2\k}\left(C_{41}\,p_1^{\mu_3}p_2^{\mu_1}p_2^{\nu_1}+C_{42}\,\delta^{\mu_1\mu_3}p_2^{\nu_1}\right)+\delta^{\mu_3\k}\left(C_{51}\,p_3^{\mu_2}p_2^{\nu_1}p_3^{\mu_2}+C_{52}\,\delta^{\mu_1\mu_2}p_2^{\nu_1}\right)\bigg]\label{StrucSWIS}.
\end{align}
At this stage, the \(C_{ij}\) represent differential equations governing the form factors \(A_1\), \(A_2\), \(A_3\), and \(A_4\) in the decomposition of the \(\langle t jj \rangle\) correlator, as described in \eqref{DecompTJJ}. These equations naturally divide into two distinct categories: the $primary$ and $secondary$ conformal Ward identities (CWIs). \\
The primary CWIs are characterized by second-order differential equations, arising as the coefficients of transverse or transverse-traceless tensors involving the momenta \(p_1^\kappa\) and \(p_2^\kappa\), where \(\kappa\) corresponds to the index of the conformal generator \(K^\kappa\). \\
In contrast, the secondary CWIs, derived from the remaining transverse or transverse-traceless components, are first-order differential equations. These secondary equations are associated with the following operators

\begin{equation}
\begin{split}
C_{31}&=-\frac{2}{p_1^2}\left[L_4 A_1+R A_3-R A_3(p_2\leftrightarrow p_3)\right],\\
C_{32}&=-\frac{2}{p_1^2}\left[L_2\,A_2-p_1^2(A_3-A_3(p_2\leftrightarrow p_3))\right],\\
C_{33}&=-\frac{1}{p_1^2}\left[L_4\,A_3-2R\,A_4\right],\\
C_{34}&=-\frac{1}{p_1^2}\left[L_4\,A_3(p_2\leftrightarrow p_3)+2R\,A_4-4p_1^2A_3(p_2\leftrightarrow p_3)\right],
\end{split}
\end{equation}
\begin{equation}
\begin{split}
C_{41}&=\frac{1}{p_2^2}\left[L'_3\,A_1-2R'A_2+2R'A_3\right],\\
C_{42}&=\frac{1}{p_2^2}\left[L'_1\,A_3(p_2\leftrightarrow p_3)+p_2^2(4A_2-2A_3)+2R'A_4\right],\\
C_{51}&=\frac{1}{p_3}\left[(L_4-L'_3)A_1-2(2d+R+R')A_2+2(2d+R+R')A_3(p_2\leftrightarrow p_3)\right],\\
C_{52}&=\frac{1}{p_3^2}\left[(L_2-L'_1)A_3-4p_3^2A_2+2p_3^2A_3(p_2\leftrightarrow p_3)+2(2d-2+R+R')A_4\right].
\end{split}
\end{equation}
while the remaining $C_{ij}$  are primary and generate the equations in \eqref{fst1}.
The secondary CWIs take the explicit form
\begin{equation}
\begin{split}
&C_{31}=C_{41}=C_{42}=C_{51}=C_{52}=0, \qquad C_{32}=\frac{16\,d\, c_{123}\,\Gamma_J}{p_1^2}\left[ \frac{1}{(p_3^2)^{\sigma_0}} - \frac{1}{(p_2^2)^{\sigma_0}} \right],\\[2ex]
& \hspace{1cm}C_{33}=\frac{16\,d\,c_{123}\,\Gamma_J}{p_1^2 (p_3^2)^{\sigma_0}}, \hspace{2.5cm} C_{34}=-\frac{16\,d\,c_{123}\,\Gamma_J }{p_1^2\,(p_2^2)^{\sigma_0}},
\end{split}
\end{equation}
where $\sigma_0=d/2-\Delta_2$. Explicitly 
 \begin{equation}
 \begin{split}
&L_4 A_1+R A_3^R-R A_3^R(p_2\leftrightarrow p_3)=0\\[1.5ex]
&L_2\,A_2^R-s\,(A_3^R-A_3^R(p_2\leftrightarrow p_3))=\sdfrac{16}{9}\pi^2e^2\left[3s_1\,B_0^R(s_1,0,0)-3s_2B_0^R(s_2,0,0)-s_1+s_2\right]+\sdfrac{24}{9}\pi^2 g^2s\\[1.5ex]
&L_4\,A^R_3-2R\,A^R_4=\sdfrac{32}{9}\pi^2\,g^2s_2\left[1-3B_0^R(s_2,0,0)\right]+\sdfrac{48}{9}\pi^2\,g^2\,s\\[1.5ex]
&L_4\,A^R_3(p_2\leftrightarrow p_3)+2R\,A^R_4-4\,sA^R_3(p_2\leftrightarrow p_3)=\sdfrac{32}{9}\,\pi^2\,g^2s_1\left[3B_0^R(s_1,0,0)-1\right]\\[1.5ex]
&L'_3\,A_1^R-2R'A_2^R+2R'A_3^R=0\\[1.5ex]
&L'_1\,A_3^R(p_2\leftrightarrow p_3)+p_2^2(4A^R_2-2A^R_3)+2R'A^R_4=\sdfrac{16}{3}\pi^2\,g^2\,s_1,\\
 \end{split}
 \end{equation}
where we have used the relations 
\begin{equation}
\sdfrac{\partial}{\partial s_i}\,B_0^R(s_j,0,0)=-\sdfrac{\d_{ij}}{s_i}\  i=0,1,2
\end{equation}
where $s_0=s$. 
\section{Relations between master integrals and 3K integrals} 
\label{dd}
Master integrals and 3K integrals are linked by the relation
\beq
\label{davy}
J(\nu_1,\nu_2,\nu_3) = \int \frac{d^d l}{(2 \pi)^d} \frac{1}{(l^2)^{\nu_3} ((l+p_1)^2)^{\nu_2} ((l-p_2)^2)^{\nu_1}}\, ,
\eeq
\beqa
J(\nu_1,\nu_2,\nu_3)
&=&\frac{\pi^{-d/2} 2^{4 - 3 d/2} }{\Gamma(\nu_1)\Gamma(\nu_2)\Gamma(\nu_3)\Gamma(d-\nu_1-\nu_2-\nu_3)} |p_1|^{d/2-\nu_2-\nu_3} |p_2|^{d/2-\nu_1-\nu_3}|p_3|^{d/2-\nu_1-\nu_2}\nonumber \\
&& \times \int_0^\infty d x x^{d/2-1}K_{d-\nu_1-\nu_3} (|p_1| x) K_{d-\nu_2-\nu_3} (|p_2| x) K_{d-\nu_1-\nu_2} (|p_3 |x) 
\eeqa
where $|p_i|$, in this case indicate the magnitudes of the momenta $p_i$, i.e. $|p_i|=\sqrt{p_i^2}$. 

\section{Form factors of the transverse traceless sector: $A_1,A_2,A_3,A_4$ for massless quarks} 
\label{ff}
We present here the explicit expressions of the coefficient functions $A_i$ in \eqref{Abar}  characterising the (tt) sector. The quark contributions are identified by the factor $n_f$, the gluon contributions by the factor $C_A$:

{\allowdisplaybreaks
\begin{eqnarray*}
	A_{10} &=& 16\, (p_1\cdot p_2)^7+36 \, p_1^2 \, (p_1\cdot p_2)^6+36 \, p_2^2 \, (p_1\cdot p_2)^6+12 \, p_1^4 \, (p_1\cdot p_2)^5+12 \, p_2^4 \, (p_1\cdot p_2)^5+240 \, p_1^2 \, p_2^2 \, (p_1\cdot p_2)^5 \notag \\ &&
	+188 \, p_1^2 \, p_2^4 \, (p_1\cdot p_2)^4+188 \, p_1^4 \, p_2^2 \, (p_1\cdot p_2)^4+46 \, p_1^2 \, p_2^6 \, (p_1\cdot p_2)^3-108 \, p_1^4 \, p_2^4 \, (p_1\cdot p_2)^3+46 \, p_1^6 \, p_2^2 \, (p_1\cdot p_2)^3 \notag \\ &&
	-204 \, p_1^4 \, p_2^6 \, (p_1\cdot p_2)^2-204 \, p_1^6 \, p_2^4 \, (p_1\cdot p_2)^2-20 \, p_1^6 \, p_2^8 \,-20 \, p_1^8 \, p_2^6 \,-58 \, p_1^4 \, p_2^8 \, (p_1\cdot p_2) \notag \\ &&
	-148 \, p_1^6 \, p_2^6 \, (p_1\cdot p_2)-58 \, p_1^8 \, p_2^4 \, (p_1\cdot p_2) \\
	A_{11} &=& -24 \, p_1^2 \,  \, (p_1\cdot p_2)^6-24 \, p_1^4 \,  \,\, (p_1\cdot p_2)^5-224 \, p_1^2 \,  \, p_2^2 \, (p_1\cdot p_2)^5-6\, p_1^6 \,  \,\, (p_1\cdot p_2)^4-228 \, p_1^2 \,  \, p_2^4 \, (p_1\cdot p_2)^4 \notag \\ &&
	-522 \, p_1^4 \,  \, p_2^2 \, (p_1\cdot p_2)^4- 60 \, p_1^2 \,  \, p_2^6 \, (p_1\cdot p_2)^3- 652 \, p_1^4 \,  \, p_2^4 \, (p_1\cdot p_2)^3-372 \, p_1^6 \,  \, p_2^2 \, (p_1\cdot p_2)^3  \notag \\ &&
	-279 \, p_1^4 \,  \, p_2^6 \, (p_1\cdot p_2)^2-474 \, p_1^6 \,  \, p_2^4 \, (p_1\cdot p_2)^2-83 \, p_1^8 \,  \, p_2^2 \, (p_1\cdot p_2)^2-18 \, p_1^6 \,  \, p_2^8 \,  \notag \\ &&
	-30 \, p_1^8 \,  \, p_2^6 \,-16 p_1^{10} \, p_2^4 \,-45 \, p_1^4 \,  \, p_2^8 \, (p_1\cdot p_2)-174 \, p_1^6 \,  \, p_2^6 \, (p_1\cdot p_2)-129 \, p_1^8 \,  \, p_2^4 \, (p_1\cdot p_2) \\
	A_{12} &=& -24 \, p_2^2 \, (p_1\cdot p_2)^6-24 \, p_2^4 \, (p_1\cdot p_2)^5-224 \, p_1^2 \,  \, p_2^2 \, (p_1\cdot p_2)^5-6 \, p_2^6 \, (p_1\cdot p_2)^4-522 \, p_1^2 \,  \, p_2^4 \, (p_1\cdot p_2)^4   \notag \\ &&
	-228 \, p_1^4 \,  \, p_2^2 \, (p_1\cdot p_2)^4-372 \, p_1^2 \,  \, p_2^6 \, (p_1\cdot p_2)^3-652 \, p_1^4 \,  \, p_2^4 \, (p_1\cdot p_2)^3-60 \, p_1^6 \,  \, p_2^2 \, (p_1\cdot p_2)^3-83 \, p_1^2 \,  \, p_2^8 \, (p_1\cdot p_2)^2   \notag \\ &&
	-474 \, p_1^4 \,  \, p_2^6 \, (p_1\cdot p_2)^2-279 \, p_1^6 \,  \, p_2^4 \, (p_1\cdot p_2)^2-16 \, p_1^4 \,  \, p_2^{10} \,-30 \, p_1^6 \,  \, p_2^8 \,-18 \, p_1^8 \,  \, p_2^6 \,-129 \, p_1^4 \,  \, p_2^8 \, (p_1\cdot p_2)   \notag \\ &&
	-174 \, p_1^6 \,  \, p_2^6 \, (p_1\cdot p_2)-45 \, p_1^8 \,  \, p_2^4 \, (p_1\cdot p_2) \\
	A_{13} &=& 24 \, p_1^2 \,  \, (p_1\cdot p_2)^6+24 \, p_2^2 \, (p_1\cdot p_2)^6+24 \, p_1^4 \,  \, (p_1\cdot p_2)^5+24 \, p_2^4 \, (p_1\cdot p_2)^5 + 448 \, p_1^2 \,  \, p_2^2 \, (p_1\cdot p_2)^5  \notag \\ &&
	+6 \, p_1^6 \,  \, (p_1\cdot p_2)^4+6 \, p_2^6 \, (p_1\cdot p_2)^4+750 \, p_1^2 \,  \, p_2^4 \, (p_1\cdot p_2)^4  +750 \, p_1^4 \,  \, p_2^2 \, (p_1\cdot p_2)^4+432 \, p_1^2 \,  \, p_2^6 \, (p_1\cdot p_2)^3  \notag \\ &&
	+ 1304 \, p_1^4 \,  \, p_2^4 \, (p_1\cdot p_2)^3+432 \, p_1^6 \,  \, p_2^2 \, (p_1\cdot p_2)^3  + 83 \, p_1^2 \,  \, p_2^8 \, (p_1\cdot p_2)^2+753 \, p_1^4 \,  \, p_2^6 \, (p_1\cdot p_2)^2+753 \, p_1^6 \,  \, p_2^4 \, (p_1\cdot p_2)^2  \notag \\ &&
	+83 \, p_1^8 \,  \, p_2^2 \, (p_1\cdot p_2)^2 + 16 \, p_1^4 \,  \, p_2^{10}+48 \, p_1^6 \,  \, p_2^8 \,+48 \, p_1^8 \,  \, p_2^6 \,+16 p_1^{10} \, p_2^4 \,  \notag \\ &&
	+174 \, p_1^4 \,  \, p_2^8 \, (p_1\cdot p_2)+348 \, p_1^6 \,  \, p_2^6 \, (p_1\cdot p_2)+174 \, p_1^8 \,  \, p_2^4 \, (p_1\cdot p_2)	\\
	A_{14} &=& 192 \, p_1^2 \,  \, p_2^2 \, (p_1\cdot p_2 )^6+432 \, p_1^2 \,  \, p_2^4 \, (p_1\cdot p_2 )^5+432 \, p_1^4 \,  \, p_2^2 \, (p_1\cdot p_2 )^5+288 \, p_1^2 \,  \, p_2^6 \, (p_1\cdot p_2 )^4+1152 \, p_1^4 \,  \, p_2^4 \, (p_1\cdot p_2 )^4  \notag \\ &&
	+288 \, p_1^6 \,  \, p_2^2 \, (p_1\cdot p_2 )^4+60 \, p_1^2 \,  \, p_2^8 \, (p_1\cdot p_2 )^3+936 \, p_1^4 \,  \, p_2^6 \, (p_1\cdot p_2 )^3+936 \, p_1^6 \,  \, p_2^4 \, (p_1\cdot p_2 )^3+60 \, p_1^8 \,  \, p_2^2 \, (p_1\cdot p_2 )^3  \notag \\ &&
	+324 \, p_1^4 \,  \, p_2^8 \, (p_1\cdot p_2 )^2+720 \, p_1^6 \,  \, p_2^6 \, (p_1\cdot p_2 )^2+324 \, p_1^8 \,  \, p_2^4 \, (p_1\cdot p_2 )^2+18 \, p_1^6 \,  \, p_2^{10} \,+36 \, p_1^8 \,  \, p_2^8 \notag \\ &&
	+18 p_1^{10} \, p_2^6 \,+45 \, p_1^4 \,  \, p_2^{10} (p_1\cdot p_2) +207 \, p_1^6 \,  \, p_2^8 \, (p_1\cdot p_2)+207 \, p_1^8 \,  \, p_2^6 \, (p_1\cdot p_2)+45 p_1^{10} \, p_2^4 \, (p_1\cdot p_2)
\end{eqnarray*}

\begin{eqnarray*}
	A_{20} &=& C_A \Big[-520 \, (p_1\cdot p_2)^6-372 \, p_1^2 \, (p_1\cdot p_2)^5-372 \, p_2^2 \, (p_1\cdot p_2)^5-60 \, p_1^4 \, (p_1\cdot p_2)^4-60 \, p_2^4 \, (p_1\cdot p_2)^4 \notag \\ &&
	+1056 \, p_1^2 \, p_2^2 \, (p_1\cdot p_2)^4+864 \, p_1^2 \, p_2^4 \, (p_1\cdot p_2)^3+864 \, p_1^4 \, p_2^2 \, (p_1\cdot p_2)^3+150 \, p_1^2 \, p_2^6 \, (p_1\cdot p_2)^2 \notag \\ &&
	-372 \, p_1^4 \, p_2^4 \, (p_1\cdot p_2)^2+150 \, p_1^6 \, p_2^2 \, (p_1\cdot p_2)^2-90 \, p_1^4 \, p_2^8 \,-164 \, p_1^6 \, p_2^6 \,-90 \, p_1^8 \, p_2^4 \notag \\ &&
	-492 \, p_1^4 \, p_2^6 \, (p_1\cdot p_2)-492 \, p_1^6 \, p_2^4 \, (p_1\cdot p_2) \Big]+ \notag \\ &&
	n_F\Big[ 40 (p_1\cdot p_2)^6+12 \, p_1^2 \, (p_1\cdot p_2)^5+12 \, p_2^2 \, (p_1\cdot p_2)^5-12 \, p_1^4 \, (p_1\cdot p_2)^4-12 \, p_2^4 \, (p_1\cdot p_2)^4 \notag \\ &&
	-192 \, p_1^2 \, p_2^2 \, (p_1\cdot p_2)^4-144 \, p_1^2 \, p_2^4 \, (p_1\cdot p_2)^3-144 \, p_1^4 \, p_2^2 \, (p_1\cdot p_2)^3-6 \, p_1^2 \, p_2^6 \, (p_1\cdot p_2)^2 \notag \\ &&
	+84 \, p_1^4 \, p_2^4 \, (p_1\cdot p_2)^2-6 \, p_1^6 \, p_2^2 \, (p_1\cdot p_2)^2+18 \, p_1^4 \, p_2^8 \,+68 \, p_1^6 \, p_2^6 \,+18 \, p_1^8 \, p_2^4 \notag \\ &&
	+132 \, p_1^4 \, p_2^6 \, (p_1\cdot p_2)+132 \, p_1^6 \, p_2^4 \, (p_1\cdot p_2) \Big] \\
	A_{21} &=& 3 \, p_1^2 \,  \, n_F \Big[-24 (p_1\cdot p_2)^5-20 \, p_2^2 \, (p_1\cdot p_2)^4+2 \, p_1^4 \,  \, (p_1\cdot p_2)^3+24 \, p_2^4 \, (p_1\cdot p_2)^3+90 \, p_1^2 \,  \, p_2^2 \, (p_1\cdot p_2)^3  \notag \\ &&
	+12 \, p_2^6 \, (p_1\cdot p_2)^2+162 \, p_1^2 \,  \, p_2^4 \, (p_1\cdot p_2)^2+54 \, p_1^4 \,  \, p_2^2 \, (p_1\cdot p_2)^2+3 \, p_1^2 \,  \, p_2^8 \,+8 \, p_1^4 \,  \, p_2^6 \,+21 \, p_1^6 \,  \, p_2^4 \,+51 \, p_1^2 \,  \, p_2^6 \, (p_1\cdot p_2)   \notag \\ &&
	+84 \, p_1^4 \,  \, p_2^4 \, (p_1\cdot p_2)+13 \, p_1^6 \,  \, p_2^2 \, (p_1\cdot p_2) \Big] \notag \\ &&
	-3 C_A \Big[ -48 (p_1\cdot p_2)^6-240 \, p_1^2 \,  \, (p_1\cdot p_2)^5-72 \, p_2^2 \, (p_1\cdot p_2)^5  \notag \\ &&
	-156 \, p_1^4 \,  \, (p_1\cdot p_2)^4 -24 \, p_2^4 \, (p_1\cdot p_2)^4-344 \, p_1^2 \,  \, p_2^2 \, (p_1\cdot p_2)^4-34 \, p_1^6 \,  \, (p_1\cdot p_2)^3-48 \, p_1^2 \,  \, p_2^4 \, (p_1\cdot p_2)^3   \notag \\ &&
	+54 \, p_1^4 \,  \, p_2^2 \, (p_1\cdot p_2)^3 +24 \, p_1^2 \,  \, p_2^6 \, (p_1\cdot p_2)^2+486 \, p_1^4 \,  \, p_2^4 \, (p_1\cdot p_2)^2+150 \, p_1^6 \,  \, p_2^2 \, (p_1\cdot p_2)^2+15 \, p_1^4 \,  \, p_2^8 \, \notag \\ &&
	+56 \, p_1^6 \,  \, p_2^6 \, +81 \, p_1^8 \,  \, p_2^4  +195 \, p_1^4 \,  \, p_2^6 \, (p_1\cdot p_2)+336 \, p_1^6 \,  \, p_2^4 \, (p_1\cdot p_2)+49 \, p_1^8 \,  \, p_2^2 \, (p_1\cdot p_2) \Big] \\
	A_{22} &=&  3 \, p_2^2 \, n_F \Big[-24 (p_1\cdot p_2)^5-20 \, p_1^2 \,  \, (p_1\cdot p_2)^4+24 \, p_1^4 \,  \, (p_1\cdot p_2)^3+2 \, p_2^4 \, (p_1\cdot p_2)^3+90 \, p_1^2 \,  \, p_2^2 \, (p_1\cdot p_2)^3  \notag \\ &&
	+12 \, p_1^6 \,  \, (p_1\cdot p_2)^2+54 \, p_1^2 \,  \, p_2^4 \, (p_1\cdot p_2)^2+162 \, p_1^4 \,  \, p_2^2 \, (p_1\cdot p_2)^2+21 \, p_1^4 \,  \, p_2^6 \,+8 \, p_1^6 \,  \, p_2^4 \,+3 \, p_1^8 \,  \, p_2^2 \,+13 \, p_1^2 \,  \, p_2^6 \, (p_1\cdot p_2)  \notag \\ &&
	+84 \, p_1^4 \,  \, p_2^4 \, (p_1\cdot p_2)+51 \, p_1^6 \,  \, p_2^2 \, (p_1\cdot p_2)\Big]-3 C_A \Big[-48 (p_1\cdot p_2)^6-72 \, p_1^2 \,  \, (p_1\cdot p_2)^5-240 \, p_2^2 \, (p_1\cdot p_2)^5  \notag \\ &&
	-24 \, p_1^4 \,  \, (p_1\cdot p_2)^4-156 \, p_2^4 \, (p_1\cdot p_2)^4-344 \, p_1^2 \,  \, p_2^2 \, (p_1\cdot p_2)^4-34 \, p_2^6 \, (p_1\cdot p_2)^3+54 \, p_1^2 \,  \, p_2^4 \, (p_1\cdot p_2)^3  \notag \\ &&
	-48 \, p_1^4 \,  \, p_2^2 \, (p_1\cdot p_2)^3+150 \, p_1^2 \,  \, p_2^6 \, (p_1\cdot p_2)^2+486 \, p_1^4 \,  \, p_2^4 \, (p_1\cdot p_2)^2+24 \, p_1^6 \,  \, p_2^2 \, (p_1\cdot p_2)^2+81 \, p_1^4 \,  \, p_2^8 \,  \notag \\ &&
	+56 \, p_1^6 \,  \, p_2^6 \,+15 \, p_1^8 \,  \, p_2^4 \,+49 \, p_1^2 \,  \, p_2^8 \, (p_1\cdot p_2)+336 \, p_1^4 \,  \, p_2^6 \, (p_1\cdot p_2)+195 \, p_1^6 \,  \, p_2^4 \, (p_1\cdot p_2)\Big] \\
	A_{23} &=& 3 C_A \, \Big(2 (p_1\cdot p_2)+\, p_1^2 \,  \,+\, p_2^2 \Big)^2 \, \Big[-44 (p_1\cdot p_2)^4-34 \, p_1^2 \,  \, (p_1\cdot p_2)^3-34 \, p_2^2 \, (p_1\cdot p_2)^3-22 \, p_1^2 \,  \, p_2^2 \, (p_1\cdot p_2)^2  \notag \\ &&
	+96 \, p_1^4 \,  \, p_2^4 \,+49 \, p_1^2 \,  \, p_2^4 \, (p_1\cdot p_2)+49 \, p_1^4 \,  \, p_2^2 \, (p_1\cdot p_2) \Big]  \notag \\ &&
	-3 n_F (2 (p_1\cdot p_2)+\, p_1^2 \,  \,+\, p_2^2 \,)^2  \Big[ -8 (p_1\cdot p_2)^4+2 \, p_1^2 \,  \, (p_1\cdot p_2)^3+2 \, p_2^2 \, (p_1\cdot p_2)^3+14 \, p_1^2 \,  \, p_2^2 \, (p_1\cdot p_2)^2  \notag \\ &&
	+24 \, p_1^4 \,  \, p_2^4 \,+13 \, p_1^2 \,  \, p_2^4 \, (p_1\cdot p_2)+13 \, p_1^4 \,  \, p_2^2 \, (p_1\cdot p_2) \Big]  \\
	A_{24} &=& 9\, C_A \Big(2 (p_1\cdot p_2)+\, p_1^2 \,  \,+\, p_2^2 \Big) \, \Big[ -32 (p_1\cdot p_2)^6-16 \, p_1^2 \,  \, (p_1\cdot p_2)^5-16 \, p_2^2 \, (p_1\cdot p_2)^5-8 \, p_1^4 \,  \, (p_1\cdot p_2)^4  \notag \\ &&
	-8 \, p_2^4 \, (p_1\cdot p_2)^4-24 \, p_1^2 \,  \, p_2^4 \, (p_1\cdot p_2)^3-24 \, p_1^4 \,  \, p_2^2 \, (p_1\cdot p_2)^3+8 \, p_1^2 \,  \, p_2^6 \, (p_1\cdot p_2)^2+20 \, p_1^4 \,  \, p_2^4 \, (p_1\cdot p_2)^2  \notag \\ &&
	+8 \, p_1^6 \,  \, p_2^2 \, (p_1\cdot p_2)^2+5 \, p_1^4 \,  \, p_2^8 \,+42 \, p_1^6 \,  \, p_2^6 \,+5 \, p_1^8 \,  \, p_2^4 \,+60 \, p_1^4 \,  \, p_2^6 \, (p_1\cdot p_2)+60 \, p_1^6 \,  \, p_2^4 \, (p_1\cdot p_2) \Big]  \notag \\ &&
	-9 \, p_1^2 \,  \, p_2^2 \, n_F  \Big( (2 (p_1\cdot p_2)+\, p_1^2 \,  \,+\, p_2^2 \, \Big)^2 \Big[-4 (p_1\cdot p_2)^3+4 \, p_1^2 \,  \, (p_1\cdot p_2)^2+4 \, p_2^2 \, (p_1\cdot p_2)^2   \notag \\ &&
	+\, p_1^2 \,  \, p_2^4 \,+\, p_1^4 \,  \, p_2^2 \,+14 \, p_1^2 \,  \, p_2^2 \, (p_1\cdot p_2) \Big]
\end{eqnarray*}

\begin{eqnarray*}
	A_{31} &=& 	3 \, p_1^2 \,  \, n_F \Big[-16 (p_1\cdot p_2)^5-4 \, p_1^2 \,  \, (p_1\cdot p_2)^4-70 \, p_2^2 \, (p_1\cdot p_2)^4-54 \, p_2^4 \, (p_1\cdot p_2)^3-44 \, p_1^2 \,  \, p_2^2 \, (p_1\cdot p_2)^3  \notag \\ &&
	-12 \, p_2^6 \, (p_1\cdot p_2)^2-15 \, p_1^2 \,  \, p_2^4 \, (p_1\cdot p_2)^2-15 \, p_1^4 \,  \, p_2^2 \, (p_1\cdot p_2)^2-3 \, p_1^2 \,  \, p_2^8 \,-5 \, p_1^4 \,  \, p_2^6 \,+4 \, p_1^6 \,  \, p_2^4 \,-6 \, p_1^2 \,  \, p_2^6 \, (p_1\cdot p_2)\Big]  \notag \\ &&
	-3 C_A\Big[-24 (p_1\cdot p_2)^6-100 \, p_1^2 \,  \, (p_1\cdot p_2)^5-12 \, p_2^2 \, (p_1\cdot p_2)^5-34 \, p_1^4 \,  \, (p_1\cdot p_2)^4-172 \, p_1^2 \,  \, p_2^2 \, (p_1\cdot p_2)^4  \notag \\ && 
	-84 \, p_1^2 \,  \, p_2^4 \, (p_1\cdot p_2)^3 -38 \, p_1^4 \,  \, p_2^2 \, (p_1\cdot p_2)^3-12 \, p_1^2 \,  \, p_2^6 \, (p_1\cdot p_2)^2+99 \, p_1^4 \,  \, p_2^4 \, (p_1\cdot p_2)^2-9 \, p_1^6 \,  \, p_2^2 \, (p_1\cdot p_2)^2-3 \, p_1^4 \,  \, p_2^8 \,  \notag \\ &&
	+7 \, p_1^6 \,  \, p_2^6 \,+28 \, p_1^8 \,  \, p_2^4 \,+36 \, p_1^4 \,  \, p_2^6 \, (p_1\cdot p_2)+78 \, p_1^6 \,  \, p_2^4 \, (p_1\cdot p_2)\Big] \\
	A_{32} &=& 3 \, p_2^2 \, n_F \Big[-24 (p_1\cdot p_2)^5-54 \, p_1^2 \,  \, (p_1\cdot p_2)^4-12 \, p_2^2 \, (p_1\cdot p_2)^4-18 \, p_1^4 \,  \, (p_1\cdot p_2)^3-2 \, p_2^4 \, (p_1\cdot p_2)^3 \notag \\ &&
	-62 \, p_1^2 \,  \, p_2^2 \, (p_1\cdot p_2)^3-55 \, p_1^2 \,  \, p_2^4 \, (p_1\cdot p_2)^2-3 \, p_1^4 \,  \, p_2^2 \, (p_1\cdot p_2)^2+7 \, p_1^4 \,  \, p_2^6 \,-3 \, p_1^6 \,  \, p_2^4 \,-13 \, p_1^2 \,  \, p_2^6 \, (p_1\cdot p_2) \notag \\ &&
	-4 \, p_1^4 \,  \, p_2^4 \, (p_1\cdot p_2)+3 \, p_1^6 \,  \, p_2^2 \, (p_1\cdot p_2)\Big] \notag \\ &&
	-3  C_A \Big[-24 (p_1\cdot p_2)^6-12 \, p_1^2 \,  \, (p_1\cdot p_2)^5-108 \, p_2^2 \, (p_1\cdot p_2)^5-42 \, p_2^4 \, (p_1\cdot p_2)^4 \notag \\ &&
	-156 \, p_1^2 \,  \, p_2^2 \, (p_1\cdot p_2)^4-2 \, p_2^6 \, (p_1\cdot p_2)^3-56 \, p_1^2 \,  \, p_2^4 \, (p_1\cdot p_2)^3-48 \, p_1^4 \,  \, p_2^2 \, (p_1\cdot p_2)^3-49 \, p_1^2 \,  \, p_2^6 \, (p_1\cdot p_2)^2 \notag \\ &&
	+111 \, p_1^4 \,  \, p_2^4 \, (p_1\cdot p_2)^2+31 \, p_1^4 \,  \, p_2^8 \,+9 \, p_1^6 \,  \, p_2^6 \,-13 \, p_1^2 \,  \, p_2^8 \, (p_1\cdot p_2)+74 \, p_1^4 \,  \, p_2^6 \, (p_1\cdot p_2)+45 \, p_1^6 \,  \, p_2^4 \, (p_1\cdot p_2)\Big] \notag \\ 
	A_{33} &=& 3 \, C_A \, \Big(2 (p_1\cdot p_2)+\, p_1^2 \,  \,+\, p_2^2 \Big) \, \Big[ -44 (p_1\cdot p_2)^5-34 \, p_1^2 \,  \, (p_1\cdot p_2)^4-38 \, p_2^2 \, (p_1\cdot p_2)^4-2 \, p_2^4 \, (p_1\cdot p_2)^3 \notag \\ &&
	-68 \, p_1^2 \,  \, p_2^2 \, (p_1\cdot p_2)^3-35 \, p_1^2 \,  \, p_2^4 \, (p_1\cdot p_2)^2-9 \, p_1^4 \,  \, p_2^2 \, (p_1\cdot p_2)^2+28 \, p_1^4 \,  \, p_2^6 \,+28 \, p_1^6 \,  \, p_2^4  \notag \\ &&
	-13 \, p_1^2 \,  \, p_2^6 \, (p_1\cdot p_2) +67 \, p_1^4 \,  \, p_2^4 \, (p_1\cdot p_2)\Big] \notag \\ &&
	-3 \, n_F \, \Big(2 (p_1\cdot p_2)+\, p_1^2 \,  \,+\, p_2^2 \Big)^2 \, \Big[-4 (p_1\cdot p_2)^4-2 \, p_2^2 \, (p_1\cdot p_2)^3-15 \, p_1^2 \,  \, p_2^2 \, (p_1\cdot p_2)^2+4 \, p_1^4 \,  \, p_2^4 \,-13 \, p_1^2 \,  \, p_2^4 \, (p_1\cdot p_2) \Big] \notag \\ 
	A_{34} &=& 9 \, C_A \, \Big(2 (p_1\cdot p_2)+\, p_1^2 \,  \,+\, p_2^2 \Big) \, \Big[-16 (p_1\cdot p_2)^6-4 \, p_1^2 \,  \, (p_1\cdot p_2)^5-4 \, p_2^2 \, (p_1\cdot p_2)^5-24 \, p_1^2 \,  \, p_2^4 \, (p_1\cdot p_2)^3 \notag \\ &&
	-16 \, p_1^4 \,  \, p_2^2 \, (p_1\cdot p_2)^3-4 \, p_1^2 \,  \, p_2^6 \, (p_1\cdot p_2)^2-14 \, p_1^4 \,  \, p_2^4 \, (p_1\cdot p_2)^2-\, p_1^4 \,  \, p_2^8 \,\notag \\ &&
	+15 \, p_1^6 \,  \, p_2^6 \,+13 \, p_1^4 \,  \, p_2^6 \, (p_1\cdot p_2)+15 \, p_1^6 \,  \, p_2^4 \, (p_1\cdot p_2)\Big] \notag \\ &&
	-9 \, p_1^2 \,  \, p_2^2 \, n_F \, \Big(2 (p_1\cdot p_2)+\, p_1^2 \,  \,+\, p_2^2 \Big)^2 \, \Big[-6 (p_1\cdot p_2)^3-4 \, p_2^2 \, (p_1\cdot p_2)^2-\, p_1^2 \,  \, p_2^4 \,+\, p_1^2 \,  \, p_2^2 \, (p_1\cdot p_2) \Big] \notag \\
\end{eqnarray*}

\begin{eqnarray*}
	A_{40} &=& C_A \Big[ -284 \, (p_1\cdot p_2)^5-132 \, p_1^2 \, (p_1\cdot p_2)^4-132 \, p_2^2 \, (p_1\cdot p_2)^4+556 \, p_1^2 \, p_2^2 \, (p_1\cdot p_2)^3 \notag \\ &&
	+258 \, p_1^2 \, p_2^4 \, (p_1\cdot p_2)^2+258 \, p_1^4 \, p_2^2 \, (p_1\cdot p_2)^2-126 \, p_1^4 \, p_2^6 \,-126 \, p_1^6 \, p_2^4 \,-272 \, p_1^4 \, p_2^4 \, (p_1\cdot p_2) \Big]  \notag \\ &&
	+ n_F \Big[ 44 (p_1\cdot p_2)^5+24 \, p_1^2 \, (p_1\cdot p_2)^4+24 \, p_2^2 \, (p_1\cdot p_2)^4-76 \, p_1^2 \, p_2^2 \, (p_1\cdot p_2)^3-42 \, p_1^2 \, p_2^4 \, (p_1\cdot p_2)^2 \notag \\ &&
	-42 \, p_1^4 \, p_2^2 \, (p_1\cdot p_2)^2+18 \, p_1^4 \, p_2^6 \,+18 \, p_1^6 \, p_2^4 \,+32 \, p_1^4 \, p_2^4 \, (p_1\cdot p_2) \Big] \\
	A_{41} &=& 3 \, p_1^2 \,  \, n_F \Big[-16 \, (p_1\cdot p_2)^4-4 \, p_1^2 \,  \, (p_1\cdot p_2)^3-18 \, p_2^2 \, (p_1\cdot p_2)^3-6 \, p_2^4 \, (p_1\cdot p_2)^2+10 \, p_1^2 \,  \, p_2^2 \, (p_1\cdot p_2)^2 \notag \\ &&
	+3 \, p_1^2 \,  \, p_2^6 \,-3 \, p_1^4 \,  \, p_2^4 \,+9 \, p_1^2 \,  \, p_2^4 \, (p_1\cdot p_2)+\, p_1^4 \,  \, p_2^2 \, (p_1\cdot p_2)\Big] \notag \\ &&
	-3 C_A\Big[-24 \, (p_1\cdot p_2)^5-76 \, p_1^2 \,  \, (p_1\cdot p_2)^4-12 \, p_2^2 \, (p_1\cdot p_2)^4-22 \, p_1^4 \,  \, (p_1\cdot p_2)^3-24 \, p_1^2 \,  \, p_2^2 \, (p_1\cdot p_2)^3 \notag \\ &&
	+76 \, p_1^4 \,  \, p_2^2 \, (p_1\cdot p_2)^2+9 \, p_1^4 \,  \, p_2^6 \,-9 \, p_1^6 \,  \, p_2^4 \,+39 \, p_1^4 \,  \, p_2^4 \, (p_1\cdot p_2)+19 \, p_1^6 \,  \, p_2^2 \, (p_1\cdot p_2)\Big] \\
	A_{42} &=&  3 \, p_2^2 \, n_F \Big[-16 (p_1\cdot p_2)^4-18 \, p_1^2 \,  \, (p_1\cdot p_2)^3-4 \, p_2^2 \, (p_1\cdot p_2)^3-6 \, p_1^4 \,  \, (p_1\cdot p_2)^2+10 \, p_1^2 \,  \, p_2^2 \, (p_1\cdot p_2)^2 \notag \\ &&
	-3 \, p_1^4 \,  \, p_2^4 \,+3 \, p_1^6 \,  \, p_2^2 \,+\, p_1^2 \,  \, p_2^4 \, (p_1\cdot p_2)+9 \, p_1^4 \,  \, p_2^2 \, (p_1\cdot p_2)\Big] \notag \\ &&
	-3 C_A\Big[-24 \, (p_1\cdot p_2)^5-12 \, p_1^2 \,  \, (p_1\cdot p_2)^4-76 \, p_2^2 \, (p_1\cdot p_2)^4-22 \, p_2^4 \, (p_1\cdot p_2)^3-24 \, p_1^2 \,  \, p_2^2 \, (p_1\cdot p_2)^3 \notag \\ &&
	+76 \, p_1^2 \,  \, p_2^4 \, (p_1\cdot p_2)^2-9 \, p_1^4 \,  \, p_2^6 \,+9 \, p_1^6 \,  \, p_2^4 \,+19 \, p_1^2 \,  \, p_2^6 \, (p_1\cdot p_2)+39 \, p_1^4 \,  \, p_2^4 \, (p_1\cdot p_2)\Big] \\
	A_{43} &=& 3 C_A \, (p_1\cdot p_2) \, \Big(2 (p_1\cdot p_2)+\, p_1^2 \,  \,+\, p_2^2 \Big)^2 \Big[ 19 \, p_1^2 \,  \, p_2^2 \,-22 (p_1\cdot p_2)^2 \Big] \notag \\ &&
	-3 n_F \, (p_1\cdot p_2) \, \Big(2 (p_1\cdot p_2)+\, p_1^2 \,  \,+\, p_2^2 \Big)^2 \Big[ p_1^2 \,  \, p_2^2 \,-4 (p_1\cdot p_2)^2 \Big] \\
	A_{44} &=& 9 C_A \Big(2 (p_1\cdot p_2)+\, p_1^2 \,  \,+\, p_2^2 \Big) \Big[-16 (p_1\cdot p_2)^5-4 \, p_1^2 \,  \, (p_1\cdot p_2)^4-4 \, p_2^2 \, (p_1\cdot p_2)^4+16 \, p_1^2 \,  \, p_2^2 \, (p_1\cdot p_2)^3 \notag \\ &&
	+3 \, p_1^4 \,  \, p_2^6 \,+3 \, p_1^6 \,  \, p_2^4 \,-2 \, p_1^4 \,  \, p_2^4 \, (p_1\cdot p_2) -9 \, p_1^2 \,  \, p_2^2 \, n_F (2 (p_1\cdot p_2)+\, p_1^2 \,  \,+\, p_2^2 \,)^2 \, p_1^2 \,  \, p_2^2 \,-2 (p_1\cdot p_2)^2 \Big]
\end{eqnarray*}
}

\subsection{Transverse traceless sector for quarks: massive contributions }\label{aiq}
For a single quark  of mass $m$ in the loop, the form factors of  \eqref{DecompTJJ} should be replaced by the expression

{\small
\begin{align}
    A^{(q)}_1= - \frac{g_s^2}{16\p^2}\Biggl(&\frac{P_1}{3 \bar{D}^3}+\frac{2 q^2 D_1(q,p_1,p_2)  \Sigma(q,m) }{3 \bar{D}^4}+\frac{P_3}{3 \bar{D}^3}-\frac{8   C_0 \left(-2 q^4+\left(p_1^2+p_2^2\right) q^2+\left(p_1^2-p_2^2\right)^2\right) m^4 }{\bar{D}^2}\nn\\ &-\frac{8   C_0 p_1^2 p_2^2 q^4  P_2}{\bar{D}^4} +m^2 \Biggl(\frac{ P_4 \Sigma(q,m) }{3 \bar{D}^3}-\frac{4  \left(-18 q^4+13 \left(p_1^2+p_2^2\right) q^2+5 \left(p_1^2-p_2^2\right)^2\right) }{3 \bar{D}^2}\nn\\&-\frac{8  q^2  C_0 P_5 }{\bar{D}^3}-\frac{4   \Sigma(p_1,m) E_1(q,p_1,p_2) }{3 p_1^2 \bar{D}^3}-\frac{4   \Sigma(p_2,m) E_1(q,p_2,p_1)}{3 p_2^2\bar{D}^3}\Biggl)\nn\\ &+\frac{2  \Sigma(p_1,m)  D_2(q,p_1,p_2)}{3 \bar{D}^4}+\frac{2    \Sigma(p_2,m)   D_2(q,p_2,p_1)}{3 \bar{D}^4}\Biggl)
\end{align}

\begin{align}
    A^{(q)}_2=- \frac{g_s^2}{16\p^2} \Biggl(& \frac{2}{3}   \log \left(\frac{\mu ^2}{m^2}\right) -\frac{4}{3}   \log (2 \pi ) +\frac{2}{3} \log (4 \pi ) +\frac{  \Sigma(p_1,m) D_3(q,p_1,p_2)}{3\bar{D}^3}+\frac{ \Sigma(p_2,m) D_3(q,p_2,p_1)}{3 \bar{D}^3}\nn\\&+\frac{ q^2  P_6 \Sigma(q,m)}{3 \bar{D}^3}-\frac{4  C_0 q^2 m^4 }{\bar{D}}-\frac{D_4(q,p_1,p_2)}{18 \bar{D}^2}-\frac{2  C_0 p_1^2 p_2^2 q^4 P_7 }{\bar{D}^3}\nn\\&+m^2 \Biggl(\frac{2   E_2(q,p_1,p_2)  \Sigma(p_1,m)}{3 p_1^2 \bar{D}^2}+\frac{2  E_2(q,p_2,p_1)  \Sigma(p_2,m)}{3 p_2^2 \bar{D}^2}+\frac{26  q^2 \left(p_1^2+p_2^2-q^2\right)   \Sigma(q,m) }{3 \bar{D}^2}
   \nn\\& -\frac{10   q^2 }{3 \bar{D}}-\frac{2  C_0 q^2 P_8}{\bar{D}^2}\Biggl)\Biggl)
\end{align}

\begin{align}
    A^{(q)}_3=- \frac{g_s^2}{16\p^2} \Biggl(& \frac{2  p_1^2 p_2^2  q^4 C_0 P_9}{\bar{D}^3}+\frac{4 m^4  \left(p_1^2-p_2^2-q^2\right)   C_0 }{\bar{D}}-\frac{2}{3}   \log \left(\frac{\mu ^2}{m^2}\right)+\frac{4}{3}  \log (2 \pi ) -\frac{2}{3}  \log (4 \pi ) \nn\\&-\frac{  \Sigma(p_1,m)   D_5(q,p_1,p_2)}{3 \bar{D}^3}-\frac{2  q^2 \Sigma(q,m) P_{10} }{3 \bar{D}^3}+m^2 \Biggl(\frac{10  \left(p_1^2-p_2^2-q^2\right) }{3 \bar{D}}\nn\\&+\frac{2   P_{11} \Sigma(p_1,m)}{3 p_1^2 \bar{D}^2}-\frac{2 q^2  C_0  P_{12}}{\bar{D}^2}-\frac{2   \Sigma(p_2,m) \left(-21 q^4+16 \left(p_1^2+p_2^2\right) q^2+5 \left(p_1^2-p_2^2\right)^2\right) }{3 \bar{D}^2}\nn\\&-\frac{2   \Sigma(q,m) \left(5 p_1^4+2 \left(8 p_2^2-5 q^2\right) p_1^2-\left(p_2^2-q^2\right) \left(21 p_2^2+5 q^2\right)\right) }{3 \bar{D}^2}\Biggl)\nn\\&-\frac{  \Sigma(p_2,m)  D_6(q,p_1,p_2)}{3 \bar{D}^3}+\frac{D_7(q,p_1,p_2)}{18 \bar{D}^2}\Biggl)
\end{align}

\begin{align}
    A^{(q)}_4=- \frac{g_s^2}{16\p^2}  \Biggl( &-\frac{2}{3}    \log (2 \pi ) \left(p_1^2+p_2^2-q^2\right) -2  m^4   C_0 +\frac{1}{3}  \left(p_1^2+p_2^2-q^2\right)   \log \left(\frac{\mu ^2}{m^2}\right) +\frac{1}{3}  \left(p_1^2+p_2^2-q^2\right)   \log (4 \pi )\nn\\&+\frac{  \Sigma(p_1,m) P_{13}}{6 \bar{D}^2}+\frac{ P_{14} \Sigma(p_2,m) }{6 \bar{D}^2}+\frac{ q^2 \left(p_1^2+p_2^2-q^2\right) \left(p_1^4+\left(p_2^2-2 q^2\right) p_1^2+\left(p_2^2-q^2\right)^2\right)  \Sigma(q,m) }{3 \bar{D}^2}\nn\\&-\frac{  C_0 p_1^2 p_2^2 q^4 \left(p_1^4-2 q^2 p_1^2+\left(p_2^2-q^2\right)^2\right)}{\bar{D}^2}-\frac{  P_{15} }{36 \bar{D}}+m^2 \Biggl(1+\frac{5  \left(p_1^2+p_2^2-q^2\right)   \Sigma(q,m) }{3 \bar{D}}\nn\\&-\frac{ C_0 \left(p_1^2+p_2^2-q^2\right)^2 q^2 }{\bar{D}}-\frac{\Sigma(p_1,m) \left(p_1^4+3 \left(p_2^2+q^2\right) p_1^2-4 \left(p_2^2-q^2\right)^2\right) }{3 p_1^2 \bar{D}}\nn\\&-\frac{ \Sigma(p_2,m) \left(-4 p_1^4+\left(3 p_2^2+8 q^2\right) p_1^2+\left(p_2^2-q^2\right) \left(p_2^2+4 q^2\right)\right) }{3 p_2^2 \bar{D}}\Biggl)
\end{align}

}

where

\begin{equation}
    \Sigma(p,m)=\log \left(\frac{2 m^2-p^2+\sqrt{p^2 \left(p^2-4 m^2\right)}}{2 m^2}\right) \sqrt{p^2 \left(p^2-4 m^2\right)}
\end{equation}

\begin{align}
    \bar{D}=p_1^4-2 \left(p_2^2+q^2\right) p_1^2+\left(p_2^2-q^2\right)^2
\end{align}

{\allowdisplaybreaks
\begin{align}
     D_1(q,p_1,p_2)=&\ 3 p_1^{10}+3 \left(27 p_2^2-4 q^2\right) p_1^8-2 \left(42 p_2^4+40 q^2 p_2^2-9 q^4\right) p_1^6-4 \left(21 p_2^6-76 q^2 p_2^4+20 q^4 p_2^2+3 q^6\right) p_1^4\nonumber\\&+\left(p_2^2-q^2\right)^2 \left(81 p_2^4+82 q^2 p_2^2+3 q^4\right) p_1^2+3 p_2^2 \left(p_2^2-q^2\right)^4
\nonumber\\
 D_2(q,p_1,p_2)=&-3 q^{12}+2 \left(6 p_1^2-19 p_2^2\right) q^{10}+\left(-18 p_1^4-41 p_2^2 p_1^2+121 p_2^4\right) q^8+4 \left(3 p_1^6+46 p_2^2 p_1^4-38 p_2^4 p_1^2-26 p_2^6\right) q^6\nonumber\\&-\left(p_1^2-p_2^2\right) \left(3 p_1^6+95 p_2^2 p_1^4+215 p_2^4 p_1^2+11 p_2^6\right) q^4-14 \left(p_1^2-p_2^2\right)^3 p_2^2 \left(p_1^2+p_2^2\right) q^2+p_2^2 \left(p_1^2-p_2^2\right)^5
\nonumber\\
    D_3(q,p_1,p_2)=&-3 q^{10}+5 \left(3 p_1^2+2 p_2^2\right) q^8+\left(-29 p_1^4+5 p_2^2 p_1^2+2 p_2^4\right) q^6+3 \left(p_1^2-p_2^2\right) \left(9 p_1^4+p_2^2 p_1^2+8 p_2^4\right) q^4\nn\\&-\left(12 p_1^2+17 p_2^2\right) \left(p_1^2-p_2^2\right)^3 q^2+2 \left(p_1^2-p_2^2\right)^5
\nonumber\\
    D_4(q,p_1,p_2)=&\ q^8(-29+12 \gamma ) -(-113+48 \gamma ) \left(p_1^2+p_2^2\right) q^6+(3 (-53+24 \gamma ) p_1^4+2 (-31+24 \gamma ) p_2^2 p_1^2\nn\\&+3 (-53+24 \gamma ) p_2^4) q^4-(-95+48 \gamma ) \left(p_1^2+p_2^2\right) \left(p_1^2-p_2^2\right)^2 q^2+4 (-5+3 \gamma ) \left(p_1^2-p_2^2\right)^4
\nonumber\\
    D_5(q,p_1,p_2)=&-4 q^{10}+3 \left(6 p_1^2-5 p_2^2\right) q^8+2 \left(-16 p_1^4+12 p_2^2 p_1^2+23 p_2^4\right) q^6+4 \left(7 p_1^6-3 p_2^2 p_1^4-12 p_2^4 p_1^2-8 p_2^6\right) q^4\nn\\&-6 \left(p_1^2-p_2^2\right)^2 \left(2 p_1^4+2 p_2^2 p_1^2-p_2^4\right) q^2+\left(p_1^2-p_2^2\right)^4 \left(2 p_1^2-p_2^2\right)
\nonumber\\
    D_6(q,p_1,p_2)=&-6 q^{10}+3 \left(p_1^2+8 p_2^2\right) q^8+2 \left(15 p_1^4-25 p_2^2 p_1^2-19 p_2^4\right) q^6+2 \left(-24 p_1^6+27 p_2^2 p_1^4+14 p_2^4 p_1^2+15 p_2^6\right) q^4\nn\\&+6 \left(p_1^2-p_2^2\right)^2 \left(4 p_1^4+p_2^2 p_1^2-2 p_2^4\right) q^2-\left(3 p_1^2-2 p_2^2\right) \left(p_1^2-p_2^2\right)^4
\nonumber\\
    D_7(q,p_1,p_2)=&\ q^8 (-35+12 \gamma )+2 \left(8 (8-3 \gamma ) p_1^2+(55-24 \gamma ) p_2^2\right) q^6+2 ((-87+36 \gamma ) p_1^4\nn\\&+(-73+24 \gamma ) p_2^2 p_1^2+6 (-11+6 \gamma ) p_2^4) q^4-2 \left(p_1^2-p_2^2\right) (4 (-13+6 \gamma ) p_1^4\nn\\&-27 p_2^2 p_1^2+(37-24 \gamma ) p_2^4) q^2+\left(p_1^2-p_2^2\right)^3 \left((-23+12 \gamma ) p_1^2+(17-12 \gamma ) p_2^2\right)
    \end{align}

\begin{align}
    E_1(q,p_1,p_2)=&\ 6 q^8+2 \left(9 p_1^2-8 p_2^2\right) q^6+3 \left(-17 p_1^4+15 p_2^2 p_1^2+4 p_2^4\right) q^4\nn\\&+6 p_1^2 \left(p_1^2-p_2^2\right) \left(4 p_1^2+11 p_2^2\right) q^2+\left(p_1^2-p_2^2\right)^3 \left(3 p_1^2+2 p_2^2\right)
\nonumber\\
    E_2(q,p_1,p_2)=& \ 6 q^6+\left(3 p_1^2-14 p_2^2\right) q^4-\left(11 p_1^2+10 p_2^2\right) \left(p_1^2-p_2^2\right) q^2+2 \left(p_1^2-p_2^2\right)^3
\end{align}

\begin{align}
    P_1=&\ 2 q^{10}-\left(p_1^2+p_2^2\right) q^8-2 \left(5 p_1^4-48 p_2^2 p_1^2+5 p_2^4\right) q^6\nn\\&+4 \left(p_1^2+p_2^2\right) \left(4 p_1^4-23 p_2^2 p_1^2+4 p_2^4\right) q^4
\nonumber\\
   P_2=&\ 3 p_1^8+\left(3 p_2^2-7 q^2\right) p_1^6+3 \left(-4 p_2^4+4 q^2 p_2^2+q^4\right) p_1^4\nn\\&+3 \left(p_2^6+4 q^2 p_2^4-6 q^4 p_2^2+q^6\right) p_1^2+\left(p_2^2-q^2\right)^3 \left(3 p_2^2+2 q^2\right)
\nonumber\\
    P_3=&-8\left( \left(p_1^2-p_2^2\right)^2 \left(p_1^4+4 p_2^2 p_1^2+p_2^4\right) q^2+\left(p_1^2-p_2^2\right)^4 \left(p_1^2+p_2^2\right)\right) 
\nonumber\\
    P_4=&\  18 q^6-15 \left(p_1^2+p_2^2\right) q^4-4 \left(6 p_1^4-25 p_2^2 p_1^2+6 p_2^4\right) q^2+21 \left(p_1^2-p_2^2\right)^2 \left(p_1^2+p_2^2\right)
\nonumber\\
    P_5=&\ 2 p_1^8+\left(4 p_2^2-5 q^2\right) p_1^6+\left(-12 p_2^4+11 q^2 p_2^2+3 q^4\right) p_1^4\nn\\&+\left(p_2^2-q^2\right) \left(4 p_2^4+15 q^2 p_2^2-q^4\right) p_1^2+\left(p_2^2-q^2\right)^3 \left(2 p_2^2+q^2\right)
\nonumber\\
    P_6=&\ 3 p_1^8+\left(24 p_2^2-11 q^2\right) p_1^6+\left(-54 p_2^4-31 q^2 p_2^2+15 q^4\right) p_1^4\nn\\&+\left(p_2^2-q^2\right) \left(24 p_2^4-7 q^2 p_2^2+9 q^4\right) p_1^2+\left(p_2^2-q^2\right)^3 \left(3 p_2^2-2 q^2\right)
\nonumber\\
    P_7=&\ -q^6+5 \left(p_1^2+p_2^2\right) q^4-7 \left(p_1^4+p_2^4\right) q^2+3 \left(p_1^2-p_2^2\right)^2 \left(p_1^2+p_2^2\right)
\nonumber\\
    P_8=&\	 p_1^6-\left(p_2^2+q^2\right) p_1^4-\left(p_2^4-6 q^2 p_2^2+q^4\right) p_1^2+\left(p_2^2-q^2\right)^2 \left(p_2^2+q^2\right)
\nonumber\\
    P_9=&\ 3 p_1^6+3 \left(p_2^2-3 q^2\right) p_1^4+\left(-5 p_2^4-8 q^2 p_2^2+9 q^4\right) p_1^2-\left(p_2^2-q^2\right)^2 \left(p_2^2+3 q^2\right)
\nonumber\\
    P_{10}=&\ p_1^8+2 \left(9 p_2^2-2 q^2\right) p_1^6+\left(-9 p_2^4-39 q^2 p_2^2+6 q^4\right) p_1^4\nn\\&-2 \left(p_2^2-q^2\right) \left(5 p_2^4+10 q^2 p_2^2-2 q^4\right) p_1^2-\left(p_2^2-q^2\right)^3 q^2
\nonumber\\
    P_{11}=&\ p_1^6+2 \left(p_2^2+q^2\right) p_1^4+\left(-7 p_2^4+34 q^2 p_2^2-7 q^4\right) p_1^2+4 \left(p_2^2-q^2\right)^2 \left(p_2^2+q^2\right)
\nonumber\\
    P_{12}=&-p_1^6+\left(3 q^2-7 p_2^2\right) p_1^4+\left(5 p_2^4+6 q^2 p_2^2-3 q^4\right) p_1^2\nn\\&+\left(p_2^2-q^2\right)^2 \left(3 p_2^2+q^2\right)
\nonumber\\
    P_{13}=&\ 4 q^8-7 \left(2 p_1^2+p_2^2\right) q^6+\left(18 p_1^4+5 p_2^2 p_1^2+3 p_2^4\right) q^4\nn\\&-\left(p_1^2-p_2^2\right) \left(10 p_1^4+p_2^2 p_1^2-p_2^4\right) q^2+\left(p_1^2-p_2^2\right)^3 \left(2 p_1^2-p_2^2\right)
\nonumber\\
    P_{14}=&\ p_1^8-\left(5 p_2^2+q^2\right) p_1^6+\left(9 p_2^4+2 q^2 p_2^2+3 q^4\right) p_1^4\nn\\&+\left(-7 p_2^6+9 q^2 p_2^4+5 q^4 p_2^2-7 q^6\right) p_1^2+2 \left(p_2^2-q^2\right)^3 \left(p_2^2-2 q^2\right)
\nonumber\\
    P_{15}=&\ q^6(35-12 \gamma ) +3 (-31+12 \gamma ) \left(p_1^2+p_2^2\right) q^4+(9 (9-4 \gamma ) p_1^4+2 (17-12 \gamma ) p_2^2 p_1^2\nn\\&+9 (9-4 \gamma ) p_2^4) q^2+(-23+12 \gamma ) \left(p_1^2-p_2^2\right)^2 \left(p_1^2+p_2^2\right).
\end{align}

}

\section{Appendix. Scalar integrals for massive quarks and on-shell gluons }
\label{scalars}
For the one-point function, or massive tadpole $\mathcal  A_0 (m^2)$, the massive bubble $ B_0 (s, m^2) $  and  the massive three-point function $\mathcal C_0 (s, s_1, s_2, m^2)$, with $s=q^2, s_1=p_1^2, s_2=p_2^2$.  We have defined
{\allowdisplaybreaks
\bea
\mathcal A_0 (m^2) &=& \frac{1}{i \pi^2}\int d^n l \, \frac{1}{l^2 - m^2}
= m^2 \left [ \frac{1}{\bar \eps} + 1 - \log \left( \frac{m^2}{\mu^2} \right )\right],\\
 B_0 (k^2, m^2) &=&  \frac{1}{i \pi^2} \int d^n l \, \frac{1}{(l^2 - m^2) \, ((l - k )^2 - m^2 )} \nn \\
 C_0 (s, s_1, s_2, m^2) &=&
 \frac{1}{i \pi^2} \int d^n l \, \frac{1}{(l^2 - m^2) \, ((l -q )^2 - m^2 ) \, ((l + p )^2 - m^2 )} \nn \\
&=&- \frac{1}{ \sqrt \sigma} \sum_{i=1}^3 \left[Li_2 \frac{b_i -1}{a_i + b_i}   - Li_2 \frac{- b_i -1}{a_i - b_i} + Li_2 \frac{-b_i +1}{a_i - b_i}  - Li_2 \frac{b_i +1}{a_i + b_i}
   \right],
\label{C0polylog}
\eea
with
\bea
a_i = \sqrt {1- \frac{4 m^2}{s_i }} \qquad \qquad
b_i = \frac{- s_i + s_j + s_k }{\sqrt{ \sigma}},
\eea
}
where $s_3=s$ and in the last equation $i=1,2,3$ and $j, k\neq i$. \\
The one-point and two-point functions  written before in  $n=4 - 2 \, \eps$ are divergent in dimensional regularization with the singular parts given by
\bea
\mathcal A_0 (m^2) ^{sing.}  \rightarrow  \frac{1}{\bar \eps} \, m^2,  \qquad \qquad
B_0 (s, m^2) ^{sing.}  \rightarrow  \frac{1}{\bar \eps} ,
\eea
with
\bea
\frac{1}{\bar \eps} = \frac{1}{\eps} - \g - \ln \pi
\label{bareps}
\eea
We use two finite combinations of scalar functions given by
\bea
&&  B_0 (s, m^2) \, m^2 - \mathcal A_0 (m^2) =  m^2 \left[ 1 - a_3 \log \frac{a_3 +1}{a_3 - 1}  \right] , \\
&& \mathcal D_i \equiv \mathcal D_i (s, s_i,  m^2) =
B_0 (s, m^2) -  B_0 (s_i, m^2) =  \left[ a_i \log\frac{a_i +1}{a_i - 1}
- a_3 \log \frac{a_3 +1}{a_3 - 1}  \right] \qquad i=1,2.
\label{D_i}
\nn \\
\eea
%---------------
The scalar integrals $ C_0 (s, 0,0,m^2) $ and $\mathcal D (s, 0, 0, m^2)$ are the $ \{ s_1 \rightarrow 0$, $s_2 \rightarrow 0 \} $ limits of the generic functions $C_0(s,s_1,s_2,m^2)$ and $\mathcal D_1(s,s_1,m^2)$
\bea
C_0 (s, 0,0,m^2) &=& \frac{1}{2 s} \log^2 \frac{a_3+1}{a_3-1}, \\
\mathcal D (s, 0, 0, m^2) &=& \mathcal D_1 (s,0,m^2)= \mathcal D_2(s,0,m^2) =
  \left[ 2 - a_3 \log \frac{a_3+1}{a_3-1}\right].
\eea
The singularities in \(1/\bar \epsilon\) and the dependence on the renormalization scale \(\mu\) cancel out when considering the difference between the two \(\mathcal{B}_0\) functions. As a result, the \(\mathcal{D}_i\)'s are well-defined, and the three-point master integral is convergent.

The renormalized scalar integrals  in the modified minimal subtraction scheme named $\overline{MS}$ are defined as
\beq
B_0^{\overline{MS}}(s,0) =  2 - L_s,\\
\eeq
\beq
 B_0^{\overline{MS}}(0,0) =  \frac{1}{\omega},\\
\label{B0masslessOSx}
\eeq
\beq
\mathcal C_0(s,0,0,0) = \frac{1}{s}\left[ \frac{1}{\omega^2} + \frac{1}{\omega} L_s + \frac{1}{2} L_s^2 - \frac{\pi^2}{12}  \right],
\label{C0masslessOS}
\eeq
where
\bea
L_s \equiv \log \left( - \frac{s}{\mu^2} \right ) \qquad \qquad s<0.
\eea
We have set  the space-time dimensions to $n = 4 + 2 \omega$ with $\omega > 0$. The $1/\omega$ and $1/\omega^2$ singularities in Eqs.~(\ref{B0masslessOSx}) and (\ref{C0masslessOS}) are infrared divergencies due to the zero mass of the gluons.

\section{The 13 Form factor decomposition}
\label{genbasis1}
\begin{table}
$$
\begin{array}{|c|c|}\hline
i & t_i^{\mu\nu\alpha\beta}(p,q)\\ \hline\hline
1 &
\left(k^2 g^{\mu\nu} - k^{\mu } k^{\nu}\right) u^{\alpha\beta}(p.q)\\ \hline
2 &
\left(k^2g^{\mu\nu} - k^{\mu} k^{\nu}\right) w^{\alpha\beta}(p.q)  \\ \hline
3 & \left(p^2 g^{\mu\nu} - 4 p^{\mu}  p^{\nu}\right)
u^{\alpha\beta}(p.q)\\ \hline
4 & \left(p^2 g^{\mu\nu} - 4 p^{\mu} p^{\nu}\right)
w^{\alpha\beta}(p.q)\\ \hline
5 & \left(q^2 g^{\mu\nu} - 4 q^{\mu} q^{\nu}\right)
u^{\alpha\beta}(p.q)\\ \hline
6 & \left(q^2 g^{\mu\nu} - 4 q^{\mu} q^{\nu}\right)
w^{\alpha\beta}(p.q) \\ \hline
7 & \left[p\cdot q\, g^{\mu\nu}
-2 (q^{\mu} p^{\nu} + p^{\mu} q^{\nu})\right] u^{\alpha\beta}(p.q) \\ \hline
8 & \left[p\cdot q\, g^{\mu\nu}
-2 (q^{\mu} p^{\nu} + p^{\mu} q^{\nu})\right] w^{\alpha\beta}(p.q)\\ \hline
9 & \left(p\cdot q \,p^{\alpha}  - p^2 q^{\alpha}\right)
\big[p^{\beta} \left(q^{\mu} p^{\nu} + p^{\mu} q^{\nu} \right) - p\cdot q\,
(g^{\beta\nu} p^{\mu} + g^{\beta\mu} p^{\nu})\big]  \\ \hline
10 & \big(p\cdot q \,q^{\beta} - q^2 p^{\beta}\big)\,
\big[q^{\alpha} \left(q^{\mu} p^{\nu} + p^{\mu} q^{\nu} \right) - p\cdot q\,
(g^{\alpha\nu} q^{\mu} + g^{\alpha\mu} q^{\nu})\big]  \\ \hline
11 & \left(p\cdot q \,p^{\alpha} - p^2 q^{\alpha}\right)
\big[2\, q^{\beta} q^{\mu} q^{\nu} - q^2 (g^{\beta\nu} q^ {\mu}
+ g^{\beta\mu} q^{\nu})\big]  \\ \hline
12 & \big(p\cdot q \,q^{\beta} - q^2 p^{\beta}\big)\,
\big[2 \, p^{\alpha} p^{\mu} p^{\nu} - p^2 (g^{\alpha\nu} p^ {\mu}
+ g^{\alpha\mu} p^{\nu})\big] \\ \hline
13 & \big(p^{\mu} q^{\nu} + p^{\nu} q^{\mu}\big)g^{\alpha\beta}
+ p\cdot q\, \big(g^{\alpha\nu} g^{\beta\mu}
+ g^{\alpha\mu} g^{\beta\nu}\big) - g^{\mu\nu} u^{\alpha\beta} \\
& -\big(g^{\beta\nu} p^{\mu}
+ g^{\beta\mu} p^{\nu}\big)q^{\alpha}
- \big (g^{\alpha\nu} q^{\mu}
+ g^{\alpha\mu} q^{\nu }\big)p^{\beta}  \\ \hline
\end{array}
$$
\caption{The basis of 13 fourth rank tensors satisfying the vector current conservation on the external lines with momenta $p$ and $q$. \label{genbasis}}
\end{table}

\label{ffdc}
The set of the $13$ tensors $t_i$ introuced in \cite{Giannotti:2008cv} is linearly independent for generic $k^2, p^2, q^2$
different from zero. Five of the $13$ are Bose symmetric, 
\be
t_i^{\mu\nu\alpha\beta}(p,q) = t_i^{\mu\nu\beta\alpha}(q,p)\,,\qquad i=1,2,7,8,13\,,
\ee
while the remaining eight tensors are Bose symmetric pairwise
\bea
\label{pair}
&&t_3^{\mu\nu\alpha\beta}(p,q) = t_5^{\mu\nu\beta\alpha}(q,p)\,,\\
&&t_4^{\mu\nu\alpha\beta}(p,q) = t_6^{\mu\nu\beta\alpha}(q,p)\,,\\
&&t_9^{\mu\nu\alpha\beta}(p,q) = t_{10}^{\mu\nu\beta\alpha}(q,p)\,,\\
&&t_{11}^{\mu\nu\alpha\beta}(p,q) = t_{12}^{\mu\nu\beta\alpha}(q,p)\,.
\eea
In the set are present two tensor structures
\bes\bea
&&u^{\alpha\beta}(p,q) \equiv (p\cdot q) g^{\alpha\beta} - q^{\alpha}p^{\beta}\,,\\
&&w^{\alpha\beta}(p,q) \equiv p^2 q^2 g^{\alpha\beta} + (p\cdot q) p^{\alpha}q^{\beta}
- q^2 p^{\alpha}p^{\beta} - p^2 q^{\alpha}q^{\beta}\,,
\eea \label{uwdef}\ees
which appear in $t_1$ and $t_2$ respectively.
Each of them satisfies the Bose symmetry  requirement,
\bes\bea
&&u^{\alpha\beta}(p,q) = u^{\beta\alpha}(q,p)\,,\\
&&w^{\alpha\beta}(p,q) = w^{\beta\alpha}(q,p)\,,
\eea\ees
and vector current conservation,
\bes\bea
&&p_{\alpha} u^{\alpha\beta}(p,q) = 0 = q_{\beta}u^{\alpha\beta}(p,q)\,,\\
&&p_{\alpha} w^{\alpha\beta}(p,q) = 0 = q_{\beta}w^{\alpha\beta}(p,q)\,.
\eea\ees
They are obtained from the variation of gauge invariant quantities
$F_{\mu\nu}F^{\mu\nu}$ and $(\partial_{\mu} F^{\mu}_{\ \,\lambda})(\partial_{\nu}F^{\nu\lambda})$

\bea
&&u^{\alpha\beta}(p,q) = -\frac{1}{4}\int\,d^4x\,\int\,d^4y\ e^{ip\cdot x + i q\cdot y}\ 
\frac{\delta^2 \{F_{\mu\nu}F^{\mu\nu}(0)\}} {\delta A_{\alpha}(x) A_{\beta}(y)} \,,
\label{one}\\
&&w^{\alpha\beta}(p,q) = \frac{1}{2} \int\,d^4x\,\int\,d^4y\ e^{ip\cdot x + i q\cdot y}\
\frac{\delta^2 \{\partial_{\mu} F^{\mu}_{\ \,\lambda}\partial_{\nu}F^{\nu\lambda}(0)\}} 
{\delta A_{\alpha}(x) A_{\beta}(y)}\,.\label{two}
\eea\label{three}
 All the $t_i$'s are transverse in their photon indices
\bea
q^\alpha t_i^{\mu\nu\alpha\beta}=0  \qquad p^\beta t_i^{\mu\nu\alpha\beta}=0.
\eea
$t_2\ldots t_{13}$ are traceless, $t_1$ and $t_2$ have trace parts in $d=4$. The corresponding form factors $F_i$ are related to the  $A^{(q)}_i$. 
They are conveniently expressed in terms of the momenta $(p_1,p_2,p_3)$ in the form
 \bea
A_1^{(q)}&=&4(F_7-F_3-F_5)-2p_2^2F_9-2p_3^2F_{10}\nn \\
A_2^{(q)}&=&2(p_1^2-p_2^2-p_3^2)(F_7-F_5-F_3)-4p_2^2p_3^2(F_6-F_8+F_4)-2F_{13}\nn\\
A_3^{(q)}&=&p_3^2(p_1^2-p_2^2-p_3^2)F_{10}-2p_2^2\,p_3^2 F_{12}-2F_{13}\nn\\
A_3^{(q)}(p_2\leftrightarrow p_3)&=&p_2^2(p_1^2-p_2^2-p_3^2)F_9-2p_2^2p_3^2F_{11}-2F_{13}\nn\\
A_4^{(q)}&=&(p_1^2-p_2^2-p_3^2)F_{13}.
\label{mapping1}
\eea
 
\providecommand{\href}[2]{#2}\begingroup\raggedright\endgroup

%\bibliographystyle{jhep}
%\bibliography{TJJdilatonHprime}
\end{document}